\documentclass[twocolumn,tighten]{aastex631}
\hypersetup{linkcolor=red,citecolor=blue,filecolor=cyan,urlcolor=magenta}
\usepackage{amsmath}
\usepackage{natbib}
\usepackage{todonotes}
\usepackage{CJK}

\newcommand{\Gaia}{\textit{Gaia}}
\newcommand{\degree}{$^{\circ}$}
\newcommand{\kmsec}{\mbox{km~s$^{\rm -1}$}}
\newcommand{\kmseckpc}{\mbox{km~s$^{\rm -1}$~kpc$^{\rm -1}$}}
\newcommand{\kmsecdeg}{\mbox{km~s$^{\rm -1}$~deg$^{\rm -1}$}}
\newcommand{\masyr}{\mbox{mas~yr$^{\rm -1}$}}
\newcommand{\msun}{\mbox{$M_{\odot}$}}
\newcommand{\lsun}{\mbox{$L_{\odot}$}}

\newcommand{\RN}[1]{%
  \textup{\uppercase\expandafter{\romannumeral#1}}%
}

\shorttitle{Signatures of tidal disruption of Hercules}
\shortauthors{Ou et al.} 

\begin{document}
\begin{CJK*}{UTF8}{gbsn}

\title{Signatures of tidal disruption of the Hercules ultra-faint dwarf galaxy\footnote{This paper includes data gathered with the 6.5 meter Magellan Telescopes located at Las Campanas Observatory, Chile.}}

\author[0000-0002-4669-9967]{Xiaowei Ou (欧筱葳)} 
\affiliation{%
Department of Physics and MIT Kavli Institute for Astrophysics and Space Research, \\ 
77 Massachusetts Avenue, Cambridge, MA 02139, USA}
\email{Email:\ xwou@mit.edu}

\author[0000-0002-7155-679X]{Anirudh Chiti}
\affil{Department of Astronomy \& Astrophysics, University of Chicago, 5640 S Ellis Avenue, Chicago, IL 60637, USA}
\affil{Kavli Institute for Cosmological Physics, University of Chicago, Chicago, IL 60637, USA}

\author[0000-0003-2497-091X]{Nora Shipp}
\affiliation{McWilliams Center for Cosmology, Department of Physics, Carnegie Mellon University, 5000 Forbes Ave, Pittsburgh, PA 15213, USA}

\author[0000-0002-4733-4994]{Joshua D.\ Simon}
\affiliation{Observatories of the Carnegie Institution for Science, \\
813 Santa Barbara Street, Pasadena, CA 91101, USA}

\author[0000-0002-7007-9725]{Marla Geha}
\affiliation{Astronomy Department, Yale University, New Haven, CT 06520, USA}

\author[0000-0002-2139-7145]{Anna Frebel}
\affiliation{Department of Physics and MIT Kavli Institute for Astrophysics and Space Research, \\ 
77 Massachusetts Avenue, Cambridge, MA 02139, USA}

\author[0000-0001-9178-3992]{Mohammad K.\ Mardini}
\affiliation{Department of Physics, Zarqa University, Zarqa 13110, Jordan}
\affiliation{Jordanian Astronomical Virtual Observatory, Zarqa University, Zarqa 13110, Jordan}
\affiliation{Department of Physics and MIT Kavli Institute for Astrophysics and Space Research, \\ 
77 Massachusetts Avenue, Cambridge, MA 02139, USA}
\affiliation{Joint Institute for Nuclear Astrophysics -- Center for the Evolution of the Elements (JINA-CEE), USA}

\author[0000-0002-8448-5505]{Denis Erkal}
\affiliation{Department of Physics, University of Surrey, Guildford GU2 7XH, UK}

\author[0000-0003-2806-1414]{Lina~Necib}
\affiliation{Department of Physics and MIT Kavli Institute for Astrophysics and Space Research, \\
77 Massachusetts Avenue, Cambridge, MA 02139, USA}
\affiliation{The NSF AI Institute for Artificial Intelligence and Fundamental Interactions, \\
77 Massachusetts Avenue, Cambridge, MA 02139, USA}

\begin{abstract}

The Hercules ultra-faint dwarf galaxy (UFD) has long been hypothesized to be tidally disrupting, yet no conclusive evidence has been found for tidal disruption owing partly to difficulties in identifying Hercules member stars. In this work, we present a homogeneous re-analysis of new and existing observations of Hercules, including the detection of a new potential member star located $\sim$1\,\degree ($\sim1.7$ kpc) west of the center of the system. In addition to measuring the line-of-sight velocity gradient, we compare predictions from dynamical models of stream formation to these observations.
We report an updated velocity dispersion measurement based on 28 stars, $1.9^{+0.6}_{-0.6}$~\kmsec, which is significantly lower than previous measurements. 
We find that the line-of-sight velocity gradient is $1.8^{+1.8}_{-1.8}$~\kmseckpc along the major axis of Hercules, consistent with zero within 1\,$\sigma$. 
Our dynamical models of stream formation, on the other hand, can reproduce the morphology of the Hercules UFD, specifically the misalignment between the elongation and the orbital motion direction. 
Additionally, these dynamical models indicate that any radial velocity gradient from tidal disruption would be too small, $0.00^{+0.97}_{-0.91}$~\kmseckpc, to be detectable with current sample sizes.
Combined with our analysis of the tidal radius evolution of the system as a function of its orbital phase, we argue that it is likely that Hercules is indeed currently undergoing tidal disruption in its extended stellar halo with a line-of-sight velocity gradient too small to be detected with current observational datasets.

\end{abstract}


\section{Introduction}

Ultra-faint dwarf galaxies (UFDs) are a class of stellar systems orbiting the Milky Way that have recently been discovered by the advent of large wide-field digital sky surveys. 
The first UFDs, Ursa Major and Willman~1, were discovered in the Sloan Digital Sky Survey (SDSS; \citealt{york20}) by \citet{willman05a,willman05b}, initiating the next two decades of further discovery. 
Efforts with subsequent sky surveys, such as the Dark Energy Survey (DES; \citealt{bechtol15,koposov15,drlicawagner15,kim15}), Pan-STARRS (\citealt{laevens15a,laevens15b}), MagLITeS (\citealt{drlicawagner16,torrealba18}), HSC (\citealt{homma16,homma18,homma19}), and DELVE (\citealt{mau20,cerny21a,cerny21b,cerny22}) have led to the current census of nearly 60 such satellites (for a recent review, see \citealt{simon19}). 

UFDs are of particular scientific interest as they are the least massive (extrapolated virial mass $\lesssim 10^9$~\msun; \citealt{simon07,strigari18}) and most dark matter dominated (M/L $> 100$~\msun/\lsun; \citealt{simon07,geha09}) galaxies known.
As such, they can be used to test predictions from models of galaxy formation (e.g., $\Lambda$CDM).
The low mass regime of the halo mass function, where the UFDs reside, is sensitive to the nature of dark matter particles \citep{jethwa18,kim18,nadler19,mau22}. 
In addition to the mass, studying the internal structure (e.g., density profile) of UFDs' dark matter halos tests dark matter models with different interactions with the baryonic matter on small scales (see, e.g., \citealt{calabrese16,errani18,bozek19,sales22,silverman23}).

Stellar tracers are key to obtaining mass estimates and interpreting the internal structures of the Milky Way's UFDs. 
Assuming dynamical equilibrium, the velocity dispersion of member stars can be used to estimate the dynamical mass of the UFD \citep[e.g.,][]{wolf10}. 
When combined with positional information, the stellar tracers can also map out the enclosed mass of the system as a function of distance from the center, thus providing a direct probe to the dark matter density profile of the UFD \citep[e.g.][]{cn+21,guerra23}.
The kinematics of stars in UFDs thus play a critical role in bridging the gap between observed and theoretical dark matter halo properties at a scale currently not as well constrained by either most other observational probes or simulations \citep{simon19, battaglia22a}. 

One of the more intriguing aspects of UFDs in the Milky Way is their potential tidal interaction with the Milky Way dark matter halo \citep{collins17,li18,fattahi18}, opening another doorway into studying their total halo mass and internal structure. 
The tidal radius of a UFD marks the distance from the center of the system beyond which its mass becomes tidally stripped.
Consequently, bound stellar tracers at large distances can better constrain the enclosed mass \citep[e.g.,][]{chiti21}, which makes it valuable to identify and derive kinematic information of stellar tracers at large radii.
A UFD with a given mass can also have a large tidal radius if its central density is high.
The general nature of central densities in UFDs is a question under debate, known as the core/cusp problem.
$\Lambda$CDM simulations generally predict density profiles with diverging central densities (``cuspy'') at all mass scales \citep{dubinski91,navarro97}, whilst observations tentatively prefer profiles with constant central densities (``cored'') in dwarf galaxies \citep{moore94,walker11,amorisco13,amorisco17b,contenta18,read19}, although cuspy dwarf galaxies are also observed and it is often difficult to distinguish whether a given dwarf galaxy is  ``cuspy'' or ``cored'' \citep{strigari10,jardel13,massari20}.
In the context of host-satellite galaxy interactions, assuming the same orbital properties, UFDs with more cuspy dark matter profiles are less likely to undergo tidal disruption, while a more cored profile has a smaller tidal radius and the system is more likely to be tidally disrupted and exhibit deformation in phase space \citep{penarrubia08}.
It is, thus, of great interest to study UFDs that show signs of currently experiencing or previously experienced tidal disruption to investigate the mass and internal structure of such small dark matter halos.

Hercules is a UFD first identified by \citet{belokurov07}, located $\sim$130 kpc away from the Sun \citep{musella12,mutlu-pakdil20}.
The system exhibits a highly elongated shape with a $3:1$ axis ratio \citep{coleman07,martin08,sand09}, which has been argued as an indicator of ongoing/past tidal disruption. 
However, the system is also predicted to have a line-of-sight velocity gradient if it has experienced tidal disruption, either in the tidal ``explosion'' scenario \citep{kupper17} or in the tidal stream scenario \citep{martin10}.
In the former case, the velocity gradient is expected to be present along the minor axis, whereas a tidal stream scenario predicts one along the major axis. 
Numerous studies have been carried out testing the two cases with no conclusive evidence of tidal disruption \citep{martin10,simon07,aden09}.
\citet{fu19} and \citet{gregory20} found a systemic orbit that is inconsistent with either of the tidal disruption scenarios using proper motion measurements from {\Gaia} DR2 for the known Hercules members. 

To better study and constrain velocity gradients, it is ideal to have member stars with spectroscopic/kinematic information at large spatial separations (more than a few times the half-light radius) from the center of Hercules. 
\citet{roderick15} identified several overdensities both along and perpendicular to the major axis of Hercules, along with eight blue horizontal branch stars identified by \citet{deason12} and three RR Lyrae stars from \citet{garling18} as potential members that are outside of the tidal radius of Hercules.
When combined with proper motion measurements, however, \citet{fu19} and \citet{gregory20} find no members of Hercules outside of the tidal radius matching the systemic proper motion of the UFD.
Moreover, the line-of-sight velocity gradient remained inconclusive, with most member stars around or within the half-light radius.
In addition to a lack of distant stars, not all member stars studied in previous works are confirmed with joint photometric, spectroscopic, and astrometric information, leaving such measurements susceptible to contamination from foreground Milky Way stars.

In this study, we resolve these issues by presenting the largest clean sample of Hercules member stars to-date, combining measurements from previous studies with new spectroscopic data obtained with Magellan/IMACS and Magellan/MagE.
Thanks to the advent of the third data release of the {\Gaia} mission \citep{gaia16,gaia21}, we compile a sample of 33 stars that are confirmed with photometric, spectroscopic (kinematics + metallicity), and proper motion data.
It is necessary to apply such selections, as Hercules is not well separated from the Galactic halo in line-of-sight velocity space and is particularly prone to foreground contamination.
We additionally remove potential binary stars using multi-epoch line-of-sight velocity measurements, since they may artificially inflate the velocity dispersion and subsequent dynamical mass estimate \citep{simon19}.
Our sample also provides the tightest constraint so far on the systemic proper motion of Hercules. 
We re-analyze the orbit of the UFD while accounting for the effect of the Large Magellanic Cloud.
Most importantly, we present two distant members of Hercules: Herc-1 and Herc-12. 
Herc-1 is located $\sim$1\,\degree ($\sim1.7$ kpc) west of the center of the system, discovered in this study. 
Herc-12 is located $\sim$1\,\degree north of the center of the system, also discovered by \citet{longeard23} as described below.
These recently identified members, given their relative locations, tighten the constraint on the line-of-sight velocity gradient.

While preparing this manuscript, \citet{longeard23} published a dedicated search for Hercules member stars in the outskirts of the galaxy through the Pristine Dwarf-Galaxy Survey.
We thus include a comparison with \citet{longeard23} in Section~\ref{sec:comp_pristine}.

This paper is structured as follows: We describe the observation and data reduction process for the new identified members from MagE and IMACS in Section~\ref{sec:observation}, along with a discussion of re-analyzed archival DEIMOS spectra. We present the full sample used for this study in Section~\ref{sec:data}, specifically highlighting the binary tests and systematics across datasets (Section~\ref{sec:binary}), membership selection process (Section~\ref{sec:membership}), and estimation of foreground contamination (Section~\ref{sec:foreground}). We present the dynamical analysis in Section~\ref{sec:analysis} and interpret our results in Section~\ref{sec:discussion}.

\section{Observations}
\label{sec:observation}

We introduce new observations from two spectrographs in this work: Magellan/IMACS \citep{dhb+06}; Magellan/MagE \citep{mbt+08}, and present a re-analysis of archival Keck/DEIMOS data studied in \citet{simon07, brown14, gregory20} via a new reduction and analysis pipeline (M. Geha et al. 2024, in prep). 
The velocity measurements from \citet{brown14} were not previously published.
In this Section, we summarize the new observations, and describe the data reduction procedures for each of the aforementioned instruments. 

\subsection{IMACS}
\label{sec:IMACS}
We observed Hercules with one multi-slit mask on July 23-25 2015 for 7.75\,hrs and April 11-13 2018 for 9.58\,hrs with Magellan/IMACS. 
Observations were obtained in series of 2-3 science exposures of 1800\,s to 3300\,s, followed by an arc frame for wavelength calibration, and a flatfield frame using a quartz lamp to trace the spectra on the chips. 
We used the 0\farcs7 slit, the 1200\,$\ell$\,mm$^{-1}$ grating, at an angle of 32\fdg4 degrees, which granted a resolution of $R\sim11,000$ covering $\sim7500\,${\AA} to $\sim9000$\,{\AA}.
In the 2015 observations, a HeNeAr arc lamp was used; whereas in the 2018 observations, a KrHeNeAr arc lamp was used due to the increased number of Kr lines at the blue end of our wavelength range. 
The IMACS data were reduced exactly following \citet{sld+17} and \citet{lsd+17}, using the COSMOS pipeline \citep{dbh+11, ock+17} for 2d spectrum extraction and an initial wavelength solution, and then an adapted version of the DEIMOS pipeline \citep{cnd+12, ncd+13} for 1d spectrum extraction and final wavelength calibration.

Targets were selected for the multislit mask by overlaying a [Fe/H] = $-2.5$, 12.5\,Gyr Dartmouth isochrone \citep{dcj+08} on a color-magnitude diagram of Hercules from public SDSS DR12 g,r photometry \citep{york20, sdss+06, sdss+11,sdss+15}\footnote{http://skyserver.sdss.org/dr12/en/home.aspx}.
Stars that were within 0.1\,mag of the isochrone and brighter than g$=22$ were identified as possible candidates for the multi-slit mask.
We note that as the slitmask was designed in 2015, we prioritized targets that were not published members at the time, and then added additional slits to include some known members to test for binarity.
No proper motion information went into the target selection, as this mask was designed before any \textit{Gaia} data releases (e.g., \citealt{gaia+16}).
In total, there were 33 slits for science targets on the mask, of which 25 produced spectra with sufficient signal-to-noise (S/N $> 3$) to be usable in 2015.
The 2018 observations resulted in 33 spectra that met that minimum S/N threshold.

\subsection{MagE}
\label{sec:MagE}

We observed five candidate members of Hercules at distances of $\sim$1 to $\sim$7 half-light radii (half-light radius assumed to be 243\,pc; \citealt{sand09}), out to {$\sim1$\,\degree} from the center of the system on July 2 2022 using the Magellan/MagE spectrograph \citep{mbt+08}.
Each candidate was observed for 30\,min to 55\,min, followed by a ThAr lamp frame to ensure a stable wavelength calibration.
We used the 1\farcs0 slit which yielded a resolution of $R\sim4000$ and a usable spectral wavelength coverage between $\sim3600$\,{\AA} and $\sim9000$\,{\AA}.
The reduced spectra had a signal-to-noise of $\sim$7 at 3900\,\AA, $\sim$22 at 6500\,\AA, and $\sim15$ at $\sim$8500\,\AA. 
These wavelengths correspond to the CaIIK line, the H$\alpha$ line, and the calcium triplet region, which were used for velocity and metallicity determinations. 
These data were reduced with Carpy \citep{k+03}\footnote{https://code.obs.carnegiescience.edu/mage-pipeline}, following standard data reduction procedures. 
One candidate was identified as a galaxy from these spectra, the other four were stars and are labeled Herc-1, Herc-3, Herc-4, and Herc-12.

These candidate members of Hercules were selected through \textit{Gaia} EDR3 proper motions \citep{gaia21} and through their low metallicities, as determined from wide-field metallicity-sensitive CaHK imaging of Hercules using a custom filter on the Magellan/IMACS f/2 camera.
One star, Herc-12, was beyond the range of our wide-field CaHK imaging and was identified solely using a proper motion selection. 
The details of the imaging and membership determination will be described in an upcoming paper (Chiti et al. 2024, in prep), but we list the relevant details here. 
Specifically, we imaged Hercules out to $\sim7$ half-light radii along its major axis and $\sim$4 half-light radii along its minor axis using the CaHK filter.
Then, we retained stars that had \textit{Gaia} EDR3 proper motions consistent within 2\,$\sigma$ of the proper motion of Hercules in \citet{mv+20}.
We further selected stars that lay within 0.1\,mag of a [Fe/H]$=-2.5$, 12\,Gyr Dartmouth isochrone \citep{dcj+08} on a color-magnitude diagram from Pan-STARRS DR2 photometry \citep{panstarrs+16}\footnote{https://catalogs.mast.stsci.edu/panstarrs/} assuming a distance modulus of 20.60 \citep{musella12}.
Then, we selected stars that occupied the same region of color-color space using the CaHK photometry (following Figure 3 in \citealt{cfj+20}) as previously known Hercules members.
This selection resulted in only one candidate, Herc-1, being identified in the far outskirts ($>4$\,$r_h$) of Hercules brighter than $g\sim20.5$, along with the other candidates (Herc-3 and Herc-4) that were closer to the center of the system.

\subsection{DEIMOS}
\label{sec:DEIMOS}

We re-reduced archival data taken between 2007-2015 with the DEIMOS spectrograph \citep{faber03} on the Keck II 10-m telescope.
Nine multislit masks were observed with the 1200G grating covering a wavelength range of $6400$-$9100$\,{\AA} with the OG550 blocking filter, yielding a resolution of $R\sim6000$.
Individual science exposures were reduced to 1D spectra using v1.10 of \texttt{PypeIt} \citep{prochaska20}.
Raw data files and associated calibration data were accessed from the Keck Observatories Archives (KOA).\footnote{\url{http://koa.ipac.caltech.edu/}}

\section{Data Analysis \& Membership Selection}
\label{sec:data}

We describe in this section the samples used in our study, the metallicity and velocity measurements, and the criteria used for identifying Hercules members. 
Section~\ref{sec:tot_sample} gives an overview of the number of radial velocities and metallicities provided by each sample. 
Sections~\ref{sec:velocities} and \ref{sec:metallicities} provide detailed descriptions of the radial velocity and metallicity measurementes.
Section~\ref{sec:binary} discusses the identification of binaries in our sample.
Sections~\ref{sec:membership} and~\ref{sec:foreground} discuss how these measurements are combined and used to select Hercules members, and assess foreground contamination.

\subsection{Description of samples}
\label{sec:tot_sample}

The final dataset that we use in our analysis is comprised of three subsamples: data from MagE and IMACS (collected and analyzed in this paper), data from DEIMOS (collected in \citealt{simon07, brown14, gregory20}, re-analyzed in this paper), and data from \citet{aden09}. 
We adopt line-of-sight velocities from FLAMES spectroscopy and metallicities from Str\"{o}mgren photometry from \citet{aden09} as published in their study. 
We opt to combine these three samples to maximize the size of our dataset and minimize statistical uncertainties in our final parameters.
Each dataset is briefly described in Sections~\ref{sec:velocities} and \ref{sec:metallicities}, but we encourage interested readers to refer to the original publications for further details on samples already in the literature. 

In total, we have 27 usable line-of-sight velocity measurements from IMACS, four from MagE, 18 from FLAMES, and 390 from DEIMOS. 
We have 20 usable EW measurements from IMACS, four metallicities from MagE, 28 from Str\"{o}mgren photometry, and 354 from DEIMOS. 
Our final parent sample contains 411 unique stars with at least one line-of-sight velocity measurement. 
In the case of measurements from different samples for the same stars, we use the overlap to characterize any systematics between different samples (see Section~\ref{sec:systematic} and \ref{sec:feh_systematic}), test for potential binaries (see Section~\ref{sec:binary}), and then combine the measurements for final membership identification (see Section~\ref{sec:membership}).

\subsection{Radial velocities}
\label{sec:velocities}

\subsubsection{Velocities from IMACS}
\label{sec:imacs_vel}
We derive velocities from the IMACS observations following procedures used in other UFD observations with our instrumental setup \citep[e.g.,][]{sld+17,lsd+17, heiger23}, which we briefly describe here.
We derived velocities from our spectra by minimizing $\chi^2$ from 8450\,{\AA} to 8680\,{\AA} relative to a template spectrum of HD122563 \footnote{The velocity for HD122563 is assumed to be $-26.51$\,\kmsec \citep{chubak12}.} that was observed with the same IMACS configuration.
We derived a telluric correction for the mis-centering of stars in their slits by first performing the same procedure with a template spectrum of HR4781 over the wavelength range 7550\,{\AA} to 7700\,{\AA}.

Random velocity uncertainties were derived by repeating the $\chi^2$-minimization with respect to HD122563 500 times after adding noise to the spectra according to their S/N.
We determined a systematic floor for the velocity precision by dividing the raw observations into two sets, separately for the 2015 and 2018 data, re-reducing each set independently, and then determining the systemic floor in the velocity uncertainties that was needed to bring the velocities in agreement (following e.g., \citealt{simon07}). 
We find that the systemic velocity uncertainty in the 2015 data was 1.2~\kmsec, and in the 2018 data was 0.9~\kmsec, which is comparable to previous UFD studies with this observational setup \citep{sld+17, lsd+17, heiger23}.
The difference in the velocity precision floor between 2015 and 2018 is likely due to the improvement in the wavelength solution resulting from the introduction of the Kr arc lamp.
The final velocity uncertainty was taken as the quadrature sum of the random and systematic velocity uncertainties.

\subsubsection{Velocities from MagE}

Given the low S/N (10 to 20) of our MagE spectra, we focused on deriving velocities from orders with the prominent Ca\,II H\&K ($\sim3950$\,\AA), H$\alpha$ ($\sim6560\,$\AA), and Calcium triplet ($\sim$8500\,\AA) absorption features.
We normalized each of these orders using the spectral analysis toolkit developed by A. Ji \footnote{\url{https://github.com/alexji/alexmods}}, and cross-correlated each of the resulting orders with a spectrum of HD122563 that was obtained using the same MagE configuration.
This gave one velocity measurement from each absorption feature for each star.
Similarly, we derived a telluric correction for each star by cross-correlating our spectra with a template spectrum of HR4781. 

The random uncertainty in each of these velocities was derived by replicating the procedure in Section~\ref{sec:imacs_vel}: by repeating the velocity measurements 500 times after adding random noise to each pixel based on S/N.
We found that velocities from the Ca\,II H\&K feature had prohibitively high uncertainties ($\sim15$~\kmsec), so we discarded this feature in our velocity analysis. 

The systematic uncertainty on the MagE velocitiy was derived independently for the order spanning H$\alpha$ and the order spanning the calcium triplet lines.
These systematic uncertainties were obtained by applying our method of deriving velocities from MagE data on metal-poor stars observed with the same MagE observing setup between 2011 and 2013. 
This sample included observations of HD122563, HD140283, CD$-$38\degree245, stars 10\_7\_442, 11\_1\_3334, 6\_5\_505 in the Sculptor dwarf galaxy presented in Table 5 of \citet{csf+18}, and an additional metal-poor red giant in the Carina dwarf galaxy.
For the latter stars in dwarf galaxies, repeat observations were taken on the same night so differences in velocity when estimating the systematic uncertainty are unlikely to be due to binarity.
Following the same procedure as in Section~\ref{sec:imacs_vel}, we find that a velocity uncertainty of 5.3\,{\kmsec} needed to be added to velocities from the H$\alpha$ order and 3.5\,{\kmsec} needed to be added to velocities from the calcium triplet order to bring velocity measurements of the same stars in agreement. 
These values were adopted as the systematic velocity uncertainties, and are added to the random uncertainties from each order in quadrature to derive a final uncertainty.
Then, the final velocity is taken as the inverse-variance weighted average of the velocities from each order.

\subsubsection{Velocities from DEIMOS}
\label{sec:keck_vels}

Stellar radial velocities and calcium triplet equivalent widths (EWs) were measured using a preliminary version of the \texttt{DMOST} package (M. Geha et al. 2024, in prep). 
In brief, \texttt{DMOST} forward models the 1D stellar spectrum for each star from a given exposure with both a stellar template from the PHOENIX library and a telluric absorption spectrum from \texttt{TelFit} \citep{gullikson14}. 

The velocity is determined for each science exposure through an MCMC procedure constraining both the radial velocity of the target star as well as a wavelength shift of the telluric spectrum needed to correct for slit miscentering (see, e.g. \citealt{sohn07}). 
The final radial velocity for each star is derived through an inverse-variance weighted average of the velocity measurements from each exposure. 
The systematic error reported by the pipeline, derived from the reproducibility of velocity measurements across masks and validated against spectroscopic surveys, is $\sim1$\,\kmsec (see M. Geha et al. 2024, in prep). 

\subsubsection{Velocities from \citet{aden09}}
\label{sec:aden_vels}

\citet{aden09} present velocities for 18 RGB stars that pass their criteria for Hercules membership using {$R\sim6500$} FLAMES spectra. These spectra spanned 8210\,{\AA} to 9400\,{\AA}, covering the prominent calcium triplet absorption feature. 
The authors performed a cross-correlation of the observed spectra against a synthetic template spectrum using the IRAF routine FXCOR, which returns uncertainties based on the Tonry-Davis R value \citep{tonry79}.
The minimum velocity uncertainty of this sample is $\sim0.6$~\kmsec.

\subsubsection{Assessing systematics across samples and combining velocity measurements}
\label{sec:systematic}

We discuss potential systematic differences resulting from the different instruments and techniques used to derive line-of-sight velocities in this section. 
We limit this comparison to stars with line-of-sight velocities consistent with membership to Hercules (described in Section~\ref{sec:membership}), and test for potential systematic offsets between the samples from different instruments. 

\begin{figure*}
    \centering
    \includegraphics[width=0.9\textwidth]{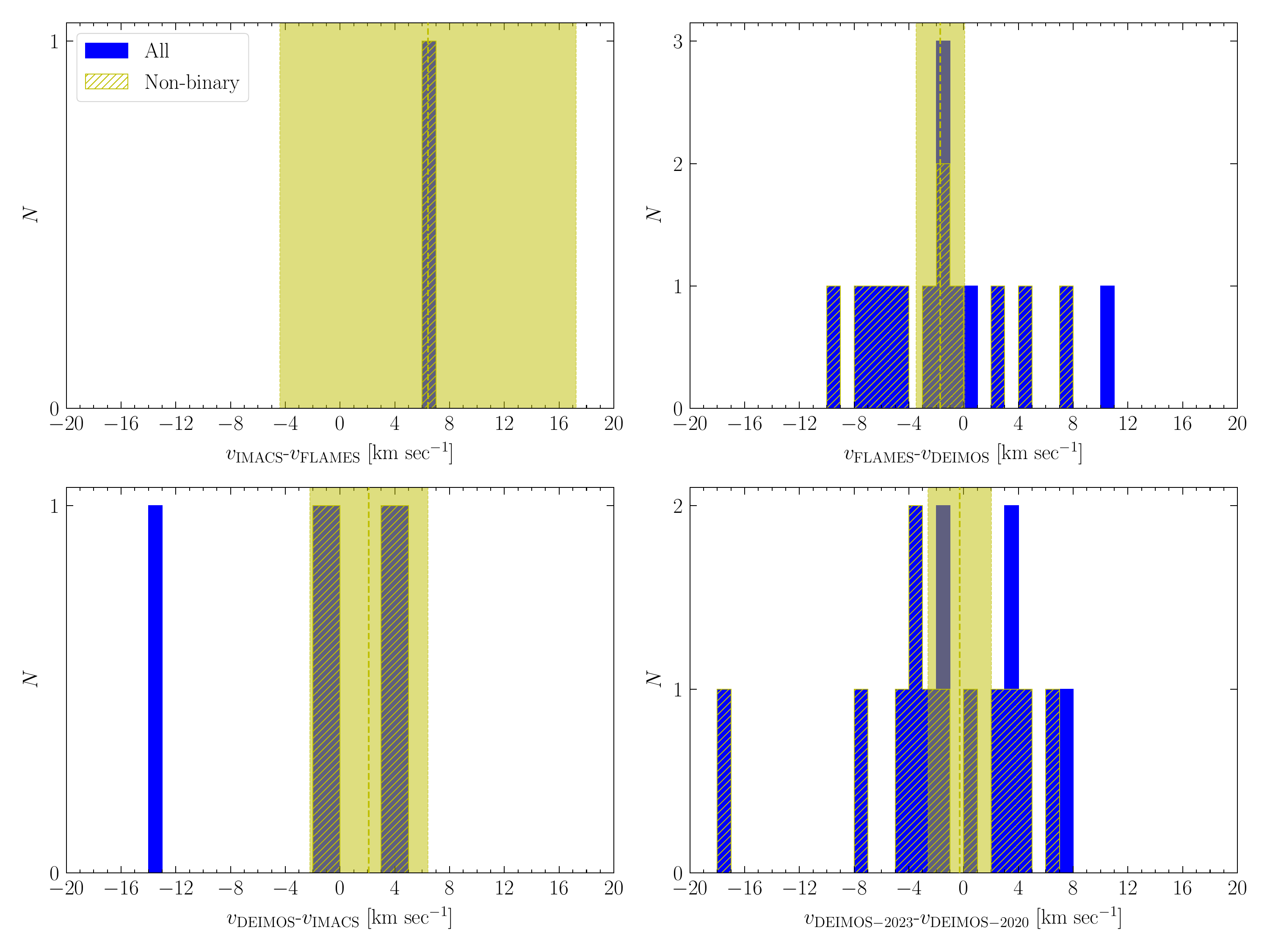}
    \caption{Differences in the velocity measurements of the same stars observed with IMACS, FLAMES, and DEIMOS. Blue histograms include all stars selected as members from their velocity and metallicity information (CD sample; see Section~\ref{sec:membership}), whereas the yellow hashed histograms only include stars flagged as not being in binary systems (see Section~\ref{sec:binary}). 
    The yellow vertical dashed lines mark the weighted mean of the differences from excluding potential binaries, while the shaded regions correspond to the 2\,$\sigma$ range. 
    The bottom right panel shows the differences between velocity measurements from \citet{gregory20} (DEIMOS-2020) and velocity in this study from the re-analysis of DEIMOS data (DEIMOS-2023). 
    }
    \label{fig:rv_sys_diff}
\end{figure*}

We first compare line-of-sight velocity measurements for the same star from any pair of instruments (i.e., IMACS, DEIMOS, and FLAMES). 
The distribution of the differences is shown in Figure~\ref{fig:rv_sys_diff}. 
The blue-shaded area corresponds to the 2\,$\sigma$ range of the mean difference between a given pair of instruments.
The weighted mean difference between IMACS and FLAMES (based on 1 star) is $6.4\pm5.4$~\kmsec. 
The weighted difference between FLAMES and DEIMOS (15 stars) is $-1.1\pm0.7$~\kmsec. 
The weighted difference between DEIMOS and IMACS (5 stars) is $-4.0\pm1.7$~\kmsec. 
We also check for systematic differences between the pipelines used in \citet{gregory20} (DEIMOS-2020) and this study (DEIMOS-2023) on DEIMOS data to test for potential systematic differences arising from the updated pipeline and addition of DEIMOS observations from \citet{brown14}. The weighted difference between DEIMOS-2023 and DEIMOS-2020 (15 stars) is $0.5\pm1.1$~\kmsec. 
The largest tension is between DEIMOS and IMACS at $2.35$\,$\sigma$.
Additionally, two stars from the MagE sample are also observed in other samples.
Herc-3 from the MagE sample overlaps with DEIMOS and FLAMES with velocity measurements consistent within 2\,$\sigma$.
Herc-4 from the MagE sample is also observed with DEIMOS, where multi-mask measurements from DEIMOS show line-of-sight velocity variation. 
The binary test performed on the MagE+DEIMOS combined multi-epoch measurements further supports the case of Herc-4 being a potential binary.

To further investigate this possible tension, we re-do this analysis after excluding stars that show evidence of radial velocity variations from being in a binary system (see Section~\ref{sec:binary}). 
After this cut, the weighted difference between IMACS and FLAMES (1 star) is $6.4\pm5.4$~\kmsec. The weighted difference between FLAMES and DEIMOS (12 stars) is $-1.7\pm0.9$~\kmsec. The weighted difference between DEIMOS and IMACS (4 stars) is $2.1\pm2.2$~\kmsec. The weighted difference between DEIMOS-2023 and DEIMOS-2020 (12 stars) is $-0.3\pm1.2$~\kmsec. 
Now, we see no strong evidence of statistically significant systematic offsets greater than $2$\,$\sigma$ between samples after excluding stars potentially in binary systems. 
Note that applying the offsets in the prior paragraph before performing the tests for binarity in Section~\ref{sec:binary} does not change the results of that analysis.
Consequently, the tension in the previous paragraph is reasonably explained by the influence of binaries. 
Conversely, it is unlikely that the binary analysis in Section~\ref{sec:binary} is affected by systematic offsets between instruments.
Given the lack of evidence for significant velocity systematics, velocity measurements from this study and the literature are combined via weighted averaging.

\subsection{Metallicities}
\label{sec:metallicities}

\subsubsection{Metallicities from IMACS}
\label{sec:metallicities_imacs}

Metallicities from the IMACS spectra are derived using the well-established calibration in \citet{cpg+13} that relates the strength of the calcium triplet lines to the overall metallicity of a star.
We apply this calibration to our IMACS spectra following previous studies of UFD stars using this observational setup \citep[e.g.][]{sld+17, lsd+17, csf+22}, which we briefly outline here.

The \citet{cpg+13} calcium triplet-metallicity calibration takes the total equivalent widths of the calcium triplet lines and the absolute $V$ magnitude as inputs. 
We compute the equivalent width of each calcium triplet line by fitting a Gaussian + Lorentzian profile to each line \citep[e.g.,][]{sld+17, lsd+17}.
The apparent $V$ magnitude was computed by converting photometry from Pan-STARRS Data Release 1 using the transformations in \citet{tsl+12}, and then converted to an absolute $V$ magnitude assuming a distance modulus of 20.60 \citep{musella12}
The random uncertainties in the equivalent width measurements were computed exactly following the Monte-Carlo re-sampling procedure that was used for the velocity uncertainties (see Section~\ref{sec:imacs_vel}).
The systematic equivalent width uncertainty floor is $0.32$\,\AA\ from \citet{sld+17} and added in quadrature to the random uncertainties.
These uncertainties were propagated to the metallicity to derive a final metallicity uncertainty. 


\subsubsection{Metallicities from MagE}
\label{sec:metallicities_mage}

We compute metallicities from the MagE spectra using the KP calibration presented in \citet{brn+99}, a well-established relationship between the strength of the CaII\,K line and the stellar metallicity.
We implement this procedure following \citet{csf+18}, which we briefly outline here.

Specifically, the KP calibration maps the pseudo-equivalent width of the CaII\,K line (denoted by the KP index) and the $B-V$ color to stellar metallicity.
The KP index is a measure of the equivalent width of the CaII\,K line at 3933.7\,{\AA}, derived by integrating over the feature using windows of 6\,\AA, 12\,\AA, or 18\,{\AA} depending on the strength of the feature. 
The $B-V$ color of each star was derived using the Pan-STARRS color transformations in Table 6 of \citet{tsl+12}, with input Pan-STARRS DR2 photometry.
The random metallicity uncertainties are adopted by varying the continuum placement, and the systematic uncertainties are provided by \citet{brn+99}.

We also compute metallicities of stars with MagE data using the calcium triplet features, exactly following Section~\ref{sec:metallicities_imacs}, as an external check on our CaII\,K metallicities. 
These metallicities are consistently well within 1\,$\sigma$ of those from the CaII\,K features, and all stars had [Fe/H] $< -2.5$, independently validating our application of the CaII\,K calibration. 
However, we do not adopt these calcium-triplet based metallicities for our MagE spectra due to the presence of scattered light in the MagE data significantly distorting the continuum in this wavelength regime ($\gtrsim8200$\,\AA).
This leads to uncertainties in the calcium triplet-based metallicities of $> 0.4\,$dex.
The effect of this scattered light is more pronounced here than in previous studies of UFD stars using MagE (e.g., \citealt{chiti21}) due to the significantly lower S/N of the MagE spectra in this study. 

\subsubsection{Metallicities from DEIMOS}
\label{sec:metallicities_keck}

\texttt{DMOST} measures the equivalent width from the calcium triplet features by fitting a Gaussian-plus-Lorentzian model to the coadded spectrum (for stars at S/N $> 15$) or a Gaussian model (for stars below S/N $< 15$). 
We then compute metallicities following \citet{cpg+13}, as described in Section~\ref{sec:metallicities_imacs}.
We assume a $0.2$ Angstrom systematic error on the total equivalent width determined from independent repeat measurements.

\subsubsection{Metallicities from \citet{aden09}}
\label{sec:metallicities_aden}

\citet{aden09} derive metallicities using Str\"{o}mgren photometry obtained from for 28 RGB stars in Hercules, following the semi-empirical calibration by \citet{calamida07}.
For the 18 stars with FLAMES spectroscopy, 15 have metallicity measurements from the calcium triplet lines using the calibration by \citet{rhs+97}.
Their comparison between the photometric and spectroscopic metallicities suggests good agreement among the 15 stars, thus the photometric metallicities for the 28 stars are reported in the original study and adopted in classifying stars as members in this study.

\begin{figure*}
    \centering
    \includegraphics[width=0.9\textwidth]{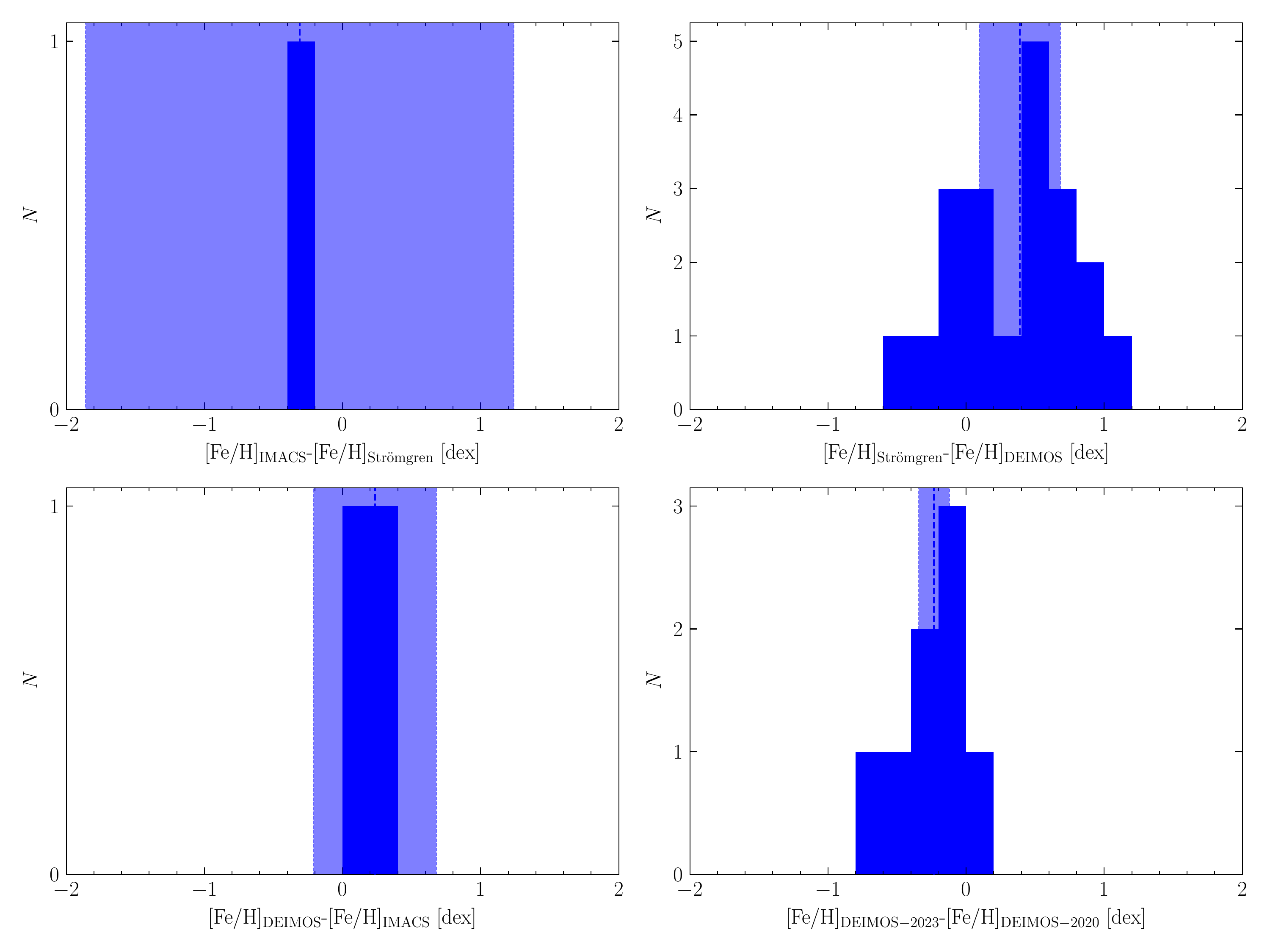}
    \caption{Differences in the metallicity measurements of the same stars from IMACS, DEIMOS, and Str\"{o}mgren photometry. 
    Blue histograms include all stars selected as members from their velocity and metallicity information (CD sample; see Section~\ref{sec:membership}). 
    The vertical dashed lines mark the weighted mean of the differences, while the shaded regions correspond to the 2\,$\sigma$ range of the weighted mean difference. 
    We see systematic offsets between the DEIMOS values derived in this work and those from \citet{gregory20}.
    We discuss the significance of this difference in Section\ref{sec:metallicities_keck}.
    }
    \label{fig:feh_sys_diff}
\end{figure*}

\subsubsection{Assessing systematics across samples and combining metallicity measurements}
\label{sec:feh_systematic}

We discuss potential systematic differences in metallicities in this section following a similar procedure in Section~\ref{sec:systematic}. 
As mentioned earlier, we only consider stars with line-of-sight velocities consistent with membership to Hercules.
The distribution of the differences is shown in Figure~\ref{fig:feh_sys_diff}. 
The weighted mean difference between metallicities from IMACS and Str\"{o}mgren photometry from \citet{aden09} (based on 1 star) is $-0.31\pm0.77$~dex.
The weighted mean difference between Str\"{o}mgren photometry and DEIMOS (based on 20 stars) is $0.39\pm0.15$~dex.
The weighted mean difference between DEIMOS and IMACS (based on 2 stars) is $0.23\pm0.22$~dex.
The weighted mean difference between DEIMOS-2023 and DEIMOS-2020 (based on 8 stars) is $-0.23\pm0.06$~dex.
For the MagE sample, Herc-3 and Herc-4 have metallicities consistent with DEIMOS measurements.

Notably, the metallicities derived from Str\"{o}mgren photometry are systematically higher than the spectroscopic DEIMOS measurements. 
The fact that \citet{aden09} found good agreement between their photometric and spectroscopic metallicities likely arises from the known bias in the older version of the calcium triplet calibration \citep{rhs+97} that overestimates the metallicity for metal-poor stars \citep[see e.g.,][]{starkenburg10}. 
Thus, the DEIMOS spectroscopic metallicities should be preferred.
For the purposes of our study, almost all stars with consistent radial velocities from \citet{aden09} already have photometric metallicities below [Fe/H]$=-2$, so the systematic difference does not affect the selection of members.
Likewise, the systematic difference between the metallicities derived in this work for the DEIMOS sample and those derived in \citet{gregory20} does not cause differences in our membership selection process, given that our metallicity selection includes all stars that plausibly have [Fe/H] $< -2$ (see Section~\ref{sec:membership}).

In general, our comparisons show evidence for slight ($\sim0.2$~dex) offsets in our samples with spectroscopically-derived metallicities (i.e., not including the photometric metallicities in \citealt{aden09}), but the small overlap sample sizes preclude any strong statements.
Thus, for the purpose of selecting plausibly metal-poor Hercules members, we compute and report the weighted average metallicity.
We advise, however, readers to be cautious with using the combined metallicity for more sophisticated applications.

\subsection{Binarity Tests}
\label{sec:binary}

Since the line-of-sight velocity measurements in our combined sample span a long baseline (2007 to 2022), we can perform a test for binarity on stars with observations that span multiple epochs. 
In total, 37 of the 62 candidate members of Hercules (see Section~\ref{sec:membership} for a description of membership selection) have at least two observations spaced by $60$ to $4000$ days that make them suitable for a test for binarity.

We perform a simple $\chi^2$ test of whether a star's velocity is constant with time to assess the likelihood that it is in a binary system. 
We flag any stars with $p$-values less than $10^{-4}$ as potential binaries and exclude them from any dynamical analysis (e.g., velocity dispersion, velocity gradient) in this study. 
These stars are indicated by having a BIN flag set to 1 in Table~\ref{tab:rv_feh_pcf}.
Five out of $33$ stars that are selected in our purest sample of Hercules members (based on metallicity, velocity, and proper motion information) are flagged as potential binaries and excluded from further dynamical analysis. 
We note that this does not guarantee that the remaining sample is all non-binaries because only $\sim 50 \%$ of the full sample (or any candidate samples described in Section~\ref{sec:membership}) have multi-epoch measurements.
We also cannot identify wide binaries with long periods beyond our baseline even with multi-epoch measurements.
The particular choice of $p$-values cut has minimal effect on the result of the test.
One additional star, PanSTARRS~ID~123322477687515309, is flagged in the final sample used for dynamical analysis if the cut is increased to $0.01$ but the results remain qualitatively unchanged.

We note that the systematic velocity offsets between datasets have minimal impact on the binary test results. 
It is thus unlikely that systematic offsets between instruments contribute to any mis-classification of non-binary systems as binaries.

\subsection{Member selection}
\label{sec:membership}

We separate stars into four samples of membership/non-membership in decreasing levels of confidence, which we describe below: proper motion confirmed members (PCF), confirmed members (CF), candidate members (CD), and non-members (NM). 
In the following analysis, line-of-sight velocity and metallicity measurements from different samples (see Sections~\ref{sec:velocities} and~\ref{sec:metallicities}) are combined via weighted averaging, while photometry is taken from Pan-STARRS \citep{chambers16}, and proper motions are taken from \Gaia~DR3 \citep{gaia21}.

\begin{figure*}
    \centering
    \includegraphics[width=0.8\textwidth]{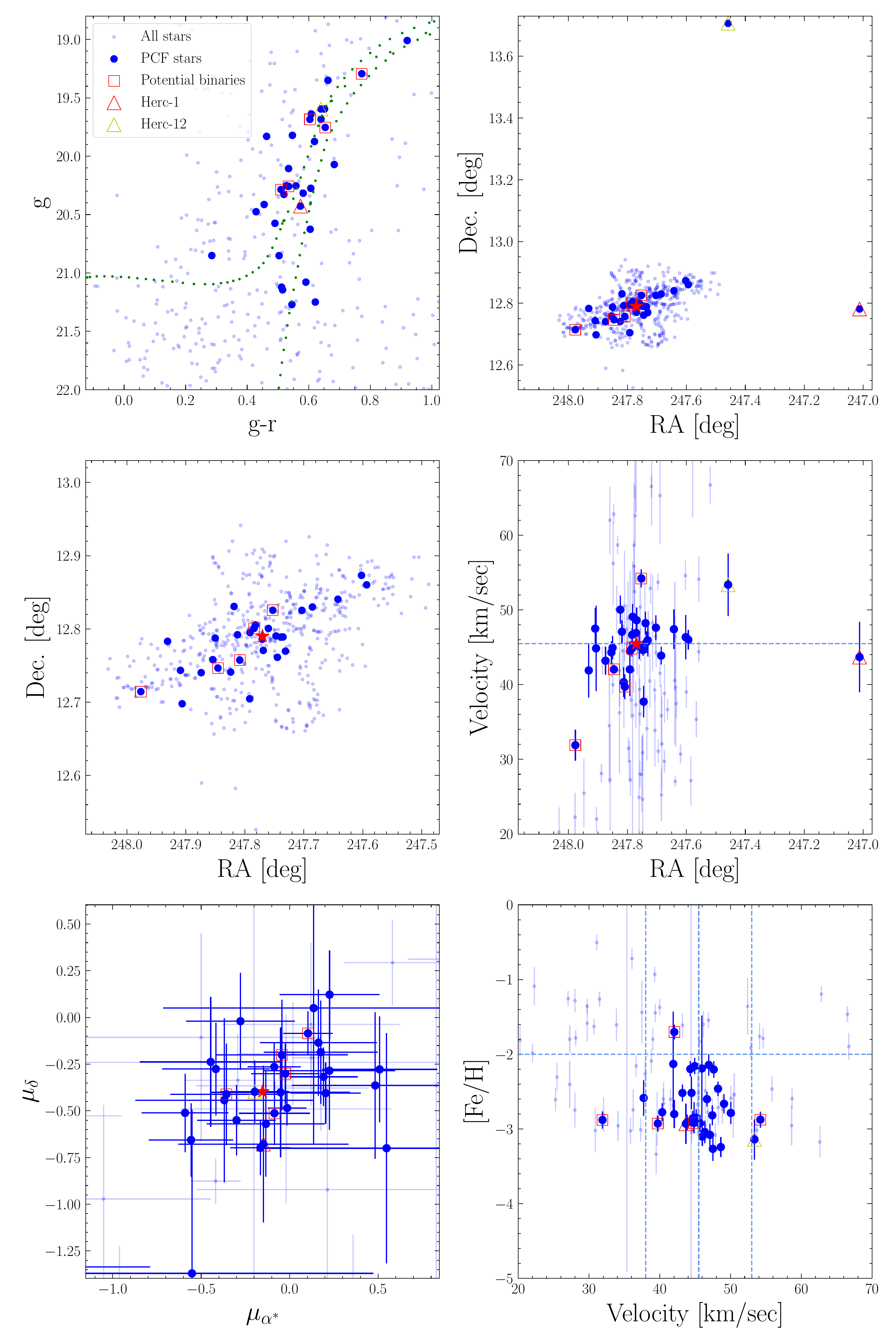}
    \caption{Summary plot of the final stars selected to be members based on metallicities, radial velocities, and proper motions (PCF sample; see Section~\ref{sec:membership}). 
    Semi-transparent blue dots represent all observed stars, whereas the blue dots represent the subset of stars satisfying all four criteria for confident membership (photometry, line-of-sight velocities, metallicities, and proper motions). 
    Red square boxes mark the stars that are flagged as potential binaries based on individual line-of-sight velocity measurements from different epochs. 
    The red (yellow) triangle marks the newly discovered Herc-1 (Herc-12).}
    \label{fig:pm_summary}
\end{figure*}

We generate our initial proper motion confirmed sample (PCF) as follows. 
We select an initial sample of candidate members by defining a line-of-sight velocity selection window at 3$\,\sigma$ around the systematic velocity (45~\kmsec) of Hercules, with 1\,$\sigma$ defined as the velocity dispersion of 5.1\kmsec\ from \citet{simon07}.
For each star, we similarly examine the range defined by three times the measurement uncertainty around its velocity measurement. 
Stars with this range overlapping the selection window are selected as having velocities consistent with membership in Hercules.
Then, we limit this sample to stars with metallicities that have 2\,$\sigma$ uncertainties consistent with being lower than [Fe/H]~$=-2$; specifically, a star is selected if the 2\,$\sigma$ lower limit on its metallicity is less than [Fe/H]~$=-2$.
We note that \citet{brown14} has shown that the metallicity distribution function of Hercules exhibits a tail towards higher metallicities of [Fe/H]~$>-2$.
For the purpose of sample purity, we still limit our metallicity selection to the above criteria.
However, we note that our selection doesn't strictly remove higher metallicity stars that might be members;
the highest metallicity star that passes our metallicity cut in the CF sample has [Fe/H] = $-1.35 \pm 0.38$ due to its large metallicity uncertainty.
After this, we require that stars have $g$ magnitudes and $g-r$ colors consistent within 0.2~dex of an isochrone generated from Padova CMD v3.7 \footnote{\url{http://stev.oapd.inaf.it/cgi-bin/cmd}} with age$=13$~Gyr and [M/H]$=-2.2$. 
We apply extinction corrections using the \citet{schlafly11} dust map from the \texttt{dustmaps} package \citep{green18} and the extinction law from \citet{tonry12}.
Lastly, we cross-match this sample with \Gaia~DR3 for proper motions, and exclude stars with proper motions that have 2\,$\sigma$ range not overlapping with that of the systemic proper motion of Hercules ($\mu_{\alpha^*}=\mu_\alpha \cos{\delta} = -0.153 \pm 0.074$~\masyr, $\mu_\delta = -0.397 \pm 0.063$~\masyr; \citealt{gregory20}). 
One star, PanSTARRS~ID~123292478456606455, is included in the PCF sample despite having [Fe/H] = $-1.7 \pm 0.10$ from DEIMOS data. 
A detailed chemical abundance study of this star by \citet{koch08} (Her-3 in their paper) indicates this star is a Hercules member, and \citet{koch14} confirmed it as a spectroscopic binary.
These criteria ensure that no stars in this sample (33 stars in total) have photometry, velocity, metallicity, and proper motion measurements inconsistent with membership, ensuring a highly pure sample.
We also exclude stars with resolved parallaxes from \Gaia.
Figure~\ref{fig:pm_summary} shows this PCF sample in spaces where selection criteria are applied. 
Table~\ref{tab:rv_feh_pcf} lists the radial velocity and metallicity measurements of the PCF sample.
Then, we loosen the criteria and define two additional samples: the CF sample for stars with no proper motion information but velocities, metallicities, and photometry consistent with membership; the CD sample for stars with no proper motion and metallicity information but velocities and photometry consistent with membership.

\begin{deluxetable*}{ccccccccccc}
\tablecaption{Radial velocity and metallicity measurements for all stars that are confirmed to be members based on their metallicities, velocities, and proper motions (PCF sample; see Section~\ref{sec:membership}) in this study. 
The BIN flag indicates the result of the binary test, where 1 means the star is in a potential binary, and 0 means the star has consistent multi-epoch line-of-sight velocity measurements. 
Stars for which no multi-epoch observations are available have no data in this column. 
Only stars that do not have 1 in BIN are included when deriving the dynamical quantities of Hercules in Section~\ref{sec:analysis}.}
\label{tab:rv_feh_pcf}
\tablehead{\colhead{PanSTARRS ID} & \colhead{RA} & \colhead{DEC} & \colhead{$g$} & \colhead{$r$} & \colhead{$v$} & \colhead{$\sigma_{v}$} & \colhead{[Fe/H]} & \colhead{$\sigma_{\rm{[Fe/H]}}$} & \colhead{MEM} & \colhead{BIN} \\
\colhead{} & \colhead{(deg)} & \colhead{(deg)} & \colhead{(mag)} & \colhead{(mag)} & \colhead{\kmsec} & \colhead{\kmsec} & \colhead{dex} & \colhead{dex} & \colhead{} & \colhead{}}
\startdata
123312477447664007 & 247.74477 & 12.76127 & 21.12 & 20.61 & 37.7 & 2.1 & $-$2.59 & 0.24 & PCF & 0 \\
123352477911384520 & 247.79115 & 12.79504 & 20.85 & 20.35 & 44.5 & 1.2 & $-$2.52 & 0.22 & PCF & 0 \\
123282478740628822 & 247.87404 & 12.7403 & 19.68 & 19.04 & 43.2 & 1.9 & $-$2.52 & 0.11 & PCF & 0 \\
123312478543710164 & 247.85432 & 12.75811 & 19.87 & 19.25 & 44.3 & 1.5 & $-$2.20 & 0.10 & PCF & 0 \\
123392478183197334 & 247.81831 & 12.8307 & 20.33 & 19.81 & 47.1 & 1.5 & $-$3.08 & 0.15 & PCF & 0 \\
123352478124740987 & 247.81247 & 12.79209 & 21.25 & 20.63 & 40.3 & 2.0 & $-$2.78 & 0.12 & PCF & 0 \\
123242477917826056 & 247.79179 & 12.70463 & 21.14 & 20.63 & 42.0 & 3.6 & $-$2.80 & 0.19 & PCF & 0 \\
123362477838422519 & 247.78386 & 12.8017 & 19.60 & 18.94 & 46.7 & 0.8 & $-$2.60 & 0.10 & PCF & 0 \\
123362477820467019 & 247.78206 & 12.80545 & 20.41 & 19.96 & 49.1 & 1.7 & $-$2.66 & 0.13 & PCF & 0 \\
123322477687515309 & 247.76877 & 12.77069 & 20.27 & 19.67 & 48.6 & 1.7 & $-$3.24 & 0.13 & PCF & 0 \\
123362477600271319 & 247.76005 & 12.80071 & 20.32 & 19.73 & 45.1 & 1.7 & $-$2.85 & 0.11 & PCF & 0 \\
123342477471739033 & 247.74718 & 12.79045 & 19.35 & 18.69 & 44.6 & 0.8 & $-$2.86 & 0.10 & PCF & 0 \\
123342477384687255 & 247.73849 & 12.78898 & 19.01 & 18.09 & 48.2 & 1.5 & $-$2.46 & 0.10 & PCF & 0 \\
123392477032240928 & 247.70322 & 12.82538 & 21.08 & 20.49 & 47.6 & 1.7 & $-$2.20 & 0.11 & PCF & 0 \\
123392476854146433 & 247.68541 & 12.82996 & 19.64 & 19.03 & 43.9 & 1.2 & $-$2.91 & 0.10 & PCF & 0 \\
123432475934522803 & 247.59341 & 12.86022 & 20.11 & 19.57 & 46.0 & 1.3 & $-$3.10 & 0.12 & PCF & 0 \\
123402476421789135 & 247.64217 & 12.84056 & 20.85 & 20.57 & 47.4 & 2.7 & $-$2.82 & 0.13 & PCF & 0 \\
123342479311100112 & 247.93108 & 12.78307 & 20.25 & 19.70 & 41.9 & 3.6 & $-$2.13 & 0.70 & PCF & -- \\
123232479063337871 & 247.90631 & 12.69785 & 21.27 & 20.73 & 44.9 & 5.7 & $-$2.85 & 0.21 & PCF & -- \\
123322477310924104 & 247.73111 & 12.76968 & 19.82 & 19.27 & 45.9 & 2.2 & $-$2.19 & 0.71 & PCF & -- \\
123292479093142603 & 247.9092 & 12.7435 & 20.48 & 20.05 & 47.5 & 2.8 & $-$3.27 & 0.16 & PCF & -- \\
124442474585078463\tablenotemark{$\dag$} & 247.45847 & 13.70663 & 19.60 & 18.96 & 53.4 & 4.2 & $-$3.14 & 0.27 & PCF & -- \\
123332470127158337\tablenotemark{$\ast$} & 247.01261 & 12.78152 & 20.43 & 19.85 & 43.7 & 4.7 & $-$2.93 & 0.27 & PCF & -- \\
123342478502635483 & 247.85026 & 12.78748 & 20.63 & 20.02 & 44.9 & 1.6 & $-$2.16 & 0.11 & PCF & -- \\
123342477706163595 & 247.77062 & 12.78592 & 20.07 & 19.39 & 46.9 & 1.4 & $-$2.14 & 0.14 & PCF & -- \\
123342477353297277 & 247.73531 & 12.78901 & 19.83 & 19.37 & 45.7 & 1.4 & $-$2.92 & 0.12 & PCF & -- \\
123442476022168188 & 247.60217 & 12.87314 & 20.57 & 20.08 & 46.4 & 3.0 & $-$3.04 & 0.15 & PCF & -- \\
123282478240779830 & 247.82408 & 12.74111 & 20.25 & 19.72 & 50.0 & 1.9 & $-$2.79 & 0.15 & PCF & -- \\
123362477843701371 & 247.78438 & 12.80076 & 20.26 & 19.72 & 44.8 & 1.1 & $-$2.92 & 0.12 & PCF & 1 \\
123292478456606455 & 247.84564 & 12.74666 & 19.29 & 18.52 & 42.0 & 0.6 & $-$1.70 & 0.10 & PCF & 1\tablenotemark{$\ddag$} \\
123302478085969362 & 247.8086 & 12.75741 & 19.75 & 19.10 & 39.7 & 1.7 & $-$2.93 & 0.11 & PCF & 1 \\
123392477525891082 & 247.75261 & 12.8255 & 19.68 & 19.08 & 54.2 & 1.2 & $-$2.87 & 0.11 & PCF & 1 \\
123252479764197747 & 247.97642 & 12.7144 & 20.29 & 19.78 & 31.9 & 2.1 & $-$2.88 & 0.12 & PCF & 1
\enddata
\tablenotetext{\ast}{Herc-1}
\tablenotetext{\dag}{Herc-12}
\tablenotetext{\ddag}{Confirmed spectroscopic binary in \citet{koch14}}
\end{deluxetable*}

Notably, the line-of-sight velocity dispersion of this PCF sample, after excluding stars in binaries, is found to be $1.9^{+0.6}_{-0.6}$~\kmsec\ (see Section~\ref{sec:dispersion}), significantly lower than the above initial selection window, and below previous dispersions in the literature \citep{gregory20,longeard23}.
The spatial distribution of the PCF sample shows no evidence of biased sampling that might naively explain this slight discrepancy (see Figure~\ref{fig:pm_summary}) relative to the sample that is confirmed via just metallicities and velocities. 
The velocity dispersion of the CF sample, on the other hand, is $6.0^{+0.9}_{-0.7}$~\kmsec. 
We note that our selection criterion in line-of-sight velocity is rather lenient with the 3\,$\sigma$ windows from both the systemic measurements of Hercules and the measurements of the individual star.
This suggests that the line-of-sight velocity dispersion may be intrinsically reduced as a result of the proper-motion selection.

We thus revise our above velocity initial line-of-sight velocity selection window to $45.5\pm 6.5$~\kmsec\ (i.e., within 3\,$\sigma$ from combining the uncertainty on the systematic uncertainty and velocity dispersion) based on the updated systemic velocity and dispersion from the PCF sample. 
Notably, this does not affect the PCF sample classification, but does remove two members from the confirmed sample (CF) and one additional member from the candidate sample (CD). 
Our final sample sizes: 33 stars with photometry, velocity, metallicity, and proper motion information consistent with membership (PCF sample); 22 additional stars with photometry, velocity, and metallicity information consistent with membership but no Gaia DR3 proper motion measurements (CF sample); seven additional stars with photometry and velocity information consistent with membership, but no metallicity or proper motion information (CD sample). 
We report the velocity dispersions derived from each of these samples (after removing potential binaries) in Section~\ref{sec:dispersion}.
All remaining stars are classified as non-members, as they would have at least one set of information (e.g., velocities, metallicities, proper motions) that are inconsistent with membership.

\subsection{Modeling Foreground Contamination}
\label{sec:foreground}

We perform a detailed analysis using the Besancon stellar population model \citep{robin03,czekaj14} to assess the likelihood that any of the stars in our PCF sample are foreground Milky Way contaminants.
The motivation for this analysis is the tentative selection of two MagE stars in our PCF sample that are $1.7$ and $2.2$~kpc distant from the center of Hercules (Herc-1 and Herc-12; see Section~\ref{sec:MagE}).
If these stars can be assessed as unlikely to be foreground Milky Way contaminants, then their large distances may provide constraints on long-standing debates regarding e.g., whether the system is tidally disrupting \citep{martin10,kupper17}.

\begin{figure*}
    \centering
    \includegraphics[width=0.90\textwidth]{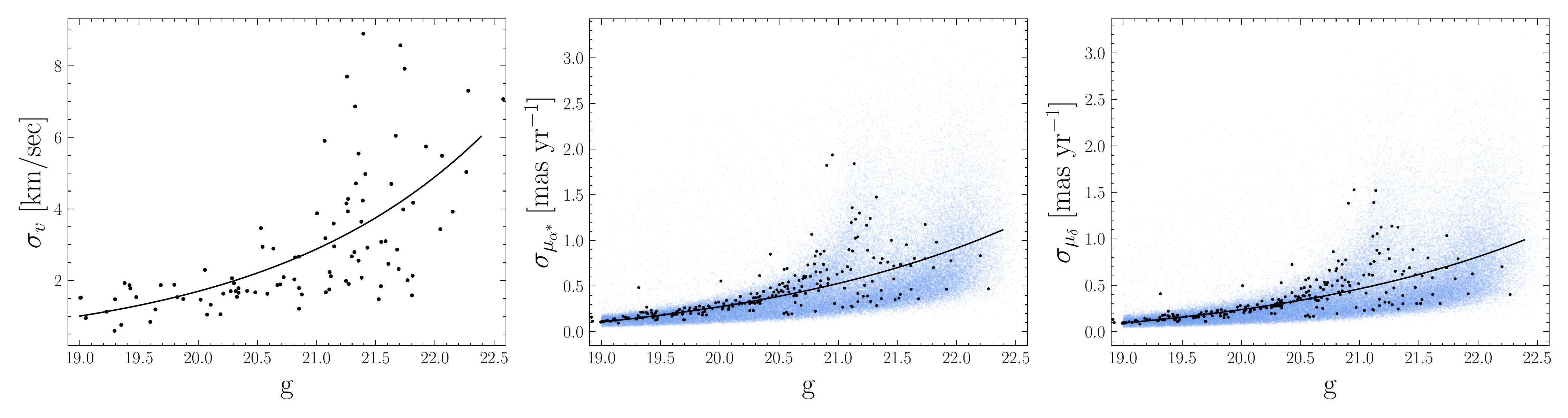}
    \caption{Error modeling of the line-of-sight velocity (left) and proper motion (middle and right) measurements as a function of $g$ magnitude.
    We adopt these error models when replicating our selection function on the simulated catalog generated by the Besancson model to assess foreground contamination. 
    Black dots in all panels correspond to the stars in this study. 
    Blue dots in the middle and right panels are all \Gaia DR3 sources within $2$\,\degree~from Hercules. 
    Black curves are the fitted error models. 
    For the line-of-sight radial velocities, the model is fitted to the weighted average velocity measurements in this study. 
    The error models for the proper motions are fitted on all \Gaia~sources (black and blue data points).}
    \label{fig:error_model}
\end{figure*}

Specifically, we queried the Besancon model to generate a simulated catalog of Milky Way stars within 10 sq. degrees of the central coordinates of Hercules (RA$=247.77$\,\degree, DEC$=12.79$\,\degree) from $g=18.5$ to $g=22$, bracketing the magnitude range considered in this study.
We replicate our selection function in Section~\ref{sec:membership} on this simulated dataset to gauge the number of Milky Way stars that may contaminate the PCF sample. 
We approximate the uncertainties on the proper motions of the simulated stars as a function of their $g$ magnitude by fitting the mean proper motion uncertainties (using \texttt{curve\_fit} from the \texttt{SciPy} package, assuming an exponential form) of all \Gaia~sources observed within $2$\,\degree~from Hercules. 
Similarly, we assign line-of-sight velocity uncertainties to the simulated stars by fitting line-of-sight velocity uncertainties as a function of $g$ magnitude in our combined sample. 
The error modeling is shown in Figure~\ref{fig:error_model}.
We note that the foreground contamination estimates listed below do not depend significantly on the error modeling of the simulated catalog; for instance, we obtain similar results when using constant characteristic uncertainties (e.g., 2~\kmsec and 1~mas~yr$^{-1}$) for line-of-sight velocity and proper motion measurements.

As mentioned in Section~\ref{sec:observation}, Herc-1 was selected based on narrow-band photometry as a potential metal-poor star with [Fe/H] $< -2$, broad-band SDSS photometry consistent with a fiducial Hercules isochrone, and proper motion consistent with the previously reported systemic proper motion of Hercules. 
Observationally, these selection criteria returned 22 candidate members-- all within $\sim3$\,$r_h$ except for Herc-1, which was located $>4$\,$r_h$\,\,west.
After applying the same metallicity, proper motion, and isochrone cuts to the simulated catalog from Besancon, we find that $\sim 5.6 \%$ of the simulated stars further satisfy the final line-of-sight velocity cut for the PCF sample. 
If we increase the simulated line-of-sight velocity error in the Besancon simulation to a floor of 4\,\kmsec, a conservative estimate of the MagE velocity precision, the fraction of simulated stars that pass the final line-of-sight velocity cut increases to $\sim7.8\,\%$.
The predicted surface density of stars passing these cuts is 0.7\,stars/sq. deg.
Notably, this is not sufficient to confidently claim membership for Herc-1, since this analysis suggests that we may expect $\sim1$ Milky Way star that passes our criteria for member out to that distance of $\sim1.7$\,kpc.

\begin{figure}
    \centering
    \includegraphics[width=0.49\textwidth]{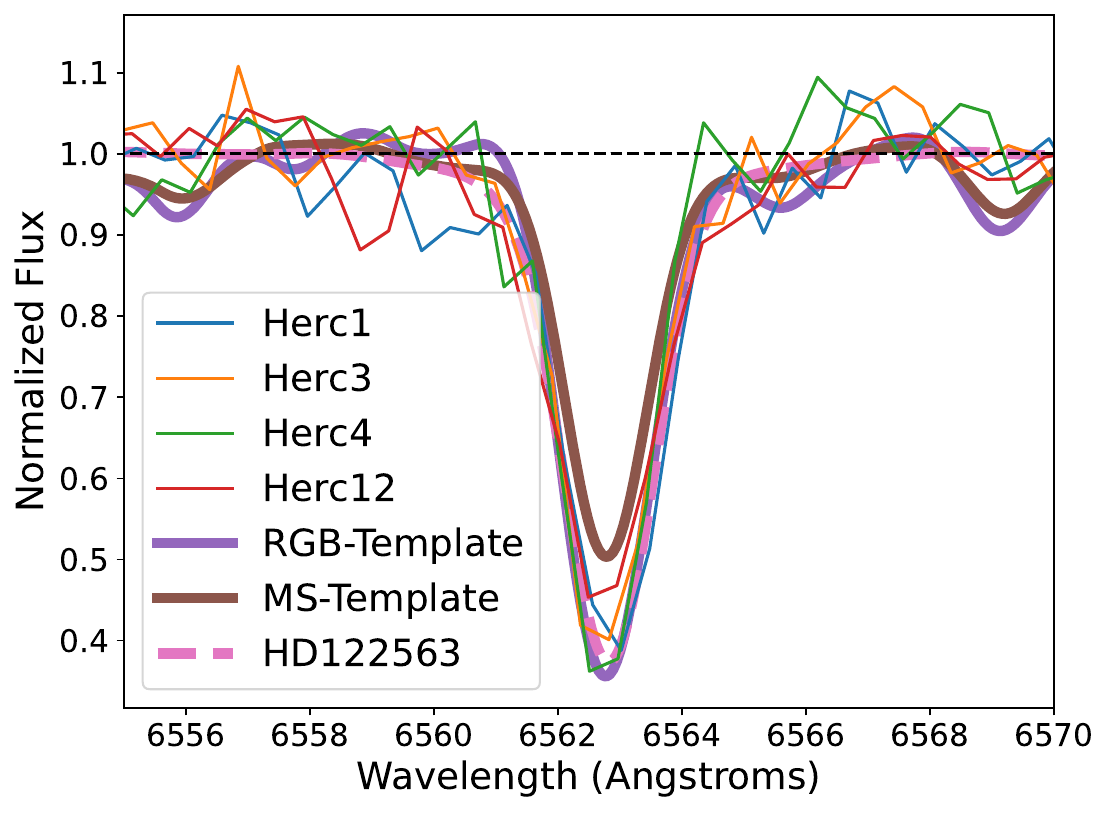}
     \caption{The H$\alpha$ absorption line 6563\,{\AA} from MagE spectra for our four candidate member stars of Hercules. 
     Spectra of K4 spectral standards (see Section~\ref{sec:foreground} smoothed to the resolution of our MagE are overplotted, along with a MagE spectrum of the metal-poor red giant HD122563. 
     Note that Herc1, Herc3, and Herc4 clearly more closely align with the H$\alpha$ feature of the HD122563 and the K4 RGB standard, suggesting they are on the red giant branch. 
     Herc12 has an H$\alpha$ feature that precludes a clear classification, but we argue it is a member in Section~\ref{sec:foreground} from foreground modeling with the Besancon simulation \citep{czekaj14}.}
    \label{fig:halpha}
\end{figure}

However, one way to further separate member stars in Hercules from Milky Way foreground stars is through their surface gravities. 
Any Hercules members that we observed with MagE should be on the red giant branch of Hercules, and thus have low surface gravities ($\log\,g < 3.0$).
We can verify whether our candidate Hercules members do indeed have the low surface gravities from their MagE spectra, using the H$\alpha$ line at $\sim6563\,${\AA}. 
This is because the level of broadening of H$\alpha$ is sensitive to the surface gravity of the star; for a fixed effective temperature H$\alpha$ will display more pressure broadening in cool main-sequence stars relative to what would be seen for stars on the red giant branch. 
We show examples of this in Figure~\ref{fig:halpha}, in which the H$\alpha$ feature of a main-sequence and a giant K4 spectral standard\footnote{http://www.astro.sunysb.edu/fwalter/SMARTS/Chiron\_Standards} are plotted along with the H$\alpha$ features of Herc-1 and Herc-12, in addition to the Hercules MagE members Herc-3 and Herc-4 that are closer to the center. 
The spectra of the K4 standard have been smoothed to match the MagE resolution.
The H$\alpha$ feature of Herc-1 clearly aligns with the giant standard as opposed to the main-sequence standard, suggesting that the it is red giant stars.
Visually, the H$\alpha$ feature of Herc-12 appears ambiguous.
We also compare the H$\alpha$ feature of these Hercules candidates with a MagE spectrum of HD122563 (a metal-poor red giant) obtained in \citet{chiti21}.
We find again that Herc-1 has a H$\alpha$ feature matching HD122563.
This information suggests that Herc-1 is very likely a member of Hercules.
Adding a $\log\,g < 3.0$ cut to the criteria in the previous paragraph removes all Milky Way stars in our Besancon query.  
This is because only 17\,\% of stars in our Besancon query that pass the initial selection criteria have $\log\,g < 3.0$ independent of radial velocity, a consequence of the relatively low density of red giant stars in the outer Milky Way halo.

For completeness, we note that Herc-12 was selected as a target based on broad-band photometry and proper motion cuts only, since it was outside the footprint of our IMACS CaII\,K imaging. 
We find that $\sim 1.4 \%$ of the stars in the Besancon catalog that satisfy the isochrone and proper motion cuts also satisfy the line-of-sight velocity criteria and have metallicity of [Fe/H] $< -2.0$. 
Observationally, the purely Gaia-selected sample of Hercules candidates included 28 stars within $1$\,\degree~from Hercules.
From this, we still expect $\lesssim 1$ Milky Way foreground star in this region to pass our selection cuts.
Accordingly, we classify Herc-12 as likely also a member, despite its H$\alpha$ feature precluding a clear classification of it as a star on the red giant branch. 
We note that this star was also identified as a member in the recent study by \citet{longeard23}.

\section{Analysis}
\label{sec:analysis}

\subsection{System Dynamics}
\label{sec:sysdynamics}

We study the dynamics of Hercules using the PCF sample. We adopt the same likelihood function used in \citet{gregory20} to perform a Markov chain Monte Carlo (MCMC) analysis using the MCMC affine invariant sampler \texttt{emcee} \citep{foremanmackey13}. Systemic velocity, velocity dispersion, line-of-sight velocity gradient, and position angle are fitted simultaneously via the likelihood function,
\begin{equation}
    \log{\mathcal{L}} = \sum^{N}_{i=1} \left( \log{\frac{1}{\sqrt{2\pi(\sigma^2+v_{\rm{err},i}^2)}}} \right) - \sum^{N}_{i=1} \left( \frac{\Delta v_{r,i}^2}{2(\sigma^2+v_{\rm{err},i}^2)} \right),
\end{equation}
where $\Delta v_{r,i}$ is the difference between the measured line-of-sight velocity of star $i$ ($v_{r,i}$) at position angle $\theta_i$ and the line-of-sight velocity calculated at an angular separation of the star from the center of the system ($R_i$) projected along the axis of the system at position angle $\theta_0$, assuming a velocity gradient $k=\frac{\partial v_r}{\partial R}$ along that axis with the systemic line-of-sight velocity ($v_0$), defined as
\begin{equation}
    \Delta v_{r,i} = v_{r,i} - (v_{0} + k R_{i} \cos{\theta_{i} - \theta_{0}}),
\end{equation}
$v_{\rm{err},i}$ is the uncertainty in the line-of-sight velocity for star $i$, and $\sigma$ is the line-of-sight velocity dispersion of the system.

The free parameters of the model thus include the systemic line-of-sight velocity $v_0$, the velocity dispersion $\sigma$, and the velocity gradient $k$ along the major axis of the system at position angle $\theta_0$, which is fixed to the major axis at $-72.6$\,\degree~or the minor axis at $17.4$\,\degree~ from \citet{sand09}. 
The prior for $v_0$ is flat between the minimum and maximum of the line-of-sight velocities of the sample.
The prior for $\sigma$ and $k$ are also flat between $[0,20]$\,\kmsec\ and $[-100,100]$\,\kmseckpc, respectively.
We summarize the results of this dynamical analysis in Table~\ref{tab:fit_res}.

\begin{deluxetable*}{cccc}
\tablecaption{Results of our dynamical analysis of Hercules in Section~\ref{sec:sysdynamics} using 28 stars in our purest sample of members without stars in potential binary systems (PCF; see Section~\ref{sec:membership}). We report results when fixing the position angle of the velocity gradient along each of the major and minor axes. Fitting result assuming no gradient is also included.}
\label{tab:fit_res}
\tablehead{\colhead{} & \colhead{No Gradient} & \colhead{Major Axis} & \colhead{Minor Axis}}
\startdata
\textbf{Systemic Velocity} & $45.7^{+0.5}_{-0.5}$~\kmsec & $45.5^{+0.5}_{-0.5}$~\kmsec & $45.6^{+0.5}_{-0.5}$~\kmsec \\
\textbf{Velocity Dispersion} & $1.8^{+0.6}_{-0.6}$~\kmsec & $1.9^{+0.6}_{-0.6}$~\kmsec & $1.8^{+0.6}_{-0.5}$~\kmsec \\
\textbf{Velocity Gradient} & \nodata & $1.8^{+1.8}_{-1.8}$~\kmseckpc & $4.5^{+2.3}_{-2.4}$~\kmseckpc \\
  & \nodata & $4.2^{+4.1}_{-4.2}$~\kmsecdeg & $10.3^{+5.3}_{-5.6}$~\kmsecdeg \\
\textbf{Position Angle of Gradient} & \nodata & $-72.6$\degree & $+17.4$\degree
\enddata
\end{deluxetable*}

\subsubsection{Systemic Line-of-sight Velocity \& Dispersion}
\label{sec:dispersion}

When fitting for the velocity gradient along the major axis, we derive a systemic line-of-sight velocity of $45.5^{+0.5}_{-0.5}$~\kmsec and a velocity dispersion of $1.9^{+0.6}_{-0.6}$~\kmsec from 28 stars. 
These values change negligibly when instead fitting for the gradient along the minor axis, or excluding the gradient term entirely (see Table~\ref{tab:fit_res}).
A likelihood ratio test suggests a non-substantial statistical preference ($K\lesssim3.2$) for the fits with a minor axis gradient over the fits with a major axis gradient. 
Both gradient fits show marginally substantial preference ($K\sim3.5$) over the no gradient fit, partly because of the extra free parameter ($k$) in the gradient fits.

The systemic velocity is consistent with the measurement from \citet{gregory20} ($46.1^{+1.3}_{-1.2}$~\kmsec) with almost half the uncertainty even though the sample size is similar (28 in this work vs. 21 in \citealt{gregory20}). 
This reduction in velocity scatter is also reflected in the velocity dispersion posterior, where our value ($1.9^{+0.6}_{-0.6}$~\kmsec) is significantly smaller than previous measurements ($4.5^{+1.4}_{-1.1}$~\kmsec; \citealt{gregory20}).
We attribute this decrease in the derived velocity dispersion to the high purity of the PCF sample, through new Gaia DR3 proper motion measurements to remove foreground contaminants, and long baseline observations to eliminate potential binaries.
We do recover the velocity dispersion measurement from \citet{gregory20} as we loosen the high-purity selection criteria used in this study.
The velocity dispersion increases to $4.9^{+0.7}_{-0.7}$\,\kmsec when potential binaries and stars with missing proper motion measurements are included.
When using a wider radial velocity selection window for membership (see paragraph 2 of Section~\ref{sec:membership}), this dispersion further increases to $6.0^{+0.9}_{-0.7}$~\kmsec, but remains stable at $1.9^{+0.6}_{-0.6}$~{\kmsec} for stars selected as having proper motion information consistent with membership.

\subsubsection{Line-of-sight Velocity Gradient}
\label{sec:rv_grad}

\begin{figure*}
    \centering
    \includegraphics[width=0.99\textwidth]{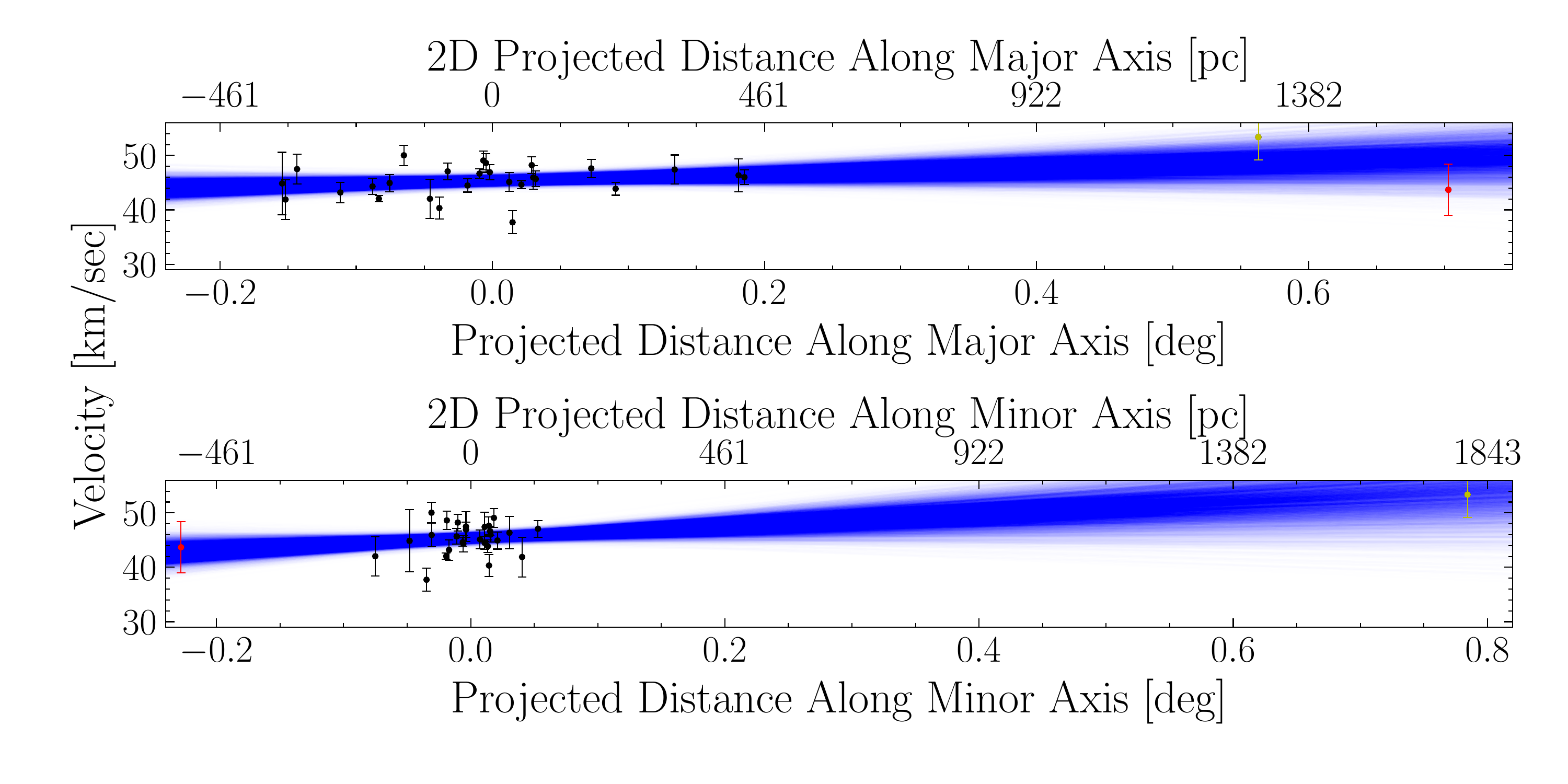}
    \caption{Line-of-sight radial velocities of our purest sample of members (PCF; see Section~\ref{sec:membership}) as a function of projected distance along the major (top panel) and minor (bottom panel) axis of Hercules. Herc-1(12) is marked red(yellow). 
    Blue lines show an ensemble of the fitted gradients as derived in Section~\ref{sec:sysdynamics}.}
    \label{fig:rv_grad}
\end{figure*}

The simultaneously fitted line-of-sight velocity gradient, $1.8^{+1.8}_{-1.8}$~\kmseckpc ($4.2^{+4.1}_{-4.2}$~\kmsecdeg), along the major-axis of Hercules provides a tighter constraint on a potential gradient caused by tidal disruption compared to previous studies \citep{gregory20,longeard23}.
Results from fitting along the minor axis are also consistent with values from \citet{gregory20} and \citet{longeard23} with a similar level of improvement in velocity gradient uncertainties.
We show the velocities of stars as a function of the projected distance along the major/minor axis of Hercules in Figure~\ref{fig:rv_grad}.

\subsection{Proper Motions}
\label{sec:analysis_pm}

The weighted average systemic proper motion of the PCF sample is $\mu_{\alpha^*}=-0.069 \pm 0.044$ and $\mu_\delta = -0.379\pm0.037$~\masyr. This is consistent with previous measurements from \citet{gregory20} at $\mu_{\alpha^*}=-0.153 \pm 0.074$ and $\mu_\delta = -0.397 \pm 0.063$~\masyr. 
The smaller statistical errors in our sample arise from the larger sample size and more precise proper motion measurements from \textit{Gaia} DR3. 
For completeness, we note that the systemic proper motion calculated from Gaia DR2 measurements for our PCF sample is $\mu_{\alpha^*}=-0.210 \pm 0.088$ and $\mu_\delta = -0.413 \pm 0.069$~\masyr, still consistent with \citet{gregory20}. 
We adopt our systemic proper motion from the whole PCF sample in the following dynamical modeling (Section~\ref{sec:dyn_model_res}) and tidal radius study (Section~\ref{sec:enclosed_mass}).

We stress that this agreement with \citet{gregory20} does not purely result from the proper motion selection criterion for membership applied in Section~\ref{sec:membership}.
If we only apply the proper motion and photometry selection criteria (44 stars), the resulting systemic proper motion is $\mu_{\alpha^*}=-0.049 \pm 0.039$ and $\mu_\delta = -0.359 \pm 0.033$~\masyr, which are further from, but still consistent with, applying the full four criteria for membership in Section~\ref{sec:membership}.
Our results are also consistent with measurements from studies that use purely photometry and proper motion information to select Hercules members. 
\citet{pace22} reported a systemic proper motion for Hercules as $\mu_{\alpha^*}=-0.035^{+0.042}_{-0.042}$ and $\mu_\delta = -0.339^{+0.035}_{-0.036}$~\masyr.

\section{Discussion and Interpretation}
\label{sec:discussion}

We evaluate the evidence for tidal disruption scenarios in Hercules in this section. We focus on the updated prediction for the tidally disrupted stream track from our dynamical modeling, highlighting the alignment between the predicted track and data.
We also address the apparent disagreement with the location of our distant members, Herc-1 and Herc-12. 
We then analyze the effect of the central dark matter (DM) density profile (i.e., core/cusp) on the tidal disruption scenario based on the evolution of the tidal radius of the system. Lastly, we consider Hercules more broadly and interpret what a tidally disrupted Hercules means to the general UFD population and what future studies should focus on when testing tidal disruption in other UFDs. 

\subsection{Dynamical model}
\label{sec:dyn_model_res}

We run a suite of dynamical models of stream formation, using the on-sky position and distance of Hercules as reported in \citet{belokurov07} and \citet{musella12} along with the proper motion and radial velocity measured in this study. In particular, we use a modified Lagrange Cloud Stripping technique developed by \citet{Gibbons:2014}, as applied in e.g. \citet{Erkal:2019, Shipp:2021, Koposov:2023}. We select initial parameters for the progenitor system by sampling the measurement uncertainties, rewind the progenitor orbit within the gravitational potential of the Milky Way (MW) and the Large Magellanic Cloud (LMC), and then simulate tidal disruption by ejecting particles at the Lagrange points of the progenitor and evolving them forward in the joint potential of the progenitor, MW, and LMC. This technique includes the reflex motion of the Milky Way in response to the LMC \citep[e.g.,][]{erkal21} which can create misalignments between tidal debris and the progenitor's motion \citep[e.g.,][]{shipp19,li21,ji21,battaglia22}. We note that while this technique is not designed to accurately model the disruption rate of the progenitor (or the progenitor's properties), it is designed to predict the resulting morphology of the tidal debris. We can therefore use this method to determine the likely on-sky orientation, velocity gradients, and distance gradients of stars tidally stripped from Hercules.

We use a realization of the Milky Way potential from \citet{McMillan:2017} with a mass of $M_{\rm MW} = 8.3 \times 10^{11} M_{\rm \odot}$, which provides the best fit to known stellar streams, as described in \citet{Shipp:2021}. The LMC potential includes a Miyamoto-Nagai \citep{Miyamoto:1975} stellar disk and a dark matter halo modelled as a Hernquist profile \citep{Hernquist:1990}, as described in \citet{Shipp:2021} and \citet{Ferguson:2022}. The Hercules system is described as a Plummer sphere \citep{Plummer:1911} with a dynamical mass of $M = 0.1 \times 10^{7} M_{\odot}$ as calculated in Section~\ref{sec:enclosed_mass} and a scale radius of $216\ \mathrm{pc}$ \citep{munoz18}.

\begin{figure}
    \centering
    \includegraphics[width=0.45\textwidth]{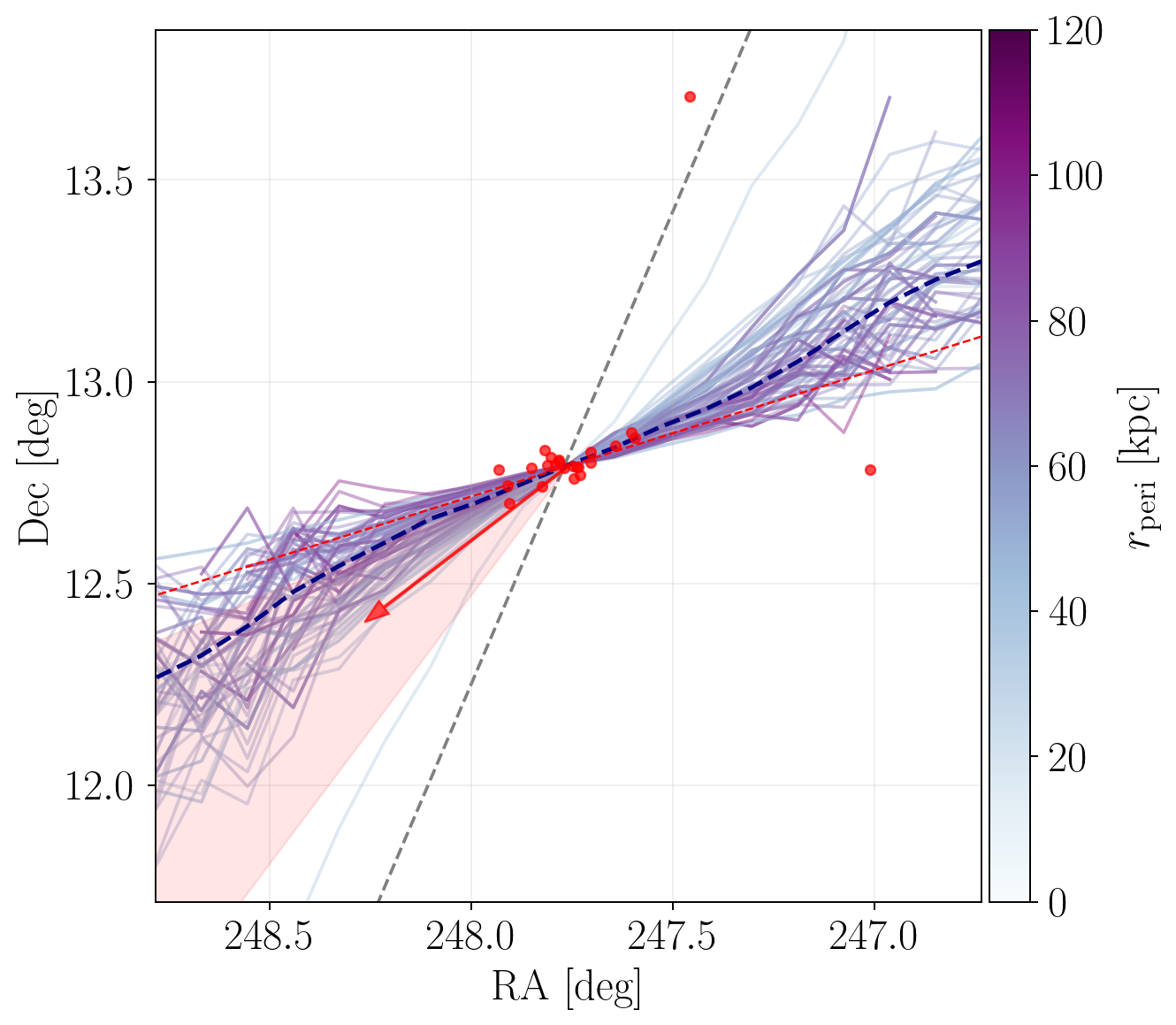}
    \caption{Dynamical modeling results of Hercules stream formation as described in Section~\ref{sec:dyn_model_res}. The individual tracks are shown in solid lines color-coded by the last pericentric passage distance of Hercules in each model. The navy dashed line represents the median track of the stream. 
    The red dots are the stars in our highest purity sample (PCF; see Section~\ref{sec:membership}).  
    The red arrow and its shaded region are the orbital direction and its uncertainty, derived from the input systemic proper motion. 
    The red dashed line is the orientation of the major axis of Hercules, which overlaps with the predicted stream within 1\,$\sigma$ uncertainty.
    The grey dashed line connects the center of Hercules (denoted by a red star) with the Galactic center.}
    \label{fig:herc_model}
\end{figure}

Our best-fit dynamical model produces a stream track aligned with the major axis of Hercules, as shown in Figure~\ref{fig:herc_model}. This is consistent with a tidal disruption scenario, suggesting that Hercules would indeed be extended along its observed major axis. In addition, we find that the model predicts a misalignment between the reflex motion-corrected weighted mean proper motion and the track of the stream. The observed offset between the proper motion direction and the major axis of Hercules is, therefore, not inconsistent with tidal disruption. 

Among the models of the tidal disruption of Hercules, we find a clear trend in the pericentric distance with respect to the orientation of the stream track, as shown in Figure~\ref{fig:herc_model}. Interestingly, we find that the models that are better-aligned with the elongation of Hercules have relatively large pericentric distances. Overall, the models predict a pericentric distance of $62.3^{+ 12.2}_{- 15.0}$ kpc and an apocentric distance of $314.9^{+ 54.1}_{-132.7}$ kpc. The models most closely aligned with the observed orientation of Hercules have pericenters close to the median value, $\sim 60$ kpc.
The orbit of Hercules is qualitatively similar with and without the presence of the LMC, which is in agreement with \citet{pace22}. We thus consider solely the MW potential in the following analysis in Section~\ref{sec:enclosed_mass}.

While the dynamical models can explain most of the stars in the PCF sample, Herc-1 and Herc-12, the distant members identified in this study, are located well off of the median predicted track. However, in some cases the predicted track does pass through the on-sky position of Herc-1, as shown in Figure~\ref{fig:herc_model}. These particular models require a closer pericentric passage, within 20 kpc of the Galactic center.   
In addition, we do find a few realizations of our models ($\sim 1\%$) in which the simulated stream shows a more complex morphology, with a component extended perpendicular to the primary stream track that would be consistent with the locations of Herc-1 or Herc-12. These models tend to have completed multiple pericentric passages and have therefore experienced multiple stripping events. However, the modified Lagrange Cloud Stripping technique used here does not fully model the behavior of stars close to the progenitor location. Therefore, further exploration of tidal disruption explanations of the positioning of Herc-1 would be best explored with full N-body simulations.

From a physical point of view, we may expect stars to appear at such positions relative to their host progenitor in general tidal stream formation. Tidal streams form first by radial distortion, then extend into an elongated stellar stream due to the resulting differential in orbital velocity \citep{kupper12}.
If Herc-1 is a signature that Hercules is in the early stages of stream formation, then its location should be aligned with the radial direction connecting Hercules and the Galactic center. 
We illustrate the direction from the center of Hercules to the Galactic center in Figure~\ref{fig:herc_model}. We find that Herc-1 is not along this direction, meaning its location is not readily explained as the remnant of an early-stage stream formation in Hercules. 
Thus, Herc-1 is likely stripped to its current location via other mechanisms, the full exploration of which is beyond the scope of this study. 
Herc-12, on the other hand, aligns with the direction towards the Galactic center.

\begin{figure}
    \centering
    \includegraphics[width=0.45\textwidth]{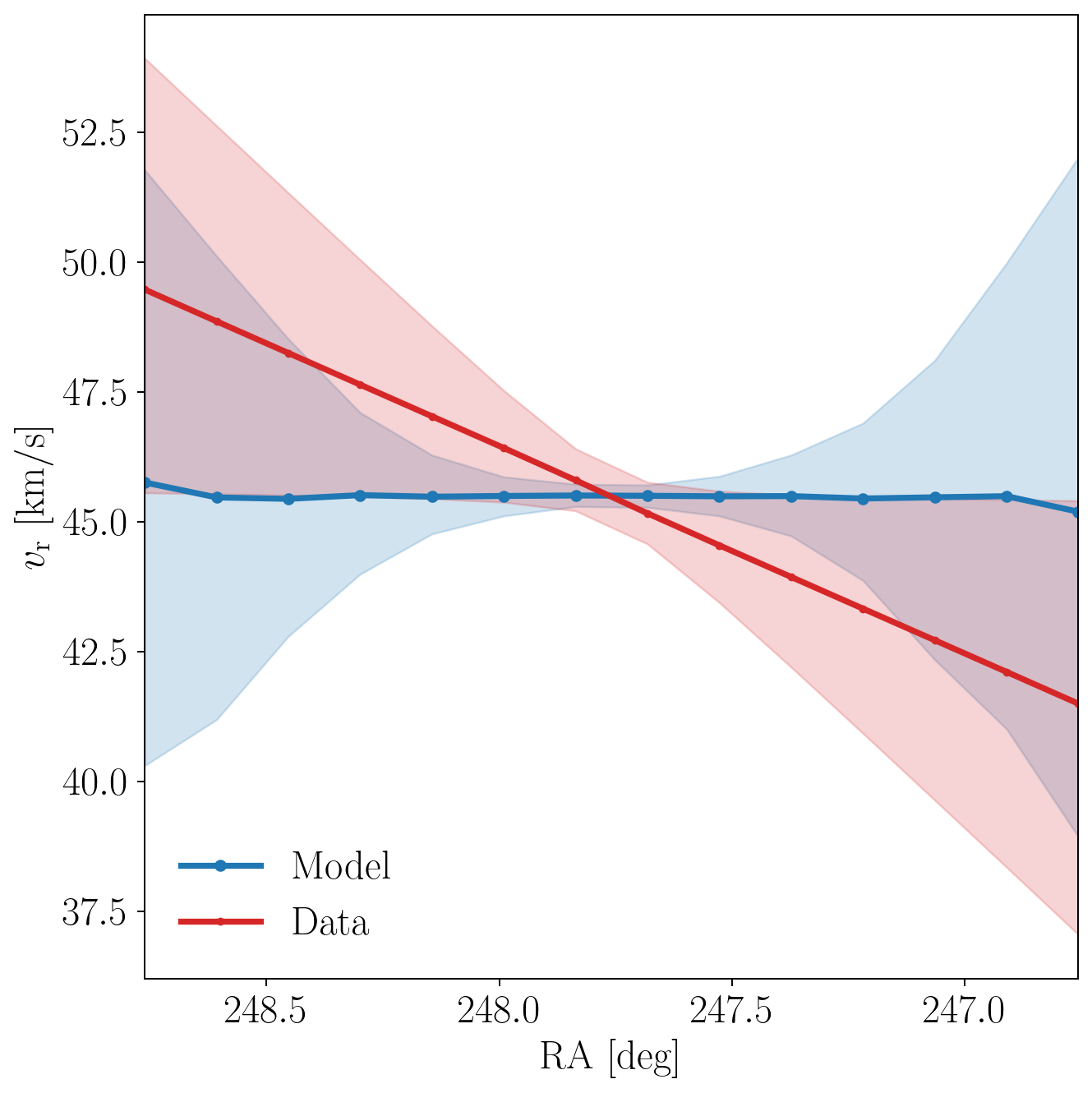}
    \caption{Line-of-sight radial velocity gradients as a function of RA. 
    Similar to Figure~\ref{fig:herc_model}, the red line and shaded region represent the observed line-of-sight velocity gradient and associated 1\,$\sigma$ uncertainty (see Section~\ref{sec:sysdynamics}), whereas the blue line and shaded region represent the predicted velocity gradient from the dynamical models in Section~\ref{sec:dyn_model_res} and associated uncertainty. 
    The model is consistent with the observed data, predicting a very weak line-of-sight radial velocity gradient over $\sim1.5$\,\degree\ in RA.}
    \label{fig:rv_model}
\end{figure}

Given the overall good agreement in the sky position between the model and the bulk of observed Hercules members, we now compare the velocity gradient predicted by the model with the observed velocity gradient. 
From Section~\ref{sec:rv_grad}, we find no evidence of line-of-sight velocity gradient along the major axis of Hercules, which is predicted to be present by \citet{martin10} if Hercules is forming a stellar stream. 
Our fitting results to the available Hercules radial velocity data along the major axis and minor axis are both consistent with zero at a 1-2\,$\sigma$ level. 
Our dynamical model of a tidally disrupting Hercules predicts a relatively flat line-of-sight velocity gradient near the center of Hercules $0.00^{+0.97}_{-0.91}$~\kmseckpc, entirely consistent with our current limits on the non-detection of a gradient as shown in Figure~\ref{fig:rv_model}. 
We see that the current observational uncertainties on the systemic line-of-sight velocity and its gradient are around two times as large as the range spanned by the models. 
Consequently, a larger sample size of Hercules members and/or smaller velocity uncertainties for stars in the outskirts are needed to conclusively test models of its tidal disruption via velocity gradient analysis.
Such a sample will more readily be accessible with the next generation of 30\,m-class telescopes.

We note that the results inferred from these dynamical models are sensitive to the input systemic proper motion of Hercules. 
The general misalignment between the orbital direction of Hercules and its major axis is less significant if we adopt the systemic proper motion from \citet{pace22}. 
A less significant misalignment allows the simulated models with larger pericentric distances to more plausibly align with the bulk of Hercules members, in contrast to the slight misalignment of models with high pericenters in Figure~\ref{fig:herc_model}. 
It is thus still plausible that the elongation in Hercules is simply aligned with its motion, as expected from a simple tidal disruption scenario \citep{martin10}.
But we highlight that predicted tidal disruption tracks can still be aligned with the elongation of Hercules, even if its proper motion is misaligned (see Figure~\ref{fig:herc_model}). 

The discussion in the previous paragraph highlights the importance of a clean sample of members with complete chemodynamic measurements in studying Hercules, and UFDs more generally. 
For the sample size of only $28$ stars in our dynamical analysis, a few foreground contaminants can significantly alter our interpretation of the system's orbit. 
Future observations may help distinguish whether or not Hercules is tidally disrupting by filling-in missing radial velocity information, supplemented by photometric metallicity techniques to flag low metallicity stars.

We additionally note that the intrinsic rotation of the progenitor system can also induce velocity gradients \citep[e.g.,][]{battaglia08,martinez-garcia21,martinez-garcia23}.
Modeling intrinsic rotation, however, is beyond the capability of our current technique.

\subsection{The enclosed mass and tidal radius of Hercules}
\label{sec:enclosed_mass}

We compute and discuss in this section the tidal radius of Hercules by assuming a generalized NFW (gNFW) dark matter profile for Hercules. We specifically test if the tidal radius is compatible with the distance of Herc-1 and Herc-12 from the center of Hercules. 
For illustrative purposes, given Herc-1 and Herc-12 are similarly $\sim$1\,\degree from the center of Hercules, we mainly consider Herc-1 below.

\begin{figure}
    \centering
    \includegraphics[width=0.45\textwidth]{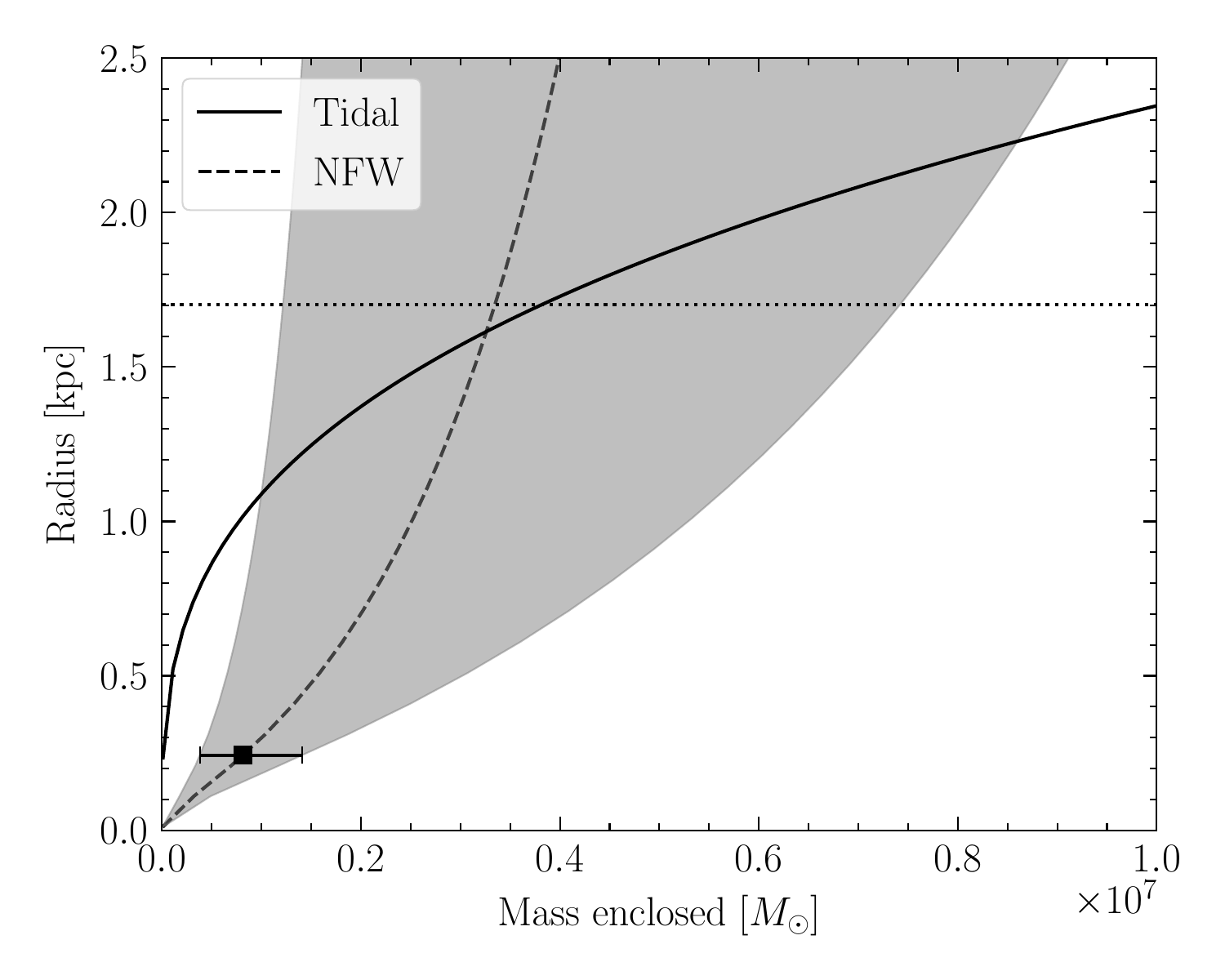}
    \caption{The tidal radius of an object in the Milky Way potential at the location of Hercules as a function of enclosed mass is shown as a solid black line. The square marks the constraint on the gNFW profile from the dynamical mass estimate, with the error bar from the uncertainty in the velocity dispersion. The radius out to which a gNFW profile ($\beta=1$) encloses a given mass is shown as a dashed line, with the shaded area corresponding to the range of the profile due to uncertainty in the velocity dispersion. The tidal radius and gNFW curve intersect between $\sim 1.1$ and $\sim 2.3$\,kpc, naively suggesting this as the range of the current tidal radius of Hercules assuming a gNFW profile. The dotted horizontal line represents the location of Herc-1 of 1.7\,kpc.
    }
    \label{fig:tidal_NFW}
\end{figure}

We compute the tidal radius of Hercules for a given enclosed mass using the \texttt{rtide} function from the \textsc{galpy} library \citep{bovy15} under the Milky Way potential \textit{MWPotential14}. Assuming the nominal values for the coordinates and distance of Hercules, we get the tidal radius of Hercules as a function of enclosed mass, shown as the solid curve in Figure~\ref{fig:tidal_NFW}.

We directly solve for the tidal radius of Hercules by further assuming an underlying dark matter halo profile. We choose a gNFW profile of the form
\begin{equation}
    \rho_{\rm{gNFW}} (r) = \frac{M_0}{4 \pi r_s^3} \frac{1}{(r/r_s)^\beta(1+r/r_s)^{3-\beta}},
    \label{eq:dens_gNFW}
\end{equation}
where $M_0$ is the mass normalization, $r_s$ is the scale radius, and $\beta$ is the characteristic power for the inner part of the potential. When $\beta=1$, we recover the standard NFW profile. We test three cases of potential gNFW dark matter profiles for Hercules with the inner power law slope ($\beta$) at $0.25$, $0.75$, $1$, and $1.25$.

The gNFW profile is required to have an enclosed mass in the half-light radius consistent with the dynamical mass computed according to \citet{wolf10} ($\sim 0.10 \times 10^7$~\msun). We further require that the concentration of the profile ($c_{200}$) is consistent with the concentration-mass relation from \citet{dutton14}, which gives $c_{200}\simeq29$ for $\beta=0.25,0.75,1,1.25$. The gNFW profile is then uniquely defined, and the resulting enclosed mass as a function of radius (for $\beta=1$ case) from the center of Hercules is shown as the dashed curve in Figure~\ref{fig:tidal_NFW}. The tidal radius of Hercules at its current location can then be approximated from the intersection of the solid and dashed curves in Figure~\ref{fig:tidal_NFW}. 

We find that the tidal radius of Hercules is between $\sim 1.1$ and $\sim 2.3$\,kpc for a gNFW profile with its inner power law slope $\beta=1$. 
This range of tidal radius is consistent with the projected separation of Herc-1 from the center of Hercules.
Naively, this would imply Herc-1 is plausibly still bound to the system if the underlying dark matter profile is intact for Hercules at its current location. 
However, this result is entirely dependent on the location of Hercules in the Milky Way potential and thus evolves as Hercules orbits the Milky Way. 
In particular, the tidal radius ought to have been smaller when Hercules was closer to the Galactic center.

\begin{figure}
    \centering
    \includegraphics[width=0.45\textwidth]{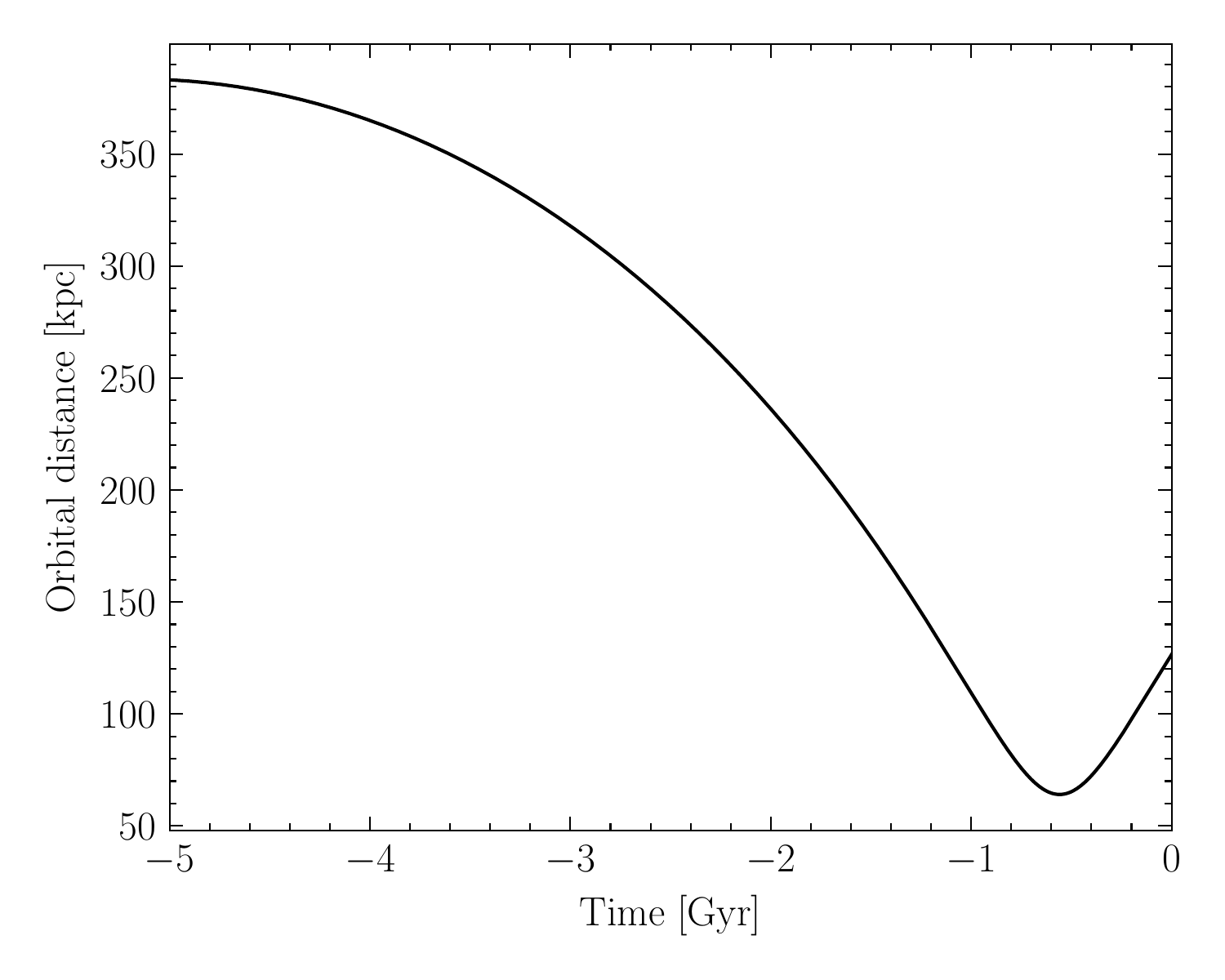}
    \caption{Orbital distance of Hercules as a function of time. The orbital is integrated backwards in time for 5\,Gyr. The shape of the curve qualitatively traces that of the tidal radius of Hercules in Figure~\ref{fig:rtidal_vs_t}, expected as the tidal radius strongly depends on the mass of the Milky Way enclosed within the orbital radius. 
    }
    \label{fig:rorbital_vs_t}
\end{figure}

\begin{figure*}
    \centering
    \includegraphics[width=0.95\textwidth]{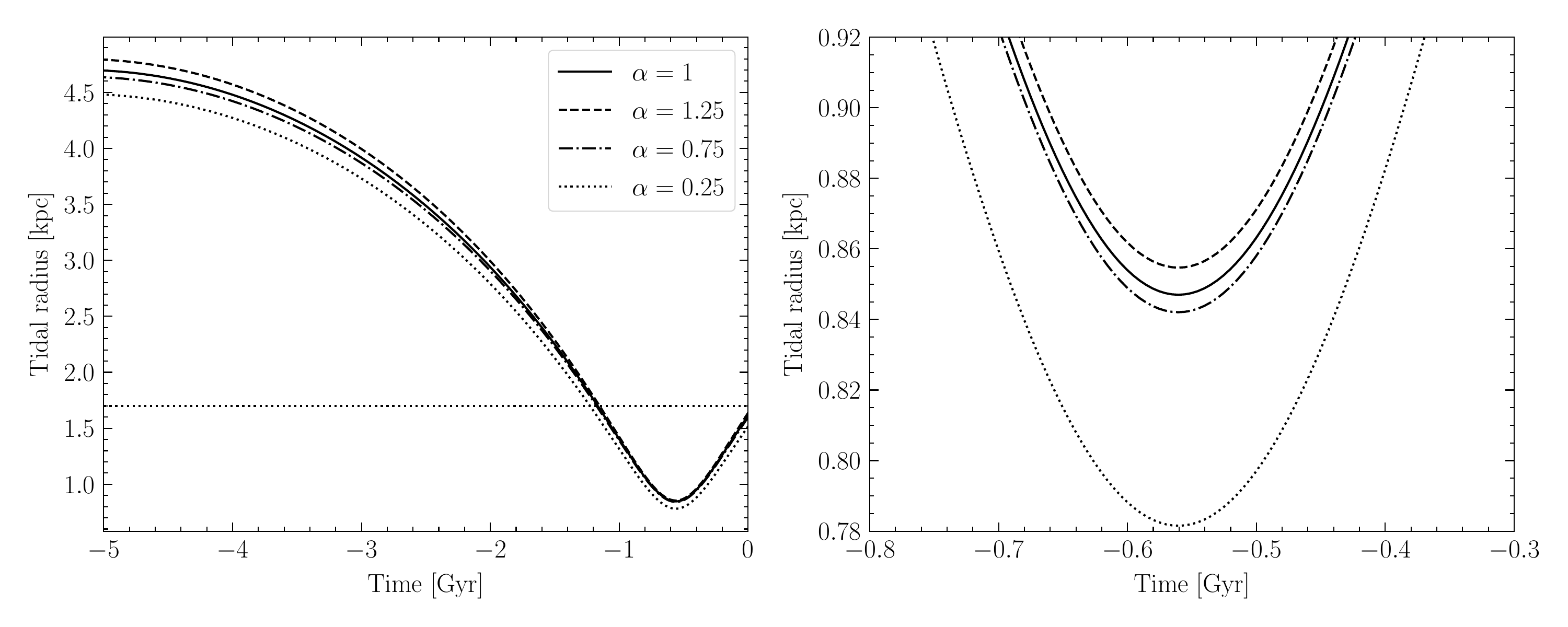}
    \caption{Tidal radius of Hercules as a function of time, assuming three cases of underlying gNFW profiles for Hercules. The horizontal dotted line in the left panel represents the location of Herc-1. The right panel shows a zoom-in around the pericenter passage of Hercules. The tidal radius at pericenter is fairly insensitive to these assumed models of the gNFW profile. 
    }
    \label{fig:rtidal_vs_t}
\end{figure*}

We thus model the tidal radius of Hercules as a function of the orbital phase, examine if Herc-1 remains within the tidal radius, and study how $\beta$ affects the tidal radius throughout the orbit. 
Assuming the same Milky Way potential \textit{MWPotential14}, we integrate the orbit of the Hercules backward in time for 5\,Gyr in steps of $0.005$~Gyr using \textsc{galpy}, as shown in Figure~\ref{fig:rorbital_vs_t}.
As pointed out in Section~\ref{sec:dyn_model_res}, the LMC potential has minimal effect on the orbit of Hercules and is thus omitted from the orbit integration here.
At each time step, we calculate the tidal radius following the above procedure so that the enclosed mass is consistent with the extrapolated gNFW profile. 
We find, for three out of four cases of gNFW profiles, that the tidal radius drops to $\sim0.85$~kpc at the pericentric passage of Hercules, which happens at $\sim0.57$~Gyr ago at $r\sim64$~kpc (see Figure~\ref{fig:rtidal_vs_t}). 
Coincidentally, this lowest tidal radius corresponds to an enclosed mass of $\sim 0.1 \times 10^7$~\msun, on the same order of magnitude with the dynamical mass within the half-light radius, $\sim 0.1 \times 10^7$~\msun, calculated from the observed velocity dispersion according to \citet{wolf10}.
In other words, when simply considering the dynamical mass as a lower mass bound for the tidal radius calculation, the tidal radius for Hercules at its current location has a lower limit of $\sim0.85$~kpc.

This lowest tidal radius is much lower than what is needed to keep Herc-1 bound at $\sim1.7$~kpc; although, its location when Hercules passed pericenter may have been closer to the center of the system. 
The previous pericentric passage of Hercules ought to have unbounded most of the mass of Hercules beyond $\sim0.85$\,kpc, including Herc-1 and Herc-12, in our simplistic modeling of its tidal radius over time (e.g., assuming the dynamical mass is constant and that the gNFW profile is invariant along the orbit). 
Accordingly, stars residing on the outskirts of Hercules may have experienced tidal disruption during its last pericentric passage, even if they are within its current tidal radius.
Herc-1 and Herc-12 could represent stars originally at $\sim0.85$\,kpc from the center of Hercules that were tidally disrupted to their current location. 
This possibility is discussed briefly in Section~\ref{sec:dyn_model_res} in the context of the early stages of stream formation.

Additionally, we note that $\beta$, the inner profile power law index, has a minimum effect on the tidal radius at the pericentric passage (see second panel of Figure~\ref{fig:rtidal_vs_t}). 
The tidal radius for the most cored case ($\beta=0.25$) is only $\sim10\%$ lower than the most cuspy case ($1.25$).
This is expected as the central density has a larger impact when the tidal radius approaches the central region of the system \citep{penarrubia08}. 
At the pericenter of Hercules, the tidal radius still only probes the outskirts of the stellar halo, and the enclosed mass at this distance does not change significantly for different $\beta$. 
It is thus hard to constrain the central density profile using stars in the outskirts of the stellar halo, in particular for Hercules.

Instead, we note that the enclosed mass (and thus, the tidal radius) needed to keep Herc-1 bound is very sensitive to the assumed Milky Way halo mass profile. 
If we increase the Milky Way halo mass by an additional factor of two, following \citet{carlin18}, the Hercules enclosed mass needed to maintain a constant tidal radius increases.
We find that we need $\sim0.71\times10^7$ solar mass to keep Herc-1 bound to the system. 
While this is still on the same order of magnitude as the dynamical mass ($\sim0.1\times10^7$ solar mass), it is meaningfully more massive than what is needed to bound Herc-1 before this correction and approaches the total enclosed mass of the gNFW profile (see Figure~\ref{fig:tidal_NFW}). 
However, constraining the halo mass of the Milky Way is beyond the scope of this work. 
We leave the matter to future studies to explore the possibility of using tidally (not) disrupted UFDs to constrain the density profile of the Milky Way.

\subsection{Hercules in the big picture}
\label{sec:big_pic}

In this section, we examine the tidal disruption of Hercules in the broader picture concerning other Milky Way UFDs. 
In particular, we discuss whether the fact that Hercules may be tidally disrupting has any implication for other Milky Way UFDs with similar orbital histories or morphologies.

\citet{pace22} studied several common diagnostics for testing the tidal influence of the Milky Way on its satellite galaxies. They concluded that the ratio of the average dwarf density within its half-light radius ($\rho_{1/2}$) to the average MW density at the dwarf's pericenter ($\rho_{\rm{MW}}(r=r_{\rm{peri}})$) is the most indicative sign of tidal disruption. In their assessment, dwarf galaxies with $\rho_{1/2}/\rho_{\rm{MW}}(r=r_{\rm{peri}}) \lesssim 10$ are likely to have experienced tidal disruption.
Hercules, with the velocity dispersion measured in this study, has a $\rho_{1/2}/\rho_{\rm{MW}}(r=r_{\rm{peri}}) \simeq 35$. While this value is higher than the proposed cut at $10$, it is nonetheless one of the lowest among dwarfs currently not showing clear tidal disruption features, shown in the left panel of Figure~6 in \citet{pace22}.

Given what we find in Section~\ref{sec:enclosed_mass}, we propose that dwarf galaxies with $\rho_{1/2}/\rho_{\rm{MW}}(r=r_{\rm{peri}}) \gtrsim 10$ could also be potentially experiencing tidal disruption. 
Specifically, the criterion proposed by \citet{pace22} is based on the expectation that tidal disruption will happen when the half-light radius is similar to the tidal radius. 
Several studies (e.g., \citealt{chiti21,filion21,longeard22,tau23,jensen24}) have shown that UFDs can have extended stellar halos up to several half-light radii away from the center. 
These stars would naturally be susceptible to tidal influences when the tidal radius is much larger than the half-light radius, which in turn allows tidal influence to shape UFDs at larger $\rho_{1/2}/\rho_{\rm{MW}}(r=r_{\rm{peri}})$.
Low surface brightness tidal features around some Milky Way satellites are predicted by recent simulations (e.g., \citealt{shipp23}).

Furthermore, we argue that the dynamical modeling of stream formation, as described in Section~\ref{sec:dyn_model_res}, is a powerful tool for evaluating potential tidal disruption in future studies of UFDs.
The apparent misalignment between the major axis and orbital direction of Hercules is still explainable with the dynamical model of stream formation. 
Additionally, the model further provides insight into the expected line-of-sight velocity gradient and the required precision for excluding tidal disruption with this particular diagnostic.

\subsection{Comparison with \citet{longeard23}}
\label{sec:comp_pristine}

During the submission phase of this study, we were made aware of a separate dedicated search of Hercules members from the Pristine survey \citep{longeard23}. Using the Anglo-Australian Telescope (AAT), the study identified three new members for Hercules: Her\_3, Her\_5, and Her\_180, as well as three stars with uncertain membership: Her\_6, Her\_10, and Her\_464. 

We compare our sample with the six (candidate) members from \citet{longeard23}. Considering the full combined sample, we have one overlapping star: labeled Herc-12 in our sample, and Her\_5 in the AAT sample. 
Interestingly, our Herc-1 is in the selection Field 1 from \citet{longeard23} but was not observed.
Additionally, three stars are identified as PCF members from the radial velocity measurements from \citet{brown14}, PanSTARRS ID~123232479063337871, 123442476022168188, and 123282478240779830.
These three stars are not observed in \citet{longeard23}.

\citet{longeard23} report a systemic line-of-sight velocity of $45.7^{+2.3}_{-3.7}$~\kmsec\ and a line-of-sight velocity gradient of $1.6^{+10.0}_{-3.8}$~\kmseckpc, both consistent with our measurements, although with larger uncertainties. We note that the larger uncertainty may be partially driven by the difference in the analysis method, as their likelihood function simultaneously factors in potential contaminants in the sample. This is complementary to our approach, as we remove foreground contaminants by applying all selections prior to the dynamical analysis.

The velocity dispersion from \citet{longeard23}, $8.0^{+1.4}_{-2.0}$~\kmsec, is significantly larger than our velocity dispersion. 
Such an inflated velocity dispersion would imply a dynamical mass of $\sim 1.4 \times 10^7$~\msun\ within the half-light radius. 
This mass implies an underlying dark matter halo with virial mass $\sim10^{10}$~\msun, assuming an NFW profile as described in Section~\ref{sec:enclosed_mass}. 
With a dark matter halo this massive, the tidal radius for Hercules would be $\sim20$~kpc. 
In Section~\ref{sec:dispersion}, we re-calculate the velocity dispersion using different criteria for membership.
We note that not excluding binaries and not limiting our study to a sample of proper motion-confirmed Hercules members (the PCF sample) when selecting an initial range of velocities for membership would have led us to derive a dispersion of $6.0^{+0.9}_{-0.7}$\,\kmsec, consistent with \citet{longeard23}.
Overall, our measurements provide tighter constraints on all commonly measured dynamical quantities. 
Differences in the measurements, when present, may be attributed to the different analysis methods of assigning membership and/or the influence of binaries.

\section{Conclusions}

We present the largest clean sample of Hercules member stars, with 33 stars (of which five exhibit evidence of binarity) confirmed with photometric, spectroscopic, and astrometric observations. 
We combine new spectroscopic observations from Magellan with literature data from \citet{simon07,aden09,brown14,gregory20} to build a catalog of 411 stars in the Hercules field with spectroscopic data. 
Foreground halo stars are removed by applying selections based on line-of-sight velocity, metallicity, color-magnitude, and proper motion information. 
We flag and remove potential binaries using multi-epoch line-of-sight velocity measurements where available. 
Our sample includes a new member (Herc-1) identified at $\sim 7$ half-light radii away from the center of Hercules. 
Foreground analysis indicates that this star is unlikely to be a Milky Way halo contaminant. 
Key takeaways are:

\begin{itemize}
    \item Our sample provides the most stringent constraint on the systemic line-of-sight velocity, velocity dispersion, and proper motion of Hercules currently in the literature based on 28 stars. 
    The systemic line-of-sight velocity, $45.5^{+0.5}_{-0.5}$~\kmsec, is consistent with previous studies.
    However, the velocity dispersion, $1.9^{+0.6}_{-0.6}$~\kmsec, is significantly lower than previous measurements. 
    Our systemic proper motion is consistent with \citet{gregory20} and \citet{pace22}. 
    We attribute the decrease in velocity dispersion to our clean selection strategy for Hercules members, which is necessary since its systemic velocity is not cleanly separated from the Milky Way foreground. 
    The reduced velocity dispersion can be anticipated after removing binaries in the sample, as they are known to artificially increase the velocity dispersion.
    
    \item 
    The long spatial baseline provided by Herc-1 and Herc-12 enables a $50\%$ tighter line-of-sight velocity gradient constraint at $1.8^{+1.8}_{-1.8}$~\kmseckpc\ compared to previous studies \citep{gregory20}.
    We find no conclusive evidence of a line-of-sight velocity gradient at a 95\% confidence interval [-1.8, 5.4].
    However, our dynamical modeling analysis in Section~\ref{sec:dyn_model_res} indicates that a tidally disrupting Hercules may still show a very small line-of-sight velocity gradient, consistent with our measured value.
    As demonstrated in this work, the radial extent of the sample is important in constraining the gradient by providing a long baseline.

    \item Our orbital integration analysis reveals that the elongation of Hercules can be reasonably explained by tidal interaction with the Milky Way. 
    The median track produced by our suite of dynamical models predicts that if Hercules is undergoing tidal disruption, then tidally displaced stars are expected to distribute along a major axis that is not necessarily aligned with its orbital direction, but well-aligned with the observed elongation of Hercules. 
    We find that the tidal radius inferred from assuming an underlying gNFW dark matter profile is only $\sim0.85$\,kpc at the pericenter, lower than the projected separation of our distant members Herc-1 and Herc-12 from the center of Hercules. 
    This makes it possible that Herc-1 and Herc-12 are tidally stripped to their current location, although the exact mechanism for their exact location with respect to the center of Hercules remains to be studied.

\end{itemize}

In summary, our study has shown principal evidence of tidal stripping in the extended stellar halo of Hercules. 
While the line-of-sight velocity gradient is still inconclusive, any reasonably predicted gradient is within the limits of our observations. 
The tidal radius analysis, given that Hercules has passed its pericenter, indicates that our distant members Herc-1 and Herc-12 are plausibly unbound.  
Moreover, the elongation of Hercules is readily explained by tidal disruption tracks, even if its orbital motion is misaligned with its elongation. 
The dynamical modeling of stream formation used in this study may be used to assess tidal disturbances in other UFDs, in particular, as stars are discovered in the outskirts of these systems.


This study has also demonstrated the importance of eliminating foreground contamination and binary stars in studying the dynamics of a UFD. 
This is especially the case for Hercules, as its systemic line-of-sight velocity is not well separated from that of the Milky Way halo, making potential member samples prone to foreground contamination.
We emphasize, again, the value of UFD member stars at large spatial distances from the center of these systems. 
While difficult to identify, they are necessary in constraining key dynamical properties of UFDs, such as the line-of-sight velocity gradient and tidal disruption signatures.
Notably, we find that the tidal radius of a UFD is quite sensitive to the assumed mass profile of the Milky Way halo, making their past/ongoing disruption a possible probe for the Milky Way potential at the location of the UFD.
With the advent of deep large astrometric and photometric surveys, combined with reliable photometric metallicity techniques, future studies will inevitably push the discovery frontier of UFD outskirts to better understand their evolution and dynamical state.

\begin{acknowledgements}

For the 2011-2013 MagE data used to determine the systematic uncertainty, we acknowledge Josh Adams for obtaining and reducing the data.
Nora Shipp is supported by an NSF Astronomy and Astrophysics Postdoctoral Fellowship under award AST-2303841.
Denis Erkal acknowledges support from ARC DP210100855.
A.C. is supported by a Brinson Prize Fellowship at UChicago/KICP.

Xiaowei~Ou thanks the LSST Discovery Alliance Data Science Fellowship Program, which is funded by LSST Discovery Alliance, NSF Cybertraining Grant \#1829740, the Brinson Foundation, and the Moore Foundation; his participation in the program has benefited this work.

This work presents results from the European Space Agency (ESA) space mission \emph{Gaia}. {\Gaia} data are being processed by the \Gaia Data Processing and Analysis Consortium (DPAC). Funding for the DPAC is provided by national institutions, in particular the institutions participating in the Gaia MultiLateral Agreement (MLA). The Gaia mission website is \url{https://www.cosmos.esa.int/gaia}. The Gaia archive website is \url{https://archives.esac.esa.int/gaia}.

This research has made use of NASA's Astrophysics Data System Bibliographic Services; the arXiv pre-print server operated by Cornell University; the SIMBAD and VizieR databases hosted by the Strasbourg Astronomical Data Center.

\end{acknowledgements}

\software{%
matplotlib \citep{hunter07},
numpy \citep{vanderwalt11},
scipy \citep{jones01},
emcee \citep{foremanmackey13},
astropy \citep{astropy:2013,astropy:2018}, and
galpy \citep{bovy15}.}

\clearpage

\bibliographystyle{aasjournal}
\bibliography{xou}

\begin{thebibliography}{}
\expandafter\ifx\csname natexlab\endcsname\relax\def\natexlab#1{#1}\fi
\providecommand{\url}[1]{\href{#1}{#1}}
\providecommand{\dodoi}[1]{doi:~\href{http://doi.org/#1}{\nolinkurl{#1}}}
\providecommand{\doeprint}[1]{\href{http://ascl.net/#1}{\nolinkurl{http://ascl.net/#1}}}
\providecommand{\doarXiv}[1]{\href{https://arxiv.org/abs/#1}{\nolinkurl{https://arxiv.org/abs/#1}}}

\bibitem[{{Ad{\'e}n} {et~al.}(2009){Ad{\'e}n}, {Feltzing}, {Koch}, {Wilkinson}, {Grebel}, {Lundstr{\"o}m}, {Gilmore}, {Zucker}, {Belokurov}, {Evans}, \& {Faria}}]{aden09}
{Ad{\'e}n}, D., {Feltzing}, S., {Koch}, A., {et~al.} 2009, \aap, 506, 1147, \dodoi{10.1051/0004-6361/200912718}

\bibitem[{{Alam} {et~al.}(2015){Alam}, {Albareti}, {Allende Prieto}, {Anders}, {Anderson}, {Anderton}, {Andrews}, {Armengaud}, {Aubourg}, {Bailey}, {Basu}, {Bautista}, {Beaton}, {Beers}, {Bender}, {Berlind}, {Beutler}, {Bhardwaj}, {Bird}, {Bizyaev}, {Blake}, {Blanton}, {Blomqvist}, {Bochanski}, {Bolton}, {Bovy}, {Shelden Bradley}, {Brandt}, {Brauer}, {Brinkmann}, {Brown}, {Brownstein}, {Burden}, {Burtin}, {Busca}, {Cai}, {Capozzi}, {Carnero Rosell}, {Carr}, {Carrera}, {Chambers}, {Chaplin}, {Chen}, {Chiappini}, {Chojnowski}, {Chuang}, {Clerc}, {Comparat}, {Covey}, {Croft}, {Cuesta}, {Cunha}, {da Costa}, {Da Rio}, {Davenport}, {Dawson}, {De Lee}, {Delubac}, {Deshpande}, {Dhital}, {Dutra-Ferreira}, {Dwelly}, {Ealet}, {Ebelke}, {Edmondson}, {Eisenstein}, {Ellsworth}, {Elsworth}, {Epstein}, {Eracleous}, {Escoffier}, {Esposito}, {Evans}, {Fan}, {Fern{\'a}ndez-Alvar}, {Feuillet}, {Filiz Ak}, {Finley}, {Finoguenov}, {Flaherty}, {Fleming}, {Font-Ribera}, {Foster}, {Frinchaboy}, {Galbraith-Frew}, {Garc{\'\i}a},
  {Garc{\'\i}a-Hern{\'a}ndez}, {Garc{\'\i}a P{\'e}rez}, {Gaulme}, {Ge}, {G{\'e}nova-Santos}, {Georgakakis}, {Ghezzi}, {Gillespie}, {Girardi}, {Goddard}, {Gontcho}, {Gonz{\'a}lez Hern{\'a}ndez}, {Grebel}, {Green}, {Grieb}, {Grieves}, {Gunn}, {Guo}, {Harding}, {Hasselquist}, {Hawley}, {Hayden}, {Hearty}, {Hekker}, {Ho}, {Hogg}, {Holley-Bockelmann}, {Holtzman}, {Honscheid}, {Huber}, {Huehnerhoff}, {Ivans}, {Jiang}, {Johnson}, {Kinemuchi}, {Kirkby}, {Kitaura}, {Klaene}, {Knapp}, {Kneib}, {Koenig}, {Lam}, {Lan}, {Lang}, {Laurent}, {Le Goff}, {Leauthaud}, {Lee}, {Lee}, {Licquia}, {Liu}, {Long}, {L{\'o}pez-Corredoira}, {Lorenzo-Oliveira}, {Lucatello}, {Lundgren}, {Lupton}, {Mack}, {Mahadevan}, {Maia}, {Majewski}, {Malanushenko}, {Malanushenko}, {Manchado}, {Manera}, {Mao}, {Maraston}, {Marchwinski}, {Margala}, {Martell}, {Martig}, {Masters}, {Mathur}, {McBride}, {McGehee}, {McGreer}, {McMahon}, {M{\'e}nard}, {Menzel}, {Merloni}, {M{\'e}sz{\'a}ros}, {Miller}, {Miralda-Escud{\'e}}, {Miyatake}, {Montero-Dorta}, {More},
  {Morganson}, {Morice-Atkinson}, {Morrison}, {Mosser}, {Muna}, {Myers}, {Nandra}, {Newman}, {Neyrinck}, {Nguyen}, {Nichol}, {Nidever}, {Noterdaeme}, {Nuza}, {O'Connell}, {O'Connell}, {O'Connell}, {Ogando}, {Olmstead}, {Oravetz}, {Oravetz}, {Osumi}, {Owen}, {Padgett}, {Padmanabhan}, {Paegert}, {Palanque-Delabrouille}, {Pan}, {Parejko}, {P{\^a}ris}, {Park}, {Pattarakijwanich}, {Pellejero-Ibanez}, {Pepper}, {Percival}, {P{\'e}rez-Fournon}, {P{\'e}rez-R{\`a}fols}, {Petitjean}, {Pieri}, {Pinsonneault}, {Porto de Mello}, {Prada}, {Prakash}, {Price-Whelan}, {Protopapas}, {Raddick}, {Rahman}, {Reid}, {Rich}, {Rix}, {Robin}, {Rockosi}, {Rodrigues}, {Rodr{\'\i}guez-Torres}, {Roe}, {Ross}, {Ross}, {Rossi}, {Ruan}, {Rubi{\~n}o-Mart{\'\i}n}, {Rykoff}, {Salazar-Albornoz}, {Salvato}, {Samushia}, {S{\'a}nchez}, {Santiago}, {Sayres}, {Schiavon}, {Schlegel}, {Schmidt}, {Schneider}, {Schultheis}, {Schwope}, {Sc{\'o}ccola}, {Scott}, {Sellgren}, {Seo}, {Serenelli}, {Shane}, {Shen}, {Shetrone}, {Shu}, {Silva Aguirre}, {Sivarani},
  {Skrutskie}, {Slosar}, {Smith}, {Sobreira}, {Souto}, {Stassun}, {Steinmetz}, {Stello}, {Strauss}, {Streblyanska}, {Suzuki}, {Swanson}, {Tan}, {Tayar}, {Terrien}, {Thakar}, {Thomas}, {Thomas}, {Thompson}, {Tinker}, {Tojeiro}, {Troup}, {Vargas-Maga{\~n}a}, {Vazquez}, {Verde}, {Viel}, {Vogt}, {Wake}, {Wang}, {Weaver}, {Weinberg}, {Weiner}, {White}, {Wilson}, {Wisniewski}, {Wood-Vasey}, {Ye`che}, {York}, {Zakamska}, {Zamora}, {Zasowski}, {Zehavi}, {Zhao}, {Zheng}, {Zhou}, {Zhou}, {Zou}, \& {Zhu}}]{sdss+15}
{Alam}, S., {Albareti}, F.~D., {Allende Prieto}, C., {et~al.} 2015, \apjs, 219, 12, \dodoi{10.1088/0067-0049/219/1/12}

\bibitem[{{Amorisco}(2017)}]{amorisco17b}
{Amorisco}, N.~C. 2017, \apj, 844, 64, \dodoi{10.3847/1538-4357/aa745f}

\bibitem[{{Amorisco} {et~al.}(2013){Amorisco}, {Agnello}, \& {Evans}}]{amorisco13}
{Amorisco}, N.~C., {Agnello}, A., \& {Evans}, N.~W. 2013, \mnras, 429, L89, \dodoi{10.1093/mnrasl/sls031}

\bibitem[{{Astropy Collaboration} {et~al.}(2013){Astropy Collaboration}, {Robitaille}, {Tollerud}, {Greenfield}, {Droettboom}, {Bray}, {Aldcroft}, {Davis}, {Ginsburg}, {Price-Whelan}, {Kerzendorf}, {Conley}, {Crighton}, {Barbary}, {Muna}, {Ferguson}, {Grollier}, {Parikh}, {Nair}, {Unther}, {Deil}, {Woillez}, {Conseil}, {Kramer}, {Turner}, {Singer}, {Fox}, {Weaver}, {Zabalza}, {Edwards}, {Azalee Bostroem}, {Burke}, {Casey}, {Crawford}, {Dencheva}, {Ely}, {Jenness}, {Labrie}, {Lim}, {Pierfederici}, {Pontzen}, {Ptak}, {Refsdal}, {Servillat}, \& {Streicher}}]{astropy:2013}
{Astropy Collaboration}, {Robitaille}, T.~P., {Tollerud}, E.~J., {et~al.} 2013, \aap, 558, A33, \dodoi{10.1051/0004-6361/201322068}

\bibitem[{{Astropy Collaboration} {et~al.}(2018){Astropy Collaboration}, {Price-Whelan}, {Sip{\H{o}}cz}, {G{\"u}nther}, {Lim}, {Crawford}, {Conseil}, {Shupe}, {Craig}, {Dencheva}, {Ginsburg}, {Vand erPlas}, {Bradley}, {P{\'e}rez-Su{\'a}rez}, {de Val-Borro}, {Aldcroft}, {Cruz}, {Robitaille}, {Tollerud}, {Ardelean}, {Babej}, {Bach}, {Bachetti}, {Bakanov}, {Bamford}, {Barentsen}, {Barmby}, {Baumbach}, {Berry}, {Biscani}, {Boquien}, {Bostroem}, {Bouma}, {Brammer}, {Bray}, {Breytenbach}, {Buddelmeijer}, {Burke}, {Calderone}, {Cano Rodr{\'\i}guez}, {Cara}, {Cardoso}, {Cheedella}, {Copin}, {Corrales}, {Crichton}, {D'Avella}, {Deil}, {Depagne}, {Dietrich}, {Donath}, {Droettboom}, {Earl}, {Erben}, {Fabbro}, {Ferreira}, {Finethy}, {Fox}, {Garrison}, {Gibbons}, {Goldstein}, {Gommers}, {Greco}, {Greenfield}, {Groener}, {Grollier}, {Hagen}, {Hirst}, {Homeier}, {Horton}, {Hosseinzadeh}, {Hu}, {Hunkeler}, {Ivezi{\'c}}, {Jain}, {Jenness}, {Kanarek}, {Kendrew}, {Kern}, {Kerzendorf}, {Khvalko}, {King}, {Kirkby}, {Kulkarni},
  {Kumar}, {Lee}, {Lenz}, {Littlefair}, {Ma}, {Macleod}, {Mastropietro}, {McCully}, {Montagnac}, {Morris}, {Mueller}, {Mumford}, {Muna}, {Murphy}, {Nelson}, {Nguyen}, {Ninan}, {N{\"o}the}, {Ogaz}, {Oh}, {Parejko}, {Parley}, {Pascual}, {Patil}, {Patil}, {Plunkett}, {Prochaska}, {Rastogi}, {Reddy Janga}, {Sabater}, {Sakurikar}, {Seifert}, {Sherbert}, {Sherwood-Taylor}, {Shih}, {Sick}, {Silbiger}, {Singanamalla}, {Singer}, {Sladen}, {Sooley}, {Sornarajah}, {Streicher}, {Teuben}, {Thomas}, {Tremblay}, {Turner}, {Terr{\'o}n}, {van Kerkwijk}, {de la Vega}, {Watkins}, {Weaver}, {Whitmore}, {Woillez}, {Zabalza}, \& {Astropy Contributors}}]{astropy:2018}
{Astropy Collaboration}, {Price-Whelan}, A.~M., {Sip{\H{o}}cz}, B.~M., {et~al.} 2018, \aj, 156, 123, \dodoi{10.3847/1538-3881/aabc4f}

\bibitem[{{Battaglia} {et~al.}(2008){Battaglia}, {Helmi}, {Tolstoy}, {Irwin}, {Hill}, \& {Jablonka}}]{battaglia08}
{Battaglia}, G., {Helmi}, A., {Tolstoy}, E., {et~al.} 2008, \apjl, 681, L13, \dodoi{10.1086/590179}

\bibitem[{{Battaglia} \& {Nipoti}(2022)}]{battaglia22a}
{Battaglia}, G., \& {Nipoti}, C. 2022, Nature Astronomy, 6, 659, \dodoi{10.1038/s41550-022-01638-7}

\bibitem[{{Battaglia} {et~al.}(2022){Battaglia}, {Taibi}, {Thomas}, \& {Fritz}}]{battaglia22}
{Battaglia}, G., {Taibi}, S., {Thomas}, G.~F., \& {Fritz}, T.~K. 2022, \aap, 657, A54, \dodoi{10.1051/0004-6361/202141528}

\bibitem[{{Bechtol} {et~al.}(2015){Bechtol}, {Drlica-Wagner}, {Balbinot}, {Pieres}, {Simon}, {Yanny}, {Santiago}, {Wechsler}, {Frieman}, {Walker}, {Williams}, {Rozo}, {Rykoff}, {Queiroz}, {Luque}, {Benoit-L{\'e}vy}, {Tucker}, {Sevilla}, {Gruendl}, {da Costa}, {Fausti Neto}, {Maia}, {Abbott}, {Allam}, {Armstrong}, {Bauer}, {Bernstein}, {Bernstein}, {Bertin}, {Brooks}, {Buckley-Geer}, {Burke}, {Carnero Rosell}, {Castander}, {Covarrubias}, {D'Andrea}, {DePoy}, {Desai}, {Diehl}, {Eifler}, {Estrada}, {Evrard}, {Fernandez}, {Finley}, {Flaugher}, {Gaztanaga}, {Gerdes}, {Girardi}, {Gladders}, {Gruen}, {Gutierrez}, {Hao}, {Honscheid}, {Jain}, {James}, {Kent}, {Kron}, {Kuehn}, {Kuropatkin}, {Lahav}, {Li}, {Lin}, {Makler}, {March}, {Marshall}, {Martini}, {Merritt}, {Miller}, {Miquel}, {Mohr}, {Neilsen}, {Nichol}, {Nord}, {Ogando}, {Peoples}, {Petravick}, {Plazas}, {Romer}, {Roodman}, {Sako}, {Sanchez}, {Scarpine}, {Schubnell}, {Smith}, {Soares-Santos}, {Sobreira}, {Suchyta}, {Swanson}, {Tarle}, {Thaler}, {Thomas},
  {Wester}, {Zuntz}, \& {DES Collaboration}}]{bechtol15}
{Bechtol}, K., {Drlica-Wagner}, A., {Balbinot}, E., {et~al.} 2015, \apj, 807, 50, \dodoi{10.1088/0004-637X/807/1/50}

\bibitem[{{Beers} {et~al.}(1999){Beers}, {Rossi}, {Norris}, {Ryan}, \& {Shefler}}]{brn+99}
{Beers}, T.~C., {Rossi}, S., {Norris}, J.~E., {Ryan}, S.~G., \& {Shefler}, T. 1999, \aj, 117, 981, \dodoi{10.1086/300727}

\bibitem[{{Belokurov} {et~al.}(2007){Belokurov}, {Zucker}, {Evans}, {Kleyna}, {Koposov}, {Hodgkin}, {Irwin}, {Gilmore}, {Wilkinson}, {Fellhauer}, {Bramich}, {Hewett}, {Vidrih}, {De Jong}, {Smith}, {Rix}, {Bell}, {Wyse}, {Newberg}, {Mayeur}, {Yanny}, {Rockosi}, {Gnedin}, {Schneider}, {Beers}, {Barentine}, {Brewington}, {Brinkmann}, {Harvanek}, {Kleinman}, {Krzesinski}, {Long}, {Nitta}, \& {Snedden}}]{belokurov07}
{Belokurov}, V., {Zucker}, D.~B., {Evans}, N.~W., {et~al.} 2007, \apj, 654, 897, \dodoi{10.1086/509718}

\bibitem[{{Bovy}(2015)}]{bovy15}
{Bovy}, J. 2015, \apjs, 216, 29, \dodoi{10.1088/0067-0049/216/2/29}

\bibitem[{{Bozek} {et~al.}(2019){Bozek}, {Fitts}, {Boylan-Kolchin}, {Garrison-Kimmel}, {Abazajian}, {Bullock}, {Kere{\v{s}}}, {Faucher-Gigu{\`e}re}, {Wetzel}, {Feldmann}, \& {Hopkins}}]{bozek19}
{Bozek}, B., {Fitts}, A., {Boylan-Kolchin}, M., {et~al.} 2019, \mnras, 483, 4086, \dodoi{10.1093/mnras/sty3300}

\bibitem[{{Brown} {et~al.}(2014){Brown}, {Tumlinson}, {Geha}, {Simon}, {Vargas}, {VandenBerg}, {Kirby}, {Kalirai}, {Avila}, {Gennaro}, {Ferguson}, {Mu{\~n}oz}, {Guhathakurta}, \& {Renzini}}]{brown14}
{Brown}, T.~M., {Tumlinson}, J., {Geha}, M., {et~al.} 2014, \apj, 796, 91, \dodoi{10.1088/0004-637X/796/2/91}

\bibitem[{{Calabrese} \& {Spergel}(2016)}]{calabrese16}
{Calabrese}, E., \& {Spergel}, D.~N. 2016, \mnras, 460, 4397, \dodoi{10.1093/mnras/stw1256}

\bibitem[{{Calamida} {et~al.}(2007){Calamida}, {Bono}, {Stetson}, {Freyhammer}, {Cassisi}, {Grundahl}, {Pietrinferni}, {Hilker}, {Primas}, {Richtler}, {Romaniello}, {Buonanno}, {Caputo}, {Castellani}, {Corsi}, {Ferraro}, {Iannicola}, \& {Pulone}}]{calamida07}
{Calamida}, A., {Bono}, G., {Stetson}, P.~B., {et~al.} 2007, \apj, 670, 400, \dodoi{10.1086/521424}

\bibitem[{{Carlin} \& {Sand}(2018)}]{carlin18}
{Carlin}, J.~L., \& {Sand}, D.~J. 2018, \apj, 865, 7, \dodoi{10.3847/1538-4357/aad8c1}

\bibitem[{{Carrera} {et~al.}(2013){Carrera}, {Pancino}, {Gallart}, \& {del Pino}}]{cpg+13}
{Carrera}, R., {Pancino}, E., {Gallart}, C., \& {del Pino}, A. 2013, \mnras, 434, 1681, \dodoi{10.1093/mnras/stt1126}

\bibitem[{{Cerny} {et~al.}(2021{\natexlab{a}}){Cerny}, {Pace}, {Drlica-Wagner}, {Ferguson}, {Mau}, {Adam{\'o}w}, {Carlin}, {Choi}, {Erkal}, {Johnson}, {Li}, {Mart{\'\i}nez-V{\'a}zquez}, {Mutlu-Pakdil}, {Nidever}, {Olsen}, {Pieres}, {Tollerud}, {Simon}, {Vivas}, {James}, {Kuropatkin}, {Majewski}, {Mart{\'\i}nez-Delgado}, {Massana}, {Miller}, {Neilsen}, {No{\"e}l}, {Riley}, {Sand}, {Santana-Silva}, {Stringfellow}, {Tucker}, \& {Delve Collaboration}}]{cerny21a}
{Cerny}, W., {Pace}, A.~B., {Drlica-Wagner}, A., {et~al.} 2021{\natexlab{a}}, \apj, 910, 18, \dodoi{10.3847/1538-4357/abe1af}

\bibitem[{{Cerny} {et~al.}(2021{\natexlab{b}}){Cerny}, {Pace}, {Drlica-Wagner}, {Koposov}, {Vivas}, {Mau}, {Riley}, {Bom}, {Carlin}, {Choi}, {Erkal}, {Ferguson}, {James}, {Li}, {Mart{\'\i}nez-Delgado}, {Mart{\'\i}nez-V{\'a}zquez}, {Munoz}, {Mutlu-Pakdil}, {Olsen}, {Pieres}, {Sakowska}, {Sand}, {Simon}, {Smercina}, {Stringfellow}, {Tollerud}, {Adam{\'o}w}, {Hernandez-Lang}, {Kuropatkin}, {Santana-Silva}, {Tucker}, {Zenteno}, \& {Delve Collaboration}}]{cerny21b}
---. 2021{\natexlab{b}}, \apjl, 920, L44, \dodoi{10.3847/2041-8213/ac2d9a}

\bibitem[{{Cerny} {et~al.}(2022){Cerny}, {Mart{\'\i}nez-V{\'a}zquez}, {Drlica-Wagner}, {Pace}, {Mutlu-Pakdil}, {Li}, {Riley}, {Crnojevi{\'c}}, {Bom}, {Carballo-Bello}, {Carlin}, {Chiti}, {Choi}, {Collins}, {Darragh-Ford}, {Ferguson}, {Geha}, {Mart{\'\i}nez-Delgado}, {Massana}, {Mau}, {Medina}, {Mu{\~n}oz}, {Nadler}, {Olsen}, {Pieres}, {Sakowska}, {Simon}, {Stringfellow}, {Vivas}, {Walker}, \& {Wechsler}}]{cerny22}
{Cerny}, W., {Mart{\'\i}nez-V{\'a}zquez}, C.~E., {Drlica-Wagner}, A., {et~al.} 2022, arXiv e-prints, arXiv:2209.12422, \dodoi{10.48550/arXiv.2209.12422}

\bibitem[{{Chambers} {et~al.}(2016{\natexlab{a}}){Chambers}, {Magnier}, {Metcalfe}, {Flewelling}, {Huber}, {Waters}, {Denneau}, {Draper}, {Farrow}, {Finkbeiner}, {Holmberg}, {Koppenhoefer}, {Price}, {Rest}, {Saglia}, {Schlafly}, {Smartt}, {Sweeney}, {Wainscoat}, {Burgett}, {Chastel}, {Grav}, {Heasley}, {Hodapp}, {Jedicke}, {Kaiser}, {Kudritzki}, {Luppino}, {Lupton}, {Monet}, {Morgan}, {Onaka}, {Shiao}, {Stubbs}, {Tonry}, {White}, {Ba{\~n}ados}, {Bell}, {Bender}, {Bernard}, {Boegner}, {Boffi}, {Botticella}, {Calamida}, {Casertano}, {Chen}, {Chen}, {Cole}, {Deacon}, {Frenk}, {Fitzsimmons}, {Gezari}, {Gibbs}, {Goessl}, {Goggia}, {Gourgue}, {Goldman}, {Grant}, {Grebel}, {Hambly}, {Hasinger}, {Heavens}, {Heckman}, {Henderson}, {Henning}, {Holman}, {Hopp}, {Ip}, {Isani}, {Jackson}, {Keyes}, {Koekemoer}, {Kotak}, {Le}, {Liska}, {Long}, {Lucey}, {Liu}, {Martin}, {Masci}, {McLean}, {Mindel}, {Misra}, {Morganson}, {Murphy}, {Obaika}, {Narayan}, {Nieto-Santisteban}, {Norberg}, {Peacock}, {Pier}, {Postman}, {Primak},
  {Rae}, {Rai}, {Riess}, {Riffeser}, {Rix}, {R{\"o}ser}, {Russel}, {Rutz}, {Schilbach}, {Schultz}, {Scolnic}, {Strolger}, {Szalay}, {Seitz}, {Small}, {Smith}, {Soderblom}, {Taylor}, {Thomson}, {Taylor}, {Thakar}, {Thiel}, {Thilker}, {Unger}, {Urata}, {Valenti}, {Wagner}, {Walder}, {Walter}, {Watters}, {Werner}, {Wood-Vasey}, \& {Wyse}}]{panstarrs+16}
{Chambers}, K.~C., {Magnier}, E.~A., {Metcalfe}, N., {et~al.} 2016{\natexlab{a}}, arXiv e-prints, arXiv:1612.05560, \dodoi{10.48550/arXiv.1612.05560}

\bibitem[{{Chambers} {et~al.}(2016{\natexlab{b}}){Chambers}, {Magnier}, {Metcalfe}, {Flewelling}, {Huber}, {Waters}, {Denneau}, {Draper}, {Farrow}, {Finkbeiner}, {Holmberg}, {Koppenhoefer}, {Price}, {Rest}, {Saglia}, {Schlafly}, {Smartt}, {Sweeney}, {Wainscoat}, {Burgett}, {Chastel}, {Grav}, {Heasley}, {Hodapp}, {Jedicke}, {Kaiser}, {Kudritzki}, {Luppino}, {Lupton}, {Monet}, {Morgan}, {Onaka}, {Shiao}, {Stubbs}, {Tonry}, {White}, {Ba{\~n}ados}, {Bell}, {Bender}, {Bernard}, {Boegner}, {Boffi}, {Botticella}, {Calamida}, {Casertano}, {Chen}, {Chen}, {Cole}, {Deacon}, {Frenk}, {Fitzsimmons}, {Gezari}, {Gibbs}, {Goessl}, {Goggia}, {Gourgue}, {Goldman}, {Grant}, {Grebel}, {Hambly}, {Hasinger}, {Heavens}, {Heckman}, {Henderson}, {Henning}, {Holman}, {Hopp}, {Ip}, {Isani}, {Jackson}, {Keyes}, {Koekemoer}, {Kotak}, {Le}, {Liska}, {Long}, {Lucey}, {Liu}, {Martin}, {Masci}, {McLean}, {Mindel}, {Misra}, {Morganson}, {Murphy}, {Obaika}, {Narayan}, {Nieto-Santisteban}, {Norberg}, {Peacock}, {Pier}, {Postman}, {Primak},
  {Rae}, {Rai}, {Riess}, {Riffeser}, {Rix}, {R{\"o}ser}, {Russel}, {Rutz}, {Schilbach}, {Schultz}, {Scolnic}, {Strolger}, {Szalay}, {Seitz}, {Small}, {Smith}, {Soderblom}, {Taylor}, {Thomson}, {Taylor}, {Thakar}, {Thiel}, {Thilker}, {Unger}, {Urata}, {Valenti}, {Wagner}, {Walder}, {Walter}, {Watters}, {Werner}, {Wood-Vasey}, \& {Wyse}}]{chambers16}
---. 2016{\natexlab{b}}, arXiv e-prints, arXiv:1612.05560, \dodoi{10.48550/arXiv.1612.05560}

\bibitem[{{Chang} \& {Necib}(2021)}]{cn+21}
{Chang}, L.~J., \& {Necib}, L. 2021, \mnras, 507, 4715, \dodoi{10.1093/mnras/stab2440}

\bibitem[{{Chiti} {et~al.}(2020){Chiti}, {Frebel}, {Jerjen}, {Kim}, \& {Norris}}]{cfj+20}
{Chiti}, A., {Frebel}, A., {Jerjen}, H., {Kim}, D., \& {Norris}, J.~E. 2020, \apj, 891, 8, \dodoi{10.3847/1538-4357/ab6d72}

\bibitem[{{Chiti} {et~al.}(2022){Chiti}, {Simon}, {Frebel}, {Pace}, {Ji}, \& {Li}}]{csf+22}
{Chiti}, A., {Simon}, J.~D., {Frebel}, A., {et~al.} 2022, \apj, 939, 41, \dodoi{10.3847/1538-4357/ac96ed}

\bibitem[{{Chiti} {et~al.}(2018){Chiti}, {Simon}, {Frebel}, {Thompson}, {Shectman}, {Mateo}, {Bailey}, {Crane}, \& {Walker}}]{csf+18}
---. 2018, \apj, 856, 142, \dodoi{10.3847/1538-4357/aab663}

\bibitem[{{Chiti} {et~al.}(2021){Chiti}, {Frebel}, {Simon}, {Erkal}, {Chang}, {Necib}, {Ji}, {Jerjen}, {Kim}, \& {Norris}}]{chiti21}
{Chiti}, A., {Frebel}, A., {Simon}, J.~D., {et~al.} 2021, Nature Astronomy, 5, 392, \dodoi{10.1038/s41550-020-01285-w}

\bibitem[{{Chubak} {et~al.}(2012){Chubak}, {Marcy}, {Fischer}, {Howard}, {Isaacson}, {Johnson}, \& {Wright}}]{chubak12}
{Chubak}, C., {Marcy}, G., {Fischer}, D.~A., {et~al.} 2012, arXiv e-prints, arXiv:1207.6212, \dodoi{10.48550/arXiv.1207.6212}

\bibitem[{{Coleman} {et~al.}(2007){Coleman}, {de Jong}, {Martin}, {Rix}, {Sand}, {Bell}, {Pogge}, {Thompson}, {Hippelein}, {Giallongo}, {Ragazzoni}, {DiPaola}, {Farinato}, {Smareglia}, {Testa}, {Bechtold}, {Hill}, {Garnavich}, \& {Green}}]{coleman07}
{Coleman}, M.~G., {de Jong}, J. T.~A., {Martin}, N.~F., {et~al.} 2007, \apjl, 668, L43, \dodoi{10.1086/522672}

\bibitem[{{Collins} {et~al.}(2017){Collins}, {Tollerud}, {Sand}, {Bonaca}, {Willman}, \& {Strader}}]{collins17}
{Collins}, M. L.~M., {Tollerud}, E.~J., {Sand}, D.~J., {et~al.} 2017, \mnras, 467, 573, \dodoi{10.1093/mnras/stx067}

\bibitem[{{Contenta} {et~al.}(2018){Contenta}, {Balbinot}, {Petts}, {Read}, {Gieles}, {Collins}, {Pe{\~n}arrubia}, {Delorme}, \& {Gualandris}}]{contenta18}
{Contenta}, F., {Balbinot}, E., {Petts}, J.~A., {et~al.} 2018, \mnras, 476, 3124, \dodoi{10.1093/mnras/sty424}

\bibitem[{{Cooper} {et~al.}(2012){Cooper}, {Newman}, {Davis}, {Finkbeiner}, \& {Gerke}}]{cnd+12}
{Cooper}, M.~C., {Newman}, J.~A., {Davis}, M., {Finkbeiner}, D.~P., \& {Gerke}, B.~F. 2012, {spec2d: DEEP2 DEIMOS Spectral Pipeline}.
\newblock \doeprint{1203.003}

\bibitem[{{Czekaj} {et~al.}(2014){Czekaj}, {Robin}, {Figueras}, {Luri}, \& {Haywood}}]{czekaj14}
{Czekaj}, M.~A., {Robin}, A.~C., {Figueras}, F., {Luri}, X., \& {Haywood}, M. 2014, \aap, 564, A102, \dodoi{10.1051/0004-6361/201322139}

\bibitem[{{Deason} {et~al.}(2012){Deason}, {Belokurov}, {Evans}, {Watkins}, \& {Fellhauer}}]{deason12}
{Deason}, A.~J., {Belokurov}, V., {Evans}, N.~W., {Watkins}, L.~L., \& {Fellhauer}, M. 2012, \mnras, 425, L101, \dodoi{10.1111/j.1745-3933.2012.01314.x}

\bibitem[{{Dotter} {et~al.}(2008){Dotter}, {Chaboyer}, {Jevremovi{\'c}}, {Kostov}, {Baron}, \& {Ferguson}}]{dcj+08}
{Dotter}, A., {Chaboyer}, B., {Jevremovi{\'c}}, D., {et~al.} 2008, \apjs, 178, 89, \dodoi{10.1086/589654}

\bibitem[{{Dressler} {et~al.}(2006){Dressler}, {Hare}, {Bigelow}, \& {Osip}}]{dhb+06}
{Dressler}, A., {Hare}, T., {Bigelow}, B.~C., \& {Osip}, D.~J. 2006, in Society of Photo-Optical Instrumentation Engineers (SPIE) Conference Series, Vol. 6269, Society of Photo-Optical Instrumentation Engineers (SPIE) Conference Series, ed. I.~S. {McLean} \& M.~{Iye}, 62690F, \dodoi{10.1117/12.670573}

\bibitem[{{Dressler} {et~al.}(2011){Dressler}, {Bigelow}, {Hare}, {Sutin}, {Thompson}, {Burley}, {Epps}, {Oemler}, {Bagish}, {Birk}, {Clardy}, {Gunnels}, {Kelson}, {Shectman}, \& {Osip}}]{dbh+11}
{Dressler}, A., {Bigelow}, B., {Hare}, T., {et~al.} 2011, \pasp, 123, 288, \dodoi{10.1086/658908}

\bibitem[{{Drlica-Wagner} {et~al.}(2015){Drlica-Wagner}, {Bechtol}, {Rykoff}, {Luque}, {Queiroz}, {Mao}, {Wechsler}, {Simon}, {Santiago}, {Yanny}, {Balbinot}, {Dodelson}, {Fausti Neto}, {James}, {Li}, {Maia}, {Marshall}, {Pieres}, {Stringer}, {Walker}, {Abbott}, {Abdalla}, {Allam}, {Benoit-L{\'e}vy}, {Bernstein}, {Bertin}, {Brooks}, {Buckley-Geer}, {Burke}, {Carnero Rosell}, {Carrasco Kind}, {Carretero}, {Crocce}, {da Costa}, {Desai}, {Diehl}, {Dietrich}, {Doel}, {Eifler}, {Evrard}, {Finley}, {Flaugher}, {Fosalba}, {Frieman}, {Gaztanaga}, {Gerdes}, {Gruen}, {Gruendl}, {Gutierrez}, {Honscheid}, {Kuehn}, {Kuropatkin}, {Lahav}, {Martini}, {Miquel}, {Nord}, {Ogando}, {Plazas}, {Reil}, {Roodman}, {Sako}, {Sanchez}, {Scarpine}, {Schubnell}, {Sevilla-Noarbe}, {Smith}, {Soares-Santos}, {Sobreira}, {Suchyta}, {Swanson}, {Tarle}, {Tucker}, {Vikram}, {Wester}, {Zhang}, {Zuntz}, \& {DES Collaboration}}]{drlicawagner15}
{Drlica-Wagner}, A., {Bechtol}, K., {Rykoff}, E.~S., {et~al.} 2015, \apj, 813, 109, \dodoi{10.1088/0004-637X/813/2/109}

\bibitem[{{Drlica-Wagner} {et~al.}(2016){Drlica-Wagner}, {Bechtol}, {Allam}, {Tucker}, {Gruendl}, {Johnson}, {Walker}, {James}, {Nidever}, {Olsen}, {Wechsler}, {Cioni}, {Conn}, {Kuehn}, {Li}, {Mao}, {Martin}, {Neilsen}, {Noel}, {Pieres}, {Simon}, {Stringfellow}, {van der Marel}, \& {Yanny}}]{drlicawagner16}
{Drlica-Wagner}, A., {Bechtol}, K., {Allam}, S., {et~al.} 2016, \apjl, 833, L5, \dodoi{10.3847/2041-8205/833/1/L5}

\bibitem[{{Dubinski} \& {Carlberg}(1991)}]{dubinski91}
{Dubinski}, J., \& {Carlberg}, R.~G. 1991, \apj, 378, 496, \dodoi{10.1086/170451}

\bibitem[{{Dutton} \& {Macci{\`o}}(2014)}]{dutton14}
{Dutton}, A.~A., \& {Macci{\`o}}, A.~V. 2014, \mnras, 441, 3359, \dodoi{10.1093/mnras/stu742}

\bibitem[{{Eisenstein} {et~al.}(2011){Eisenstein}, {Weinberg}, {Agol}, {Aihara}, {Allende Prieto}, {Anderson}, {Arns}, {Aubourg}, {Bailey}, {Balbinot}, {Barkhouser}, {Beers}, {Berlind}, {Bickerton}, {Bizyaev}, {Blanton}, {Bochanski}, {Bolton}, {Bosman}, {Bovy}, {Brandt}, {Breslauer}, {Brewington}, {Brinkmann}, {Brown}, {Brownstein}, {Burger}, {Busca}, {Campbell}, {Cargile}, {Carithers}, {Carlberg}, {Carr}, {Chang}, {Chen}, {Chiappini}, {Comparat}, {Connolly}, {Cortes}, {Croft}, {Cunha}, {da Costa}, {Davenport}, {Dawson}, {De Lee}, {Porto de Mello}, {de Simoni}, {Dean}, {Dhital}, {Ealet}, {Ebelke}, {Edmondson}, {Eiting}, {Escoffier}, {Esposito}, {Evans}, {Fan}, {Femen{\'\i}a Castell{\'a}}, {Dutra Ferreira}, {Fitzgerald}, {Fleming}, {Font-Ribera}, {Ford}, {Frinchaboy}, {Garc{\'\i}a P{\'e}rez}, {Gaudi}, {Ge}, {Ghezzi}, {Gillespie}, {Gilmore}, {Girardi}, {Gott}, {Gould}, {Grebel}, {Gunn}, {Hamilton}, {Harding}, {Harris}, {Hawley}, {Hearty}, {Hennawi}, {Gonz{\'a}lez Hern{\'a}ndez}, {Ho}, {Hogg}, {Holtzman},
  {Honscheid}, {Inada}, {Ivans}, {Jiang}, {Jiang}, {Johnson}, {Jordan}, {Jordan}, {Kauffmann}, {Kazin}, {Kirkby}, {Klaene}, {Knapp}, {Kneib}, {Kochanek}, {Koesterke}, {Kollmeier}, {Kron}, {Lampeitl}, {Lang}, {Lawler}, {Le Goff}, {Lee}, {Lee}, {Leisenring}, {Lin}, {Liu}, {Long}, {Loomis}, {Lucatello}, {Lundgren}, {Lupton}, {Ma}, {Ma}, {MacDonald}, {Mack}, {Mahadevan}, {Maia}, {Majewski}, {Makler}, {Malanushenko}, {Malanushenko}, {Mandelbaum}, {Maraston}, {Margala}, {Maseman}, {Masters}, {McBride}, {McDonald}, {McGreer}, {McMahon}, {Mena Requejo}, {M{\'e}nard}, {Miralda-Escud{\'e}}, {Morrison}, {Mullally}, {Muna}, {Murayama}, {Myers}, {Naugle}, {Neto}, {Nguyen}, {Nichol}, {Nidever}, {O'Connell}, {Ogando}, {Olmstead}, {Oravetz}, {Padmanabhan}, {Paegert}, {Palanque-Delabrouille}, {Pan}, {Pandey}, {Parejko}, {P{\^a}ris}, {Pellegrini}, {Pepper}, {Percival}, {Petitjean}, {Pfaffenberger}, {Pforr}, {Phleps}, {Pichon}, {Pieri}, {Prada}, {Price-Whelan}, {Raddick}, {Ramos}, {Reid}, {Reyle}, {Rich}, {Richards}, {Rieke},
  {Rieke}, {Rix}, {Robin}, {Rocha-Pinto}, {Rockosi}, {Roe}, {Rollinde}, {Ross}, {Ross}, {Rossetto}, {S{\'a}nchez}, {Santiago}, {Sayres}, {Schiavon}, {Schlegel}, {Schlesinger}, {Schmidt}, {Schneider}, {Sellgren}, {Shelden}, {Sheldon}, {Shetrone}, {Shu}, {Silverman}, {Simmerer}, {Simmons}, {Sivarani}, {Skrutskie}, {Slosar}, {Smee}, {Smith}, {Snedden}, {Stassun}, {Steele}, {Steinmetz}, {Stockett}, {Stollberg}, {Strauss}, {Szalay}, {Tanaka}, {Thakar}, {Thomas}, {Tinker}, {Tofflemire}, {Tojeiro}, {Tremonti}, {Vargas Maga{\~n}a}, {Verde}, {Vogt}, {Wake}, {Wan}, {Wang}, {Weaver}, {White}, {White}, {Wilson}, {Wisniewski}, {Wood-Vasey}, {Yanny}, {Yasuda}, {Y{\`e}che}, {York}, {Young}, {Zasowski}, {Zehavi}, \& {Zhao}}]{sdss+11}
{Eisenstein}, D.~J., {Weinberg}, D.~H., {Agol}, E., {et~al.} 2011, \aj, 142, 72, \dodoi{10.1088/0004-6256/142/3/72}

\bibitem[{{Erkal} {et~al.}(2019){Erkal}, {Belokurov}, {Laporte}, {Koposov}, {Li}, {Grillmair}, {Kallivayalil}, {Price-Whelan}, {Evans}, {Hawkins}, {Hendel}, {Mateu}, {Navarro}, {del Pino}, {Slater}, {Sohn}, \& {Orphan Aspen Treasury Collaboration}}]{Erkal:2019}
{Erkal}, D., {Belokurov}, V., {Laporte}, C.~F.~P., {et~al.} 2019, \mnras, 487, 2685, \dodoi{10.1093/mnras/stz1371}

\bibitem[{{Erkal} {et~al.}(2021){Erkal}, {Deason}, {Belokurov}, {Xue}, {Koposov}, {Bird}, {Liu}, {Simion}, {Yang}, {Zhang}, \& {Zhao}}]{erkal21}
{Erkal}, D., {Deason}, A.~J., {Belokurov}, V., {et~al.} 2021, \mnras, 506, 2677, \dodoi{10.1093/mnras/stab1828}

\bibitem[{{Errani} {et~al.}(2018){Errani}, {Pe{\~n}arrubia}, \& {Walker}}]{errani18}
{Errani}, R., {Pe{\~n}arrubia}, J., \& {Walker}, M.~G. 2018, \mnras, 481, 5073, \dodoi{10.1093/mnras/sty2505}

\bibitem[{{Faber} {et~al.}(2003){Faber}, {Phillips}, {Kibrick}, {Alcott}, {Allen}, {Burrous}, {Cantrall}, {Clarke}, {Coil}, {Cowley}, {Davis}, {Deich}, {Dietsch}, {Gilmore}, {Harper}, {Hilyard}, {Lewis}, {McVeigh}, {Newman}, {Osborne}, {Schiavon}, {Stover}, {Tucker}, {Wallace}, {Wei}, {Wirth}, \& {Wright}}]{faber03}
{Faber}, S.~M., {Phillips}, A.~C., {Kibrick}, R.~I., {et~al.} 2003, in Society of Photo-Optical Instrumentation Engineers (SPIE) Conference Series, Vol. 4841, Instrument Design and Performance for Optical/Infrared Ground-based Telescopes, ed. M.~{Iye} \& A.~F.~M. {Moorwood}, 1657--1669, \dodoi{10.1117/12.460346}

\bibitem[{{Fattahi} {et~al.}(2018){Fattahi}, {Navarro}, {Frenk}, {Oman}, {Sawala}, \& {Schaller}}]{fattahi18}
{Fattahi}, A., {Navarro}, J.~F., {Frenk}, C.~S., {et~al.} 2018, \mnras, 476, 3816, \dodoi{10.1093/mnras/sty408}

\bibitem[{{Ferguson} {et~al.}(2022){Ferguson}, {Shipp}, {Drlica-Wagner}, {Li}, {Cerny}, {Tavangar}, {Pace}, {Marshall}, {Riley}, {Adam{\'o}w}, {Carlin}, {Choi}, {Erkal}, {James}, {Koposov}, {Kuropatkin}, {Mart{\'\i}nez-V{\'a}zquez}, {Mau}, {Mutlu-Pakdil}, {Olsen}, {Sakowska}, {Stringfellow}, {Yanny}, \& {Yanny}}]{Ferguson:2022}
{Ferguson}, P.~S., {Shipp}, N., {Drlica-Wagner}, A., {et~al.} 2022, \aj, 163, 18, \dodoi{10.3847/1538-3881/ac3492}

\bibitem[{{Filion} \& {Wyse}(2021)}]{filion21}
{Filion}, C., \& {Wyse}, R. F.~G. 2021, \apj, 923, 218, \dodoi{10.3847/1538-4357/ac2df1}

\bibitem[{{Foreman-Mackey} {et~al.}(2013){Foreman-Mackey}, {Hogg}, {Lang}, \& {Goodman}}]{foremanmackey13}
{Foreman-Mackey}, D., {Hogg}, D.~W., {Lang}, D., \& {Goodman}, J. 2013, \pasp, 125, 306, \dodoi{10.1086/670067}

\bibitem[{{Fu} {et~al.}(2019){Fu}, {Simon}, \& {Alarc{\'o}n Jara}}]{fu19}
{Fu}, S.~W., {Simon}, J.~D., \& {Alarc{\'o}n Jara}, A.~G. 2019, \apj, 883, 11, \dodoi{10.3847/1538-4357/ab3658}

\bibitem[{{Gaia Collaboration} {et~al.}(2016{\natexlab{a}}){Gaia Collaboration}, {Prusti}, {de Bruijne}, {Brown}, {Vallenari}, {Babusiaux}, {Bailer-Jones}, {Bastian}, {Biermann}, {Evans}, {Eyer}, {Jansen}, {Jordi}, {Klioner}, {Lammers}, {Lindegren}, {Luri}, {Mignard}, {Milligan}, {Panem}, {Poinsignon}, {Pourbaix}, {Randich}, {Sarri}, {Sartoretti}, {Siddiqui}, {Soubiran}, {Valette}, {van Leeuwen}, {Walton}, {Aerts}, {Arenou}, {Cropper}, {Drimmel}, {H{\o}g}, {Katz}, {Lattanzi}, {O'Mullane}, {Grebel}, {Holland}, {Huc}, {Passot}, {Bramante}, {Cacciari}, {Casta{\~n}eda}, {Chaoul}, {Cheek}, {De Angeli}, {Fabricius}, {Guerra}, {Hern{\'a}ndez}, {Jean-Antoine-Piccolo}, {Masana}, {Messineo}, {Mowlavi}, {Nienartowicz}, {Ord{\'o}{\~n}ez-Blanco}, {Panuzzo}, {Portell}, {Richards}, {Riello}, {Seabroke}, {Tanga}, {Th{\'e}venin}, {Torra}, {Els}, {Gracia-Abril}, {Comoretto}, {Garcia-Reinaldos}, {Lock}, {Mercier}, {Altmann}, {Andrae}, {Astraatmadja}, {Bellas-Velidis}, {Benson}, {Berthier}, {Blomme}, {Busso}, {Carry}, {Cellino},
  {Clementini}, {Cowell}, {Creevey}, {Cuypers}, {Davidson}, {De Ridder}, {de Torres}, {Delchambre}, {Dell'Oro}, {Ducourant}, {Fr{\'e}mat}, {Garc{\'\i}a-Torres}, {Gosset}, {Halbwachs}, {Hambly}, {Harrison}, {Hauser}, {Hestroffer}, {Hodgkin}, {Huckle}, {Hutton}, {Jasniewicz}, {Jordan}, {Kontizas}, {Korn}, {Lanzafame}, {Manteiga}, {Moitinho}, {Muinonen}, {Osinde}, {Pancino}, {Pauwels}, {Petit}, {Recio-Blanco}, {Robin}, {Sarro}, {Siopis}, {Smith}, {Smith}, {Sozzetti}, {Thuillot}, {van Reeven}, {Viala}, {Abbas}, {Abreu Aramburu}, {Accart}, {Aguado}, {Allan}, {Allasia}, {Altavilla}, {{\'A}lvarez}, {Alves}, {Anderson}, {Andrei}, {Anglada Varela}, {Antiche}, {Antoja}, {Ant{\'o}n}, {Arcay}, {Atzei}, {Ayache}, {Bach}, {Baker}, {Balaguer-N{\'u}{\~n}ez}, {Barache}, {Barata}, {Barbier}, {Barblan}, {Baroni}, {Barrado y Navascu{\'e}s}, {Barros}, {Barstow}, {Becciani}, {Bellazzini}, {Bellei}, {Bello Garc{\'\i}a}, {Belokurov}, {Bendjoya}, {Berihuete}, {Bianchi}, {Bienaym{\'e}}, {Billebaud}, {Blagorodnova}, {Blanco-Cuaresma},
  {Boch}, {Bombrun}, {Borrachero}, {Bouquillon}, {Bourda}, {Bouy}, {Bragaglia}, {Breddels}, {Brouillet}, {Br{\"u}semeister}, {Bucciarelli}, {Budnik}, {Burgess}, {Burgon}, {Burlacu}, {Busonero}, {Buzzi}, {Caffau}, {Cambras}, {Campbell}, {Cancelliere}, {Cantat-Gaudin}, {Carlucci}, {Carrasco}, {Castellani}, {Charlot}, {Charnas}, {Charvet}, {Chassat}, {Chiavassa}, {Clotet}, {Cocozza}, {Collins}, {Collins}, {Costigan}, {Crifo}, {Cross}, {Crosta}, {Crowley}, {Dafonte}, {Damerdji}, {Dapergolas}, {David}, {David}, {De Cat}, {de Felice}, {de Laverny}, {De Luise}, {De March}, {de Martino}, {de Souza}, {Debosscher}, {del Pozo}, {Delbo}, {Delgado}, {Delgado}, {di Marco}, {Di Matteo}, {Diakite}, {Distefano}, {Dolding}, {Dos Anjos}, {Drazinos}, {Dur{\'a}n}, {Dzigan}, {Ecale}, {Edvardsson}, {Enke}, {Erdmann}, {Escolar}, {Espina}, {Evans}, {Eynard Bontemps}, {Fabre}, {Fabrizio}, {Faigler}, {Falc{\~a}o}, {Farr{\`a}s Casas}, {Faye}, {Federici}, {Fedorets}, {Fern{\'a}ndez-Hern{\'a}ndez}, {Fernique}, {Fienga}, {Figueras},
  {Filippi}, {Findeisen}, {Fonti}, {Fouesneau}, {Fraile}, {Fraser}, {Fuchs}, {Furnell}, {Gai}, {Galleti}, {Galluccio}, {Garabato}, {Garc{\'\i}a-Sedano}, {Gar{\'e}}, {Garofalo}, {Garralda}, {Gavras}, {Gerssen}, {Geyer}, {Gilmore}, {Girona}, {Giuffrida}, {Gomes}, {Gonz{\'a}lez-Marcos}, {Gonz{\'a}lez-N{\'u}{\~n}ez}, {Gonz{\'a}lez-Vidal}, {Granvik}, {Guerrier}, {Guillout}, {Guiraud}, {G{\'u}rpide}, {Guti{\'e}rrez-S{\'a}nchez}, {Guy}, {Haigron}, {Hatzidimitriou}, {Haywood}, {Heiter}, {Helmi}, {Hobbs}, {Hofmann}, {Holl}, {Holland}, {Hunt}, {Hypki}, {Icardi}, {Irwin}, {Jevardat de Fombelle}, {Jofr{\'e}}, {Jonker}, {Jorissen}, {Julbe}, {Karampelas}, {Kochoska}, {Kohley}, {Kolenberg}, {Kontizas}, {Koposov}, {Kordopatis}, {Koubsky}, {Kowalczyk}, {Krone-Martins}, {Kudryashova}, {Kull}, {Bachchan}, {Lacoste-Seris}, {Lanza}, {Lavigne}, {Le Poncin-Lafitte}, {Lebreton}, {Lebzelter}, {Leccia}, {Leclerc}, {Lecoeur-Taibi}, {Lemaitre}, {Lenhardt}, {Leroux}, {Liao}, {Licata}, {Lindstr{\o}m}, {Lister}, {Livanou}, {Lobel},
  {L{\"o}ffler}, {L{\'o}pez}, {Lopez-Lozano}, {Lorenz}, {Loureiro}, {MacDonald}, {Magalh{\~a}es Fernandes}, {Managau}, {Mann}, {Mantelet}, {Marchal}, {Marchant}, {Marconi}, {Marie}, {Marinoni}, {Marrese}, {Marschalk{\'o}}, {Marshall}, {Mart{\'\i}n-Fleitas}, {Martino}, {Mary}, {Matijevi{\v{c}}}, {Mazeh}, {McMillan}, {Messina}, {Mestre}, {Michalik}, {Millar}, {Miranda}, {Molina}, {Molinaro}, {Molinaro}, {Moln{\'a}r}, {Moniez}, {Montegriffo}, {Monteiro}, {Mor}, {Mora}, {Morbidelli}, {Morel}, {Morgenthaler}, {Morley}, {Morris}, {Mulone}, {Muraveva}, {Musella}, {Narbonne}, {Nelemans}, {Nicastro}, {Noval}, {Ord{\'e}novic}, {Ordieres-Mer{\'e}}, {Osborne}, {Pagani}, {Pagano}, {Pailler}, {Palacin}, {Palaversa}, {Parsons}, {Paulsen}, {Pecoraro}, {Pedrosa}, {Pentik{\"a}inen}, {Pereira}, {Pichon}, {Piersimoni}, {Pineau}, {Plachy}, {Plum}, {Poujoulet}, {Pr{\v{s}}a}, {Pulone}, {Ragaini}, {Rago}, {Rambaux}, {Ramos-Lerate}, {Ranalli}, {Rauw}, {Read}, {Regibo}, {Renk}, {Reyl{\'e}}, {Ribeiro}, {Rimoldini}, {Ripepi}, {Riva},
  {Rixon}, {Roelens}, {Romero-G{\'o}mez}, {Rowell}, {Royer}, {Rudolph}, {Ruiz-Dern}, {Sadowski}, {Sagrist{\`a} Sell{\'e}s}, {Sahlmann}, {Salgado}, {Salguero}, {Sarasso}, {Savietto}, {Schnorhk}, {Schultheis}, {Sciacca}, {Segol}, {Segovia}, {Segransan}, {Serpell}, {Shih}, {Smareglia}, {Smart}, {Smith}, {Solano}, {Solitro}, {Sordo}, {Soria Nieto}, {Souchay}, {Spagna}, {Spoto}, {Stampa}, {Steele}, {Steidelm{\"u}ller}, {Stephenson}, {Stoev}, {Suess}, {S{\"u}veges}, {Surdej}, {Szabados}, {Szegedi-Elek}, {Tapiador}, {Taris}, {Tauran}, {Taylor}, {Teixeira}, {Terrett}, {Tingley}, {Trager}, {Turon}, {Ulla}, {Utrilla}, {Valentini}, {van Elteren}, {Van Hemelryck}, {van Leeuwen}, {Varadi}, {Vecchiato}, {Veljanoski}, {Via}, {Vicente}, {Vogt}, {Voss}, {Votruba}, {Voutsinas}, {Walmsley}, {Weiler}, {Weingrill}, {Werner}, {Wevers}, {Whitehead}, {Wyrzykowski}, {Yoldas}, {{\v{Z}}erjal}, {Zucker}, {Zurbach}, {Zwitter}, {Alecu}, {Allen}, {Allende Prieto}, {Amorim}, {Anglada-Escud{\'e}}, {Arsenijevic}, {Azaz}, {Balm}, {Beck},
  {Bernstein}, {Bigot}, {Bijaoui}, {Blasco}, {Bonfigli}, {Bono}, {Boudreault}, {Bressan}, {Brown}, {Brunet}, {Bunclark}, {Buonanno}, {Butkevich}, {Carret}, {Carrion}, {Chemin}, {Ch{\'e}reau}, {Corcione}, {Darmigny}, {de Boer}, {de Teodoro}, {de Zeeuw}, {Delle Luche}, {Domingues}, {Dubath}, {Fodor}, {Fr{\'e}zouls}, {Fries}, {Fustes}, {Fyfe}, {Gallardo}, {Gallegos}, {Gardiol}, {Gebran}, {Gomboc}, {G{\'o}mez}, {Grux}, {Gueguen}, {Heyrovsky}, {Hoar}, {Iannicola}, {Isasi Parache}, {Janotto}, {Joliet}, {Jonckheere}, {Keil}, {Kim}, {Klagyivik}, {Klar}, {Knude}, {Kochukhov}, {Kolka}, {Kos}, {Kutka}, {Lainey}, {LeBouquin}, {Liu}, {Loreggia}, {Makarov}, {Marseille}, {Martayan}, {Martinez-Rubi}, {Massart}, {Meynadier}, {Mignot}, {Munari}, {Nguyen}, {Nordlander}, {Ocvirk}, {O'Flaherty}, {Olias Sanz}, {Ortiz}, {Osorio}, {Oszkiewicz}, {Ouzounis}, {Palmer}, {Park}, {Pasquato}, {Peltzer}, {Peralta}, {P{\'e}turaud}, {Pieniluoma}, {Pigozzi}, {Poels}, {Prat}, {Prod'homme}, {Raison}, {Rebordao}, {Risquez}, {Rocca-Volmerange},
  {Rosen}, {Ruiz-Fuertes}, {Russo}, {Sembay}, {Serraller Vizcaino}, {Short}, {Siebert}, {Silva}, {Sinachopoulos}, {Slezak}, {Soffel}, {Sosnowska}, {Strai{\v{z}}ys}, {ter Linden}, {Terrell}, {Theil}, {Tiede}, {Troisi}, {Tsalmantza}, {Tur}, {Vaccari}, {Vachier}, {Valles}, {Van Hamme}, {Veltz}, {Virtanen}, {Wallut}, {Wichmann}, {Wilkinson}, {Ziaeepour}, \& {Zschocke}}]{gaia16}
{Gaia Collaboration}, {Prusti}, T., {de Bruijne}, J.~H.~J., {et~al.} 2016{\natexlab{a}}, \aap, 595, A1, \dodoi{10.1051/0004-6361/201629272}

\bibitem[{{Gaia Collaboration} {et~al.}(2016{\natexlab{b}}){Gaia Collaboration}, {Prusti}, {de Bruijne}, {Brown}, {Vallenari}, {Babusiaux}, {Bailer-Jones}, {Bastian}, {Biermann}, {Evans}, \& et~al.}]{gaia+16}
---. 2016{\natexlab{b}}, \aap, 595, A1, \dodoi{10.1051/0004-6361/201629272}

\bibitem[{{Gaia Collaboration} {et~al.}(2021){Gaia Collaboration}, {Brown}, {Vallenari}, {Prusti}, {de Bruijne}, {Babusiaux}, {Biermann}, {Creevey}, {Evans}, {Eyer}, {Hutton}, {Jansen}, {Jordi}, {Klioner}, {Lammers}, {Lindegren}, {Luri}, {Mignard}, {Panem}, {Pourbaix}, {Randich}, {Sartoretti}, {Soubiran}, {Walton}, {Arenou}, {Bailer-Jones}, {Bastian}, {Cropper}, {Drimmel}, {Katz}, {Lattanzi}, {van Leeuwen}, {Bakker}, {Cacciari}, {Casta{\~n}eda}, {De Angeli}, {Ducourant}, {Fabricius}, {Fouesneau}, {Fr{\'e}mat}, {Guerra}, {Guerrier}, {Guiraud}, {Jean-Antoine Piccolo}, {Masana}, {Messineo}, {Mowlavi}, {Nicolas}, {Nienartowicz}, {Pailler}, {Panuzzo}, {Riclet}, {Roux}, {Seabroke}, {Sordo}, {Tanga}, {Th{\'e}venin}, {Gracia-Abril}, {Portell}, {Teyssier}, {Altmann}, {Andrae}, {Bellas-Velidis}, {Benson}, {Berthier}, {Blomme}, {Brugaletta}, {Burgess}, {Busso}, {Carry}, {Cellino}, {Cheek}, {Clementini}, {Damerdji}, {Davidson}, {Delchambre}, {Dell'Oro}, {Fern{\'a}ndez-Hern{\'a}ndez}, {Galluccio}, {Garc{\'\i}a-Lario},
  {Garcia-Reinaldos}, {Gonz{\'a}lez-N{\'u}{\~n}ez}, {Gosset}, {Haigron}, {Halbwachs}, {Hambly}, {Harrison}, {Hatzidimitriou}, {Heiter}, {Hern{\'a}ndez}, {Hestroffer}, {Hodgkin}, {Holl}, {Jan{\ss}en}, {Jevardat de Fombelle}, {Jordan}, {Krone-Martins}, {Lanzafame}, {L{\"o}ffler}, {Lorca}, {Manteiga}, {Marchal}, {Marrese}, {Moitinho}, {Mora}, {Muinonen}, {Osborne}, {Pancino}, {Pauwels}, {Petit}, {Recio-Blanco}, {Richards}, {Riello}, {Rimoldini}, {Robin}, {Roegiers}, {Rybizki}, {Sarro}, {Siopis}, {Smith}, {Sozzetti}, {Ulla}, {Utrilla}, {van Leeuwen}, {van Reeven}, {Abbas}, {Abreu Aramburu}, {Accart}, {Aerts}, {Aguado}, {Ajaj}, {Altavilla}, {{\'A}lvarez}, {{\'A}lvarez Cid-Fuentes}, {Alves}, {Anderson}, {Anglada Varela}, {Antoja}, {Audard}, {Baines}, {Baker}, {Balaguer-N{\'u}{\~n}ez}, {Balbinot}, {Balog}, {Barache}, {Barbato}, {Barros}, {Barstow}, {Bartolom{\'e}}, {Bassilana}, {Bauchet}, {Baudesson-Stella}, {Becciani}, {Bellazzini}, {Bernet}, {Bertone}, {Bianchi}, {Blanco-Cuaresma}, {Boch}, {Bombrun}, {Bossini},
  {Bouquillon}, {Bragaglia}, {Bramante}, {Breedt}, {Bressan}, {Brouillet}, {Bucciarelli}, {Burlacu}, {Busonero}, {Butkevich}, {Buzzi}, {Caffau}, {Cancelliere}, {C{\'a}novas}, {Cantat-Gaudin}, {Carballo}, {Carlucci}, {Carnerero}, {Carrasco}, {Casamiquela}, {Castellani}, {Castro-Ginard}, {Castro Sampol}, {Chaoul}, {Charlot}, {Chemin}, {Chiavassa}, {Cioni}, {Comoretto}, {Cooper}, {Cornez}, {Cowell}, {Crifo}, {Crosta}, {Crowley}, {Dafonte}, {Dapergolas}, {David}, {David}, {de Laverny}, {De Luise}, {De March}, {De Ridder}, {de Souza}, {de Teodoro}, {de Torres}, {del Peloso}, {del Pozo}, {Delbo}, {Delgado}, {Delgado}, {Delisle}, {Di Matteo}, {Diakite}, {Diener}, {Distefano}, {Dolding}, {Eappachen}, {Edvardsson}, {Enke}, {Esquej}, {Fabre}, {Fabrizio}, {Faigler}, {Fedorets}, {Fernique}, {Fienga}, {Figueras}, {Fouron}, {Fragkoudi}, {Fraile}, {Franke}, {Gai}, {Garabato}, {Garcia-Gutierrez}, {Garc{\'\i}a-Torres}, {Garofalo}, {Gavras}, {Gerlach}, {Geyer}, {Giacobbe}, {Gilmore}, {Girona}, {Giuffrida}, {Gomel}, {Gomez},
  {Gonzalez-Santamaria}, {Gonz{\'a}lez-Vidal}, {Granvik}, {Guti{\'e}rrez-S{\'a}nchez}, {Guy}, {Hauser}, {Haywood}, {Helmi}, {Hidalgo}, {Hilger}, {H{\l}adczuk}, {Hobbs}, {Holland}, {Huckle}, {Jasniewicz}, {Jonker}, {Juaristi Campillo}, {Julbe}, {Karbevska}, {Kervella}, {Khanna}, {Kochoska}, {Kontizas}, {Kordopatis}, {Korn}, {Kostrzewa-Rutkowska}, {Kruszy{\'n}ska}, {Lambert}, {Lanza}, {Lasne}, {Le Campion}, {Le Fustec}, {Lebreton}, {Lebzelter}, {Leccia}, {Leclerc}, {Lecoeur-Taibi}, {Liao}, {Licata}, {Lindstr{\o}m}, {Lister}, {Livanou}, {Lobel}, {Madrero Pardo}, {Managau}, {Mann}, {Marchant}, {Marconi}, {Marcos Santos}, {Marinoni}, {Marocco}, {Marshall}, {Martin Polo}, {Mart{\'\i}n-Fleitas}, {Masip}, {Massari}, {Mastrobuono-Battisti}, {Mazeh}, {McMillan}, {Messina}, {Michalik}, {Millar}, {Mints}, {Molina}, {Molinaro}, {Moln{\'a}r}, {Montegriffo}, {Mor}, {Morbidelli}, {Morel}, {Morris}, {Mulone}, {Munoz}, {Muraveva}, {Murphy}, {Musella}, {Noval}, {Ord{\'e}novic}, {Orr{\`u}}, {Osinde}, {Pagani}, {Pagano},
  {Palaversa}, {Palicio}, {Panahi}, {Pawlak}, {Pe{\~n}alosa Esteller}, {Penttil{\"a}}, {Piersimoni}, {Pineau}, {Plachy}, {Plum}, {Poggio}, {Poretti}, {Poujoulet}, {Pr{\v{s}}a}, {Pulone}, {Racero}, {Ragaini}, {Rainer}, {Raiteri}, {Rambaux}, {Ramos}, {Ramos-Lerate}, {Re Fiorentin}, {Regibo}, {Reyl{\'e}}, {Ripepi}, {Riva}, {Rixon}, {Robichon}, {Robin}, {Roelens}, {Rohrbasser}, {Romero-G{\'o}mez}, {Rowell}, {Royer}, {Rybicki}, {Sadowski}, {Sagrist{\`a} Sell{\'e}s}, {Sahlmann}, {Salgado}, {Salguero}, {Samaras}, {Sanchez Gimenez}, {Sanna}, {Santove{\~n}a}, {Sarasso}, {Schultheis}, {Sciacca}, {Segol}, {Segovia}, {S{\'e}gransan}, {Semeux}, {Shahaf}, {Siddiqui}, {Siebert}, {Siltala}, {Slezak}, {Smart}, {Solano}, {Solitro}, {Souami}, {Souchay}, {Spagna}, {Spoto}, {Steele}, {Steidelm{\"u}ller}, {Stephenson}, {S{\"u}veges}, {Szabados}, {Szegedi-Elek}, {Taris}, {Tauran}, {Taylor}, {Teixeira}, {Thuillot}, {Tonello}, {Torra}, {Torra}, {Turon}, {Unger}, {Vaillant}, {van Dillen}, {Vanel}, {Vecchiato}, {Viala}, {Vicente},
  {Voutsinas}, {Weiler}, {Wevers}, {Wyrzykowski}, {Yoldas}, {Yvard}, {Zhao}, {Zorec}, {Zucker}, {Zurbach}, \& {Zwitter}}]{gaia21}
{Gaia Collaboration}, {Brown}, A.~G.~A., {Vallenari}, A., {et~al.} 2021, \aap, 649, A1, \dodoi{10.1051/0004-6361/202039657}

\bibitem[{{Garling} {et~al.}(2018){Garling}, {Willman}, {Sand}, {Hargis}, {Crnojevi{\'c}}, {Bechtol}, {Carlin}, {Strader}, {Zou}, {Zhou}, {Nie}, {Zhang}, {Zhou}, \& {Peng}}]{garling18}
{Garling}, C., {Willman}, B., {Sand}, D.~J., {et~al.} 2018, \apj, 852, 44, \dodoi{10.3847/1538-4357/aa9bf1}

\bibitem[{{Geha} {et~al.}(2009){Geha}, {Willman}, {Simon}, {Strigari}, {Kirby}, {Law}, \& {Strader}}]{geha09}
{Geha}, M., {Willman}, B., {Simon}, J.~D., {et~al.} 2009, \apj, 692, 1464, \dodoi{10.1088/0004-637X/692/2/1464}

\bibitem[{{Gibbons} {et~al.}(2014){Gibbons}, {Belokurov}, \& {Evans}}]{Gibbons:2014}
{Gibbons}, S.~L.~J., {Belokurov}, V., \& {Evans}, N.~W. 2014, \mnras, 445, 3788, \dodoi{10.1093/mnras/stu1986}

\bibitem[{{Green}(2018)}]{green18}
{Green}, G. 2018, The Journal of Open Source Software, 3, 695, \dodoi{10.21105/joss.00695}

\bibitem[{{Gregory} {et~al.}(2020){Gregory}, {Collins}, {Erkal}, {Tollerud}, {Delorme}, {Hill}, {Sand}, {Strader}, \& {Willman}}]{gregory20}
{Gregory}, A.~L., {Collins}, M. L.~M., {Erkal}, D., {et~al.} 2020, \mnras, 496, 1092, \dodoi{10.1093/mnras/staa1553}

\bibitem[{{Guerra} {et~al.}(2023){Guerra}, {Geha}, \& {Strigari}}]{guerra23}
{Guerra}, J., {Geha}, M., \& {Strigari}, L.~E. 2023, \apj, 943, 121, \dodoi{10.3847/1538-4357/aca8a5}

\bibitem[{{Gullikson} {et~al.}(2014){Gullikson}, {Dodson-Robinson}, \& {Kraus}}]{gullikson14}
{Gullikson}, K., {Dodson-Robinson}, S., \& {Kraus}, A. 2014, \aj, 148, 53, \dodoi{10.1088/0004-6256/148/3/53}

\bibitem[{{Gunn} {et~al.}(2006){Gunn}, {Siegmund}, {Mannery}, {Owen}, {Hull}, {Leger}, {Carey}, {Knapp}, {York}, {Boroski}, {Kent}, {Lupton}, {Rockosi}, {Evans}, {Waddell}, {Anderson}, {Annis}, {Barentine}, {Bartoszek}, {Bastian}, {Bracker}, {Brewington}, {Briegel}, {Brinkmann}, {Brown}, {Carr}, {Czarapata}, {Drennan}, {Dombeck}, {Federwitz}, {Gillespie}, {Gonzales}, {Hansen}, {Harvanek}, {Hayes}, {Jordan}, {Kinney}, {Klaene}, {Kleinman}, {Kron}, {Kresinski}, {Lee}, {Limmongkol}, {Lindenmeyer}, {Long}, {Loomis}, {McGehee}, {Mantsch}, {Neilsen}, {Neswold}, {Newman}, {Nitta}, {Peoples}, {Pier}, {Prieto}, {Prosapio}, {Rivetta}, {Schneider}, {Snedden}, \& {Wang}}]{sdss+06}
{Gunn}, J.~E., {Siegmund}, W.~A., {Mannery}, E.~J., {et~al.} 2006, \aj, 131, 2332, \dodoi{10.1086/500975}

\bibitem[{{Heiger} {et~al.}(2023){Heiger}, {Li}, {Pace}, {Simon}, {Ji}, {Chiti}, {Bom}, {Carballo-Bello}, {Carlin}, {Cerny}, {Choi}, {Drlica-Wagner}, {James}, {Mart{\'\i}nez-V{\'a}zquez}, {Medina}, {Mutlu-Pakdil}, {Navabi}, {No{\"e}l}, {Sakowska}, \& {Stringfellow}}]{heiger23}
{Heiger}, M.~E., {Li}, T.~S., {Pace}, A.~B., {et~al.} 2023, arXiv e-prints, arXiv:2308.08602, \dodoi{10.48550/arXiv.2308.08602}

\bibitem[{{Hernquist}(1990)}]{Hernquist:1990}
{Hernquist}, L. 1990, \apj, 356, 359, \dodoi{10.1086/168845}

\bibitem[{{Homma} {et~al.}(2016){Homma}, {Chiba}, {Okamoto}, {Komiyama}, {Tanaka}, {Tanaka}, {Ishigaki}, {Akiyama}, {Arimoto}, {Garmilla}, {Lupton}, {Strauss}, {Furusawa}, {Miyazaki}, {Murayama}, {Nishizawa}, {Takada}, {Usuda}, \& {Wang}}]{homma16}
{Homma}, D., {Chiba}, M., {Okamoto}, S., {et~al.} 2016, \apj, 832, 21, \dodoi{10.3847/0004-637X/832/1/21}

\bibitem[{{Homma} {et~al.}(2018){Homma}, {Chiba}, {Okamoto}, {Komiyama}, {Tanaka}, {Tanaka}, {Ishigaki}, {Hayashi}, {Arimoto}, {Garmilla}, {Lupton}, {Strauss}, {Miyazaki}, {Wang}, \& {Murayama}}]{homma18}
---. 2018, \pasj, 70, S18, \dodoi{10.1093/pasj/psx050}

\bibitem[{{Homma} {et~al.}(2019){Homma}, {Chiba}, {Komiyama}, {Tanaka}, {Okamoto}, {Tanaka}, {Ishigaki}, {Hayashi}, {Arimoto}, {Carlsten}, {Lupton}, {Strauss}, {Miyazaki}, {Torrealba}, {Wang}, \& {Murayama}}]{homma19}
{Homma}, D., {Chiba}, M., {Komiyama}, Y., {et~al.} 2019, \pasj, 71, 94, \dodoi{10.1093/pasj/psz076}

\bibitem[{{Hunter}(2007)}]{hunter07}
{Hunter}, J.~D. 2007, Computing in Science and Engineering, 9, 90, \dodoi{10.1109/MCSE.2007.55}

\bibitem[{{Jardel} \& {Gebhardt}(2013)}]{jardel13}
{Jardel}, J.~R., \& {Gebhardt}, K. 2013, \apjl, 775, L30, \dodoi{10.1088/2041-8205/775/1/L30}

\bibitem[{{Jensen} {et~al.}(2024){Jensen}, {Hayes}, {Sestito}, {McConnachie}, {Waller}, {Smith}, {Navarro}, \& {Venn}}]{jensen24}
{Jensen}, J., {Hayes}, C.~R., {Sestito}, F., {et~al.} 2024, \mnras, 527, 4209, \dodoi{10.1093/mnras/stad3322}

\bibitem[{{Jethwa} {et~al.}(2018){Jethwa}, {Erkal}, \& {Belokurov}}]{jethwa18}
{Jethwa}, P., {Erkal}, D., \& {Belokurov}, V. 2018, \mnras, 473, 2060, \dodoi{10.1093/mnras/stx2330}

\bibitem[{{Ji} {et~al.}(2021){Ji}, {Koposov}, {Li}, {Erkal}, {Pace}, {Simon}, {Belokurov}, {Cullinane}, {Da Costa}, {Kuehn}, {Lewis}, {Mackey}, {Shipp}, {Simpson}, {Zucker}, {Hansen}, {Bland-Hawthorn}, \& {S5 Collaboration}}]{ji21}
{Ji}, A.~P., {Koposov}, S.~E., {Li}, T.~S., {et~al.} 2021, \apj, 921, 32, \dodoi{10.3847/1538-4357/ac1869}

\bibitem[{{Jones} {et~al.}(2001){Jones}, {Oliphant}, {Peterson}, \& {et al.}}]{jones01}
{Jones}, E., {Oliphant}, T., {Peterson}, P., \& {et al.} 2001, {SciPy}: Open source scientific tools for {Python},  online.
\newblock \url{http://www.scipy.org/}

\bibitem[{{Kelson}(2003)}]{k+03}
{Kelson}, D.~D. 2003, \pasp, 115, 688, \dodoi{10.1086/375502}

\bibitem[{{Kim} \& {Jerjen}(2015)}]{kim15}
{Kim}, D., \& {Jerjen}, H. 2015, \apjl, 808, L39, \dodoi{10.1088/2041-8205/808/2/L39}

\bibitem[{{Kim} {et~al.}(2018){Kim}, {Peter}, \& {Hargis}}]{kim18}
{Kim}, S.~Y., {Peter}, A. H.~G., \& {Hargis}, J.~R. 2018, \prl, 121, 211302, \dodoi{10.1103/PhysRevLett.121.211302}

\bibitem[{{Koch} {et~al.}(2014){Koch}, {Hansen}, {Feltzing}, \& {Wilkinson}}]{koch14}
{Koch}, A., {Hansen}, T., {Feltzing}, S., \& {Wilkinson}, M.~I. 2014, \apj, 780, 91, \dodoi{10.1088/0004-637X/780/1/91}

\bibitem[{{Koch} {et~al.}(2008){Koch}, {McWilliam}, {Grebel}, {Zucker}, \& {Belokurov}}]{koch08}
{Koch}, A., {McWilliam}, A., {Grebel}, E.~K., {Zucker}, D.~B., \& {Belokurov}, V. 2008, \apjl, 688, L13, \dodoi{10.1086/595001}

\bibitem[{{Koposov} {et~al.}(2015){Koposov}, {Belokurov}, {Torrealba}, \& {Evans}}]{koposov15}
{Koposov}, S.~E., {Belokurov}, V., {Torrealba}, G., \& {Evans}, N.~W. 2015, \apj, 805, 130, \dodoi{10.1088/0004-637X/805/2/130}

\bibitem[{{Koposov} {et~al.}(2023){Koposov}, {Erkal}, {Li}, {Da Costa}, {Cullinane}, {Ji}, {Kuehn}, {Lewis}, {Pace}, {Shipp}, {Zucker}, {Bland-Hawthorn}, {Lilleengen}, \& {Martell}}]{Koposov:2023}
{Koposov}, S.~E., {Erkal}, D., {Li}, T.~S., {et~al.} 2023, \mnras, \dodoi{10.1093/mnras/stad551}

\bibitem[{{K{\"u}pper} {et~al.}(2017){K{\"u}pper}, {Johnston}, {Mieske}, {Collins}, \& {Tollerud}}]{kupper17}
{K{\"u}pper}, A. H.~W., {Johnston}, K.~V., {Mieske}, S., {Collins}, M. L.~M., \& {Tollerud}, E.~J. 2017, \apj, 834, 112, \dodoi{10.3847/1538-4357/834/2/112}

\bibitem[{{K{\"u}pper} {et~al.}(2012){K{\"u}pper}, {Lane}, \& {Heggie}}]{kupper12}
{K{\"u}pper}, A. H.~W., {Lane}, R.~R., \& {Heggie}, D.~C. 2012, \mnras, 420, 2700, \dodoi{10.1111/j.1365-2966.2011.20242.x}

\bibitem[{{Laevens} {et~al.}(2015{\natexlab{a}}){Laevens}, {Martin}, {Ibata}, {Rix}, {Bernard}, {Bell}, {Sesar}, {Ferguson}, {Schlafly}, {Slater}, {Burgett}, {Chambers}, {Flewelling}, {Hodapp}, {Kaiser}, {Kudritzki}, {Lupton}, {Magnier}, {Metcalfe}, {Morgan}, {Price}, {Tonry}, {Wainscoat}, \& {Waters}}]{laevens15a}
{Laevens}, B. P.~M., {Martin}, N.~F., {Ibata}, R.~A., {et~al.} 2015{\natexlab{a}}, \apjl, 802, L18, \dodoi{10.1088/2041-8205/802/2/L18}

\bibitem[{{Laevens} {et~al.}(2015{\natexlab{b}}){Laevens}, {Martin}, {Bernard}, {Schlafly}, {Sesar}, {Rix}, {Bell}, {Ferguson}, {Slater}, {Sweeney}, {Wyse}, {Huxor}, {Burgett}, {Chambers}, {Draper}, {Hodapp}, {Kaiser}, {Magnier}, {Metcalfe}, {Tonry}, {Wainscoat}, \& {Waters}}]{laevens15b}
{Laevens}, B. P.~M., {Martin}, N.~F., {Bernard}, E.~J., {et~al.} 2015{\natexlab{b}}, \apj, 813, 44, \dodoi{10.1088/0004-637X/813/1/44}

\bibitem[{{Li} {et~al.}(2017){Li}, {Simon}, {Drlica-Wagner}, {Bechtol}, {Wang}, {Garc{\'\i}a-Bellido}, {Frieman}, {Marshall}, {James}, {Strigari}, {Pace}, {Balbinot}, {Zhang}, {Abbott}, {Allam}, {Benoit-L{\'e}vy}, {Bernstein}, {Bertin}, {Brooks}, {Burke}, {Carnero Rosell}, {Carrasco Kind}, {Carretero}, {Cunha}, {D'Andrea}, {da Costa}, {DePoy}, {Desai}, {Diehl}, {Eifler}, {Flaugher}, {Goldstein}, {Gruen}, {Gruendl}, {Gschwend}, {Gutierrez}, {Krause}, {Kuehn}, {Lin}, {Maia}, {March}, {Menanteau}, {Miquel}, {Plazas}, {Romer}, {Sanchez}, {Santiago}, {Schubnell}, {Sevilla-Noarbe}, {Smith}, {Sobreira}, {Suchyta}, {Tarle}, {Thomas}, {Tucker}, {Walker}, {Wechsler}, {Wester}, {Yanny}, \& {DES Collaboration}}]{lsd+17}
{Li}, T.~S., {Simon}, J.~D., {Drlica-Wagner}, A., {et~al.} 2017, \apj, 838, 8, \dodoi{10.3847/1538-4357/aa6113}

\bibitem[{{Li} {et~al.}(2018){Li}, {Simon}, {Kuehn}, {Pace}, {Erkal}, {Bechtol}, {Yanny}, {Drlica-Wagner}, {Marshall}, {Lidman}, {Balbinot}, {Carollo}, {Jenkins}, {Mart{\'\i}nez-V{\'a}zquez}, {Shipp}, {Stringer}, {Vivas}, {Walker}, {Wechsler}, {Abdalla}, {Allam}, {Annis}, {Avila}, {Bertin}, {Brooks}, {Buckley-Geer}, {Burke}, {Carnero Rosell}, {Carrasco Kind}, {Carretero}, {Cunha}, {D'Andrea}, {da Costa}, {Davis}, {De Vicente}, {Doel}, {Eifler}, {Evrard}, {Flaugher}, {Frieman}, {Garc{\'\i}a-Bellido}, {Gaztanaga}, {Gerdes}, {Gruen}, {Gruendl}, {Gschwend}, {Gutierrez}, {Hartley}, {Hollowood}, {Honscheid}, {James}, {Krause}, {Maia}, {March}, {Menanteau}, {Miquel}, {Plazas}, {Sanchez}, {Santiago}, {Scarpine}, {Schindler}, {Schubnell}, {Sevilla-Noarbe}, {Smith}, {Smith}, {Soares-Santos}, {Sobreira}, {Suchyta}, {Swanson}, {Tarle}, {Tucker}, \& {DES Collaboration}}]{li18}
{Li}, T.~S., {Simon}, J.~D., {Kuehn}, K., {et~al.} 2018, \apj, 866, 22, \dodoi{10.3847/1538-4357/aadf91}

\bibitem[{{Li} {et~al.}(2021){Li}, {Koposov}, {Erkal}, {Ji}, {Shipp}, {Pace}, {Hilmi}, {Kuehn}, {Lewis}, {Mackey}, {Simpson}, {Wan}, {Zucker}, {Bland-Hawthorn}, {Cullinane}, {Da Costa}, {Drlica-Wagner}, {Hattori}, {Martell}, {Sharma}, \& {S5 Collaboration}}]{li21}
{Li}, T.~S., {Koposov}, S.~E., {Erkal}, D., {et~al.} 2021, \apj, 911, 149, \dodoi{10.3847/1538-4357/abeb18}

\bibitem[{{Longeard} {et~al.}(2022){Longeard}, {Jablonka}, {Arentsen}, {Thomas}, {Aguado}, {Carlberg}, {Lucchesi}, {Malhan}, {Martin}, {McConnachie}, {Navarro}, {S{\'a}nchez-Janssen}, {Sestito}, {Starkenburg}, \& {Yuan}}]{longeard22}
{Longeard}, N., {Jablonka}, P., {Arentsen}, A., {et~al.} 2022, \mnras, 516, 2348, \dodoi{10.1093/mnras/stac1827}

\bibitem[{{Longeard} {et~al.}(2023){Longeard}, {Jablonka}, {Battaglia}, {Malhan}, {Martin}, {S{\'a}nchez-Janssen}, {Sestito}, {Starkenburg}, \& {Venn}}]{longeard23}
{Longeard}, N., {Jablonka}, P., {Battaglia}, G., {et~al.} 2023, \mnras, 525, 3086, \dodoi{10.1093/mnras/stad2227}

\bibitem[{{Marshall} {et~al.}(2008){Marshall}, {Burles}, {Thompson}, {Shectman}, {Bigelow}, {Burley}, {Birk}, {Estrada}, {Jones}, {Smith}, {Kowal}, {Castillo}, {Storts}, \& {Ortiz}}]{mbt+08}
{Marshall}, J.~L., {Burles}, S., {Thompson}, I.~B., {et~al.} 2008, in \procspie, Vol. 7014, Ground-based and Airborne Instrumentation for Astronomy II, 701454, \dodoi{10.1117/12.789972}

\bibitem[{{Martin} {et~al.}(2008){Martin}, {de Jong}, \& {Rix}}]{martin08}
{Martin}, N.~F., {de Jong}, J. T.~A., \& {Rix}, H.-W. 2008, \apj, 684, 1075, \dodoi{10.1086/590336}

\bibitem[{{Martin} \& {Jin}(2010)}]{martin10}
{Martin}, N.~F., \& {Jin}, S. 2010, \apj, 721, 1333, \dodoi{10.1088/0004-637X/721/2/1333}

\bibitem[{{Mart{\'\i}nez-Garc{\'\i}a} {et~al.}(2021){Mart{\'\i}nez-Garc{\'\i}a}, {del Pino}, {Aparicio}, {van der Marel}, \& {Watkins}}]{martinez-garcia21}
{Mart{\'\i}nez-Garc{\'\i}a}, A.~M., {del Pino}, A., {Aparicio}, A., {van der Marel}, R.~P., \& {Watkins}, L.~L. 2021, \mnras, 505, 5884, \dodoi{10.1093/mnras/stab1568}

\bibitem[{{Mart{\'\i}nez-Garc{\'\i}a} {et~al.}(2023){Mart{\'\i}nez-Garc{\'\i}a}, {del Pino}, {{\L}okas}, {van der Marel}, \& {Aparicio}}]{martinez-garcia23}
{Mart{\'\i}nez-Garc{\'\i}a}, A.~M., {del Pino}, A., {{\L}okas}, E.~L., {van der Marel}, R.~P., \& {Aparicio}, A. 2023, \mnras, 526, 3589, \dodoi{10.1093/mnras/stad2941}

\bibitem[{{Massari} {et~al.}(2020){Massari}, {Helmi}, {Mucciarelli}, {Sales}, {Spina}, \& {Tolstoy}}]{massari20}
{Massari}, D., {Helmi}, A., {Mucciarelli}, A., {et~al.} 2020, \aap, 633, A36, \dodoi{10.1051/0004-6361/201935613}

\bibitem[{{Mau} {et~al.}(2020){Mau}, {Cerny}, {Pace}, {Choi}, {Drlica-Wagner}, {Santana-Silva}, {Riley}, {Erkal}, {Stringfellow}, {Adam{\'o}w}, {Carlin}, {Gruendl}, {Hernandez-Lang}, {Kuropatkin}, {Li}, {Mart{\'\i}nez-V{\'a}zquez}, {Morganson}, {Mutlu-Pakdil}, {Neilsen}, {Nidever}, {Olsen}, {Sand}, {Tollerud}, {Tucker}, {Yanny}, {Zenteno}, {Allam}, {Barkhouse}, {Bechtol}, {Bell}, {Balaji}, {Crnojevi{\'c}}, {Esteves}, {Ferguson}, {Gallart}, {Hughes}, {James}, {Jethwa}, {Johnson}, {Kuehn}, {Majewski}, {Mao}, {Massana}, {McNanna}, {Monachesi}, {Nadler}, {No{\"e}l}, {Palmese}, {Paz-Chinchon}, {Pieres}, {Sanchez}, {Shipp}, {Simon}, {Soares-Santos}, {Tavangar}, {van der Marel}, {Vivas}, {Walker}, \& {Wechsler}}]{mau20}
{Mau}, S., {Cerny}, W., {Pace}, A.~B., {et~al.} 2020, \apj, 890, 136, \dodoi{10.3847/1538-4357/ab6c67}

\bibitem[{{Mau} {et~al.}(2022){Mau}, {Nadler}, {Wechsler}, {Drlica-Wagner}, {Bechtol}, {Green}, {Huterer}, {Li}, {Mao}, {Mart{\'\i}nez-V{\'a}zquez}, {McNanna}, {Mutlu-Pakdil}, {Pace}, {Peter}, {Riley}, {Strigari}, {Wang}, {Aguena}, {Allam}, {Annis}, {Bacon}, {Bertin}, {Bocquet}, {Brooks}, {Burke}, {Carnero Rosell}, {Carrasco Kind}, {Carretero}, {Costanzi}, {Crocce}, {Pereira}, {Davis}, {De Vicente}, {Desai}, {Doel}, {Ferrero}, {Flaugher}, {Frieman}, {Garc{\'\i}a-Bellido}, {Gatti}, {Giannini}, {Gruen}, {Gruendl}, {Gschwend}, {Gutierrez}, {Hinton}, {Hollowood}, {Honscheid}, {James}, {Kuehn}, {Lahav}, {Maia}, {Marshall}, {Miquel}, {Mohr}, {Morgan}, {Ogando}, {Paz-Chinch{\'o}n}, {Pieres}, {Rodriguez-Monroy}, {Sanchez}, {Scarpine}, {Serrano}, {Sevilla-Noarbe}, {Suchyta}, {Tarle}, {To}, {Tucker}, {Weller}, \& {DES Collaboration}}]{mau22}
{Mau}, S., {Nadler}, E.~O., {Wechsler}, R.~H., {et~al.} 2022, \apj, 932, 128, \dodoi{10.3847/1538-4357/ac6e65}

\bibitem[{{McConnachie} \& {Venn}(2020)}]{mv+20}
{McConnachie}, A.~W., \& {Venn}, K.~A. 2020, Research Notes of the American Astronomical Society, 4, 229, \dodoi{10.3847/2515-5172/abd18b}

\bibitem[{{McMillan}(2017)}]{McMillan:2017}
{McMillan}, P.~J. 2017, \mnras, 465, 76, \dodoi{10.1093/mnras/stw2759}

\bibitem[{{Miyamoto} \& {Nagai}(1975)}]{Miyamoto:1975}
{Miyamoto}, M., \& {Nagai}, R. 1975, Publications of the Astronomical Society of Japan, 27, 533

\bibitem[{{Moore}(1994)}]{moore94}
{Moore}, B. 1994, \nat, 370, 629, \dodoi{10.1038/370629a0}

\bibitem[{{Mu{\~n}oz} {et~al.}(2018){Mu{\~n}oz}, {C{\^o}t{\'e}}, {Santana}, {Geha}, {Simon}, {Oyarz{\'u}n}, {Stetson}, \& {Djorgovski}}]{munoz18}
{Mu{\~n}oz}, R.~R., {C{\^o}t{\'e}}, P., {Santana}, F.~A., {et~al.} 2018, \apj, 860, 66, \dodoi{10.3847/1538-4357/aac16b}

\bibitem[{{Musella} {et~al.}(2012){Musella}, {Ripepi}, {Marconi}, {Clementini}, {Dall'Ora}, {Scowcroft}, {Moretti}, {Di Fabrizio}, {Greco}, {Coppola}, {Bersier}, {Catelan}, {Grado}, {Limatola}, {Smith}, \& {Kinemuchi}}]{musella12}
{Musella}, I., {Ripepi}, V., {Marconi}, M., {et~al.} 2012, \apj, 756, 121, \dodoi{10.1088/0004-637X/756/2/121}

\bibitem[{{Mutlu-Pakdil} {et~al.}(2020){Mutlu-Pakdil}, {Sand}, {Crnojevi{\'c}}, {Olszewski}, {Zaritsky}, {Strader}, {Collins}, {Seth}, \& {Willman}}]{mutlu-pakdil20}
{Mutlu-Pakdil}, B., {Sand}, D.~J., {Crnojevi{\'c}}, D., {et~al.} 2020, \apj, 902, 106, \dodoi{10.3847/1538-4357/abb40b}

\bibitem[{{Nadler} {et~al.}(2019){Nadler}, {Gluscevic}, {Boddy}, \& {Wechsler}}]{nadler19}
{Nadler}, E.~O., {Gluscevic}, V., {Boddy}, K.~K., \& {Wechsler}, R.~H. 2019, \apjl, 878, L32, \dodoi{10.3847/2041-8213/ab1eb2}

\bibitem[{{Navarro} {et~al.}(1997){Navarro}, {Frenk}, \& {White}}]{navarro97}
{Navarro}, J.~F., {Frenk}, C.~S., \& {White}, S. D.~M. 1997, \apj, 490, 493, \dodoi{10.1086/304888}

\bibitem[{{Newman} {et~al.}(2013){Newman}, {Cooper}, {Davis}, {Faber}, {Coil}, {Guhathakurta}, {Koo}, {Phillips}, {Conroy}, {Dutton}, {Finkbeiner}, {Gerke}, {Rosario}, {Weiner}, {Willmer}, {Yan}, {Harker}, {Kassin}, {Konidaris}, {Lai}, {Madgwick}, {Noeske}, {Wirth}, {Connolly}, {Kaiser}, {Kirby}, {Lemaux}, {Lin}, {Lotz}, {Luppino}, {Marinoni}, {Matthews}, {Metevier}, \& {Schiavon}}]{ncd+13}
{Newman}, J.~A., {Cooper}, M.~C., {Davis}, M., {et~al.} 2013, \apjs, 208, 5, \dodoi{10.1088/0067-0049/208/1/5}

\bibitem[{{Oemler} {et~al.}(2017){Oemler}, {Clardy}, {Kelson}, {Walth}, \& {Villanueva}}]{ock+17}
{Oemler}, A., {Clardy}, K., {Kelson}, D., {Walth}, G., \& {Villanueva}, E. 2017, {COSMOS: Carnegie Observatories System for MultiObject Spectroscopy}.
\newblock \doeprint{1705.001}

\bibitem[{{Pace} {et~al.}(2022){Pace}, {Erkal}, \& {Li}}]{pace22}
{Pace}, A.~B., {Erkal}, D., \& {Li}, T.~S. 2022, arXiv e-prints, arXiv:2205.05699.
\newblock \doarXiv{2205.05699}

\bibitem[{{Pe{\~n}arrubia} {et~al.}(2008){Pe{\~n}arrubia}, {Navarro}, \& {McConnachie}}]{penarrubia08}
{Pe{\~n}arrubia}, J., {Navarro}, J.~F., \& {McConnachie}, A.~W. 2008, Astronomische Nachrichten, 329, 934, \dodoi{10.1002/asna.200811078}

\bibitem[{{Plummer}(1911)}]{Plummer:1911}
{Plummer}, H.~C. 1911, \mnras, 71, 460

\bibitem[{{Prochaska} {et~al.}(2020){Prochaska}, {Hennawi}, {Westfall}, {Cooke}, {Wang}, {Hsyu}, {Davies}, {Farina}, \& {Pelliccia}}]{prochaska20}
{Prochaska}, J., {Hennawi}, J., {Westfall}, K., {et~al.} 2020, The Journal of Open Source Software, 5, 2308, \dodoi{10.21105/joss.02308}

\bibitem[{{Read} {et~al.}(2019){Read}, {Walker}, \& {Steger}}]{read19}
{Read}, J.~I., {Walker}, M.~G., \& {Steger}, P. 2019, \mnras, 484, 1401, \dodoi{10.1093/mnras/sty3404}

\bibitem[{{Robin} {et~al.}(2003){Robin}, {Reyl{\'e}}, {Derri{\`e}re}, \& {Picaud}}]{robin03}
{Robin}, A.~C., {Reyl{\'e}}, C., {Derri{\`e}re}, S., \& {Picaud}, S. 2003, \aap, 409, 523, \dodoi{10.1051/0004-6361:20031117}

\bibitem[{{Roderick} {et~al.}(2015){Roderick}, {Jerjen}, {Mackey}, \& {Da Costa}}]{roderick15}
{Roderick}, T.~A., {Jerjen}, H., {Mackey}, A.~D., \& {Da Costa}, G.~S. 2015, \apj, 804, 134, \dodoi{10.1088/0004-637X/804/2/134}

\bibitem[{{Rutledge} {et~al.}(1997){Rutledge}, {Hesser}, {Stetson}, {Mateo}, {Simard}, {Bolte}, {Friel}, \& {Copin}}]{rhs+97}
{Rutledge}, G.~A., {Hesser}, J.~E., {Stetson}, P.~B., {et~al.} 1997, \pasp, 109, 883, \dodoi{10.1086/133958}

\bibitem[{{Sales} {et~al.}(2022){Sales}, {Wetzel}, \& {Fattahi}}]{sales22}
{Sales}, L.~V., {Wetzel}, A., \& {Fattahi}, A. 2022, Nature Astronomy, 6, 897, \dodoi{10.1038/s41550-022-01689-w}

\bibitem[{{Sand} {et~al.}(2009){Sand}, {Olszewski}, {Willman}, {Zaritsky}, {Seth}, {Harris}, {Piatek}, \& {Saha}}]{sand09}
{Sand}, D.~J., {Olszewski}, E.~W., {Willman}, B., {et~al.} 2009, \apj, 704, 898, \dodoi{10.1088/0004-637X/704/2/898}

\bibitem[{{Schlafly} \& {Finkbeiner}(2011)}]{schlafly11}
{Schlafly}, E.~F., \& {Finkbeiner}, D.~P. 2011, \apj, 737, 103, \dodoi{10.1088/0004-637X/737/2/103}

\bibitem[{{Shipp} {et~al.}(2019){Shipp}, {Li}, {Pace}, {Erkal}, {Drlica-Wagner}, {Yanny}, {Belokurov}, {Wester}, {Koposov}, {Kuehn}, {Lewis}, {Simpson}, {Wan}, {Zucker}, {Martell}, {Wang}, \& {S5 Collaboration}}]{shipp19}
{Shipp}, N., {Li}, T.~S., {Pace}, A.~B., {et~al.} 2019, \apj, 885, 3, \dodoi{10.3847/1538-4357/ab44bf}

\bibitem[{{Shipp} {et~al.}(2021){Shipp}, {Erkal}, {Drlica-Wagner}, {Li}, {Pace}, {Koposov}, {Cullinane}, {Da Costa}, {Ji}, {Kuehn}, {Lewis}, {Mackey}, {Simpson}, {Wan}, {Zucker}, {Bland-Hawthorn}, {Ferguson}, {Lilleengen}, \& {Lilleengen}}]{Shipp:2021}
{Shipp}, N., {Erkal}, D., {Drlica-Wagner}, A., {et~al.} 2021, \apj, 923, 149, \dodoi{10.3847/1538-4357/ac2e93}

\bibitem[{{Shipp} {et~al.}(2023){Shipp}, {Panithanpaisal}, {Necib}, {Sanderson}, {Erkal}, {Li}, {Santistevan}, {Wetzel}, {Cullinane}, {Ji}, {Koposov}, {Kuehn}, {Lewis}, {Pace}, {Zucker}, {Bland-Hawthorn}, {Cunningham}, {Kim}, {Lilleengen}, {Moreno}, {Sharma}, {S Collaboration}, \& {FIRE Collaboration}}]{shipp23}
{Shipp}, N., {Panithanpaisal}, N., {Necib}, L., {et~al.} 2023, \apj, 949, 44, \dodoi{10.3847/1538-4357/acc582}

\bibitem[{{Silverman} {et~al.}(2023){Silverman}, {Bullock}, {Kaplinghat}, {Robles}, \& {Valli}}]{silverman23}
{Silverman}, M., {Bullock}, J.~S., {Kaplinghat}, M., {Robles}, V.~H., \& {Valli}, M. 2023, \mnras, 518, 2418, \dodoi{10.1093/mnras/stac3232}

\bibitem[{{Simon}(2019)}]{simon19}
{Simon}, J.~D. 2019, \araa, 57, 375, \dodoi{10.1146/annurev-astro-091918-104453}

\bibitem[{{Simon} \& {Geha}(2007)}]{simon07}
{Simon}, J.~D., \& {Geha}, M. 2007, \apj, 670, 313, \dodoi{10.1086/521816}

\bibitem[{{Simon} {et~al.}(2017){Simon}, {Li}, {Drlica-Wagner}, {Bechtol}, {Marshall}, {James}, {Wang}, {Strigari}, {Balbinot}, {Kuehn}, {Walker}, {Abbott}, {Allam}, {Annis}, {Benoit-L{\'e}vy}, {Brooks}, {Buckley-Geer}, {Burke}, {Carnero Rosell}, {Carrasco Kind}, {Carretero}, {Cunha}, {D'Andrea}, {da Costa}, {DePoy}, {Desai}, {Doel}, {Fernandez}, {Flaugher}, {Frieman}, {Garc{\'\i}a-Bellido}, {Gaztanaga}, {Goldstein}, {Gruen}, {Gutierrez}, {Kuropatkin}, {Maia}, {Martini}, {Menanteau}, {Miller}, {Miquel}, {Neilsen}, {Nord}, {Ogando}, {Plazas}, {Romer}, {Rykoff}, {Sanchez}, {Santiago}, {Scarpine}, {Schubnell}, {Sevilla-Noarbe}, {Smith}, {Sobreira}, {Suchyta}, {Swanson}, {Tarle}, {Whiteway}, {Yanny}, \& {DES Collaboration}}]{sld+17}
{Simon}, J.~D., {Li}, T.~S., {Drlica-Wagner}, A., {et~al.} 2017, \apj, 838, 11, \dodoi{10.3847/1538-4357/aa5be7}

\bibitem[{{Sohn} {et~al.}(2007){Sohn}, {Majewski}, {Mu{\~n}oz}, {Kunkel}, {Johnston}, {Ostheimer}, {Guhathakurta}, {Patterson}, {Siegel}, \& {Cooper}}]{sohn07}
{Sohn}, S.~T., {Majewski}, S.~R., {Mu{\~n}oz}, R.~R., {et~al.} 2007, \apj, 663, 960, \dodoi{10.1086/518302}

\bibitem[{{Starkenburg} {et~al.}(2010){Starkenburg}, {Hill}, {Tolstoy}, {Gonz{\'a}lez Hern{\'a}ndez}, {Irwin}, {Helmi}, {Battaglia}, {Jablonka}, {Tafelmeyer}, {Shetrone}, {Venn}, \& {de Boer}}]{starkenburg10}
{Starkenburg}, E., {Hill}, V., {Tolstoy}, E., {et~al.} 2010, \aap, 513, A34, \dodoi{10.1051/0004-6361/200913759}

\bibitem[{{Strigari}(2018)}]{strigari18}
{Strigari}, L.~E. 2018, Reports on Progress in Physics, 81, 056901, \dodoi{10.1088/1361-6633/aaae16}

\bibitem[{{Strigari} {et~al.}(2010){Strigari}, {Frenk}, \& {White}}]{strigari10}
{Strigari}, L.~E., {Frenk}, C.~S., \& {White}, S. D.~M. 2010, \mnras, 408, 2364, \dodoi{10.1111/j.1365-2966.2010.17287.x}

\bibitem[{{Tau} {et~al.}(2023){Tau}, {Vivas}, \& {Mart{\'\i}nez-V{\'a}zquez}}]{tau23}
{Tau}, E.~A., {Vivas}, A.~K., \& {Mart{\'\i}nez-V{\'a}zquez}, C.~E. 2023, arXiv e-prints, arXiv:2312.07279, \dodoi{10.48550/arXiv.2312.07279}

\bibitem[{{Tonry} \& {Davis}(1979)}]{tonry79}
{Tonry}, J., \& {Davis}, M. 1979, \aj, 84, 1511, \dodoi{10.1086/112569}

\bibitem[{{Tonry} {et~al.}(2012{\natexlab{a}}){Tonry}, {Stubbs}, {Lykke}, {Doherty}, {Shivvers}, {Burgett}, {Chambers}, {Hodapp}, {Kaiser}, {Kudritzki}, {Magnier}, {Morgan}, {Price}, \& {Wainscoat}}]{tsl+12}
{Tonry}, J.~L., {Stubbs}, C.~W., {Lykke}, K.~R., {et~al.} 2012{\natexlab{a}}, \apj, 750, 99, \dodoi{10.1088/0004-637X/750/2/99}

\bibitem[{{Tonry} {et~al.}(2012{\natexlab{b}}){Tonry}, {Stubbs}, {Lykke}, {Doherty}, {Shivvers}, {Burgett}, {Chambers}, {Hodapp}, {Kaiser}, {Kudritzki}, {Magnier}, {Morgan}, {Price}, \& {Wainscoat}}]{tonry12}
---. 2012{\natexlab{b}}, \apj, 750, 99, \dodoi{10.1088/0004-637X/750/2/99}

\bibitem[{{Torrealba} {et~al.}(2018){Torrealba}, {Belokurov}, {Koposov}, {Bechtol}, {Drlica-Wagner}, {Olsen}, {Vivas}, {Yanny}, {Jethwa}, {Walker}, {Li}, {Allam}, {Conn}, {Gallart}, {Gruendl}, {James}, {Johnson}, {Kuehn}, {Kuropatkin}, {Martin}, {Martinez-Delgado}, {Nidever}, {No{\"e}l}, {Simon}, {Stringfellow}, \& {Tucker}}]{torrealba18}
{Torrealba}, G., {Belokurov}, V., {Koposov}, S.~E., {et~al.} 2018, \mnras, 475, 5085, \dodoi{10.1093/mnras/sty170}

\bibitem[{{van der Walt} {et~al.}(2011){van der Walt}, {Colbert}, \& {Varoquaux}}]{vanderwalt11}
{van der Walt}, S., {Colbert}, S.~C., \& {Varoquaux}, G. 2011, Computing in Science Engineering, 13, 22, \dodoi{10.1109/MCSE.2011.37}

\bibitem[{{Walker} \& {Pe{\~n}arrubia}(2011)}]{walker11}
{Walker}, M.~G., \& {Pe{\~n}arrubia}, J. 2011, \apj, 742, 20, \dodoi{10.1088/0004-637X/742/1/20}

\bibitem[{{Willman} {et~al.}(2005{\natexlab{a}}){Willman}, {Dalcanton}, {Martinez-Delgado}, {West}, {Blanton}, {Hogg}, {Barentine}, {Brewington}, {Harvanek}, {Kleinman}, {Krzesinski}, {Long}, {Neilsen}, {Nitta}, \& {Snedden}}]{willman05a}
{Willman}, B., {Dalcanton}, J.~J., {Martinez-Delgado}, D., {et~al.} 2005{\natexlab{a}}, \apjl, 626, L85, \dodoi{10.1086/431760}

\bibitem[{{Willman} {et~al.}(2005{\natexlab{b}}){Willman}, {Blanton}, {West}, {Dalcanton}, {Hogg}, {Schneider}, {Wherry}, {Yanny}, \& {Brinkmann}}]{willman05b}
{Willman}, B., {Blanton}, M.~R., {West}, A.~A., {et~al.} 2005{\natexlab{b}}, \aj, 129, 2692, \dodoi{10.1086/430214}

\bibitem[{{Wolf} {et~al.}(2010){Wolf}, {Martinez}, {Bullock}, {Kaplinghat}, {Geha}, {Mu{\~n}oz}, {Simon}, \& {Avedo}}]{wolf10}
{Wolf}, J., {Martinez}, G.~D., {Bullock}, J.~S., {et~al.} 2010, \mnras, 406, 1220, \dodoi{10.1111/j.1365-2966.2010.16753.x}

\bibitem[{{York} {et~al.}(2000){York}, {Adelman}, {Anderson}, {Anderson}, {Annis}, {Bahcall}, {Bakken}, {Barkhouser}, {Bastian}, {Berman}, {Boroski}, {Bracker}, {Briegel}, {Briggs}, {Brinkmann}, {Brunner}, {Burles}, {Carey}, {Carr}, {Castander}, {Chen}, {Colestock}, {Connolly}, {Crocker}, {Csabai}, {Czarapata}, {Davis}, {Doi}, {Dombeck}, {Eisenstein}, {Ellman}, {Elms}, {Evans}, {Fan}, {Federwitz}, {Fiscelli}, {Friedman}, {Frieman}, {Fukugita}, {Gillespie}, {Gunn}, {Gurbani}, {de Haas}, {Haldeman}, {Harris}, {Hayes}, {Heckman}, {Hennessy}, {Hindsley}, {Holm}, {Holmgren}, {Huang}, {Hull}, {Husby}, {Ichikawa}, {Ichikawa}, {Ivezi{\'c}}, {Kent}, {Kim}, {Kinney}, {Klaene}, {Kleinman}, {Kleinman}, {Knapp}, {Korienek}, {Kron}, {Kunszt}, {Lamb}, {Lee}, {Leger}, {Limmongkol}, {Lindenmeyer}, {Long}, {Loomis}, {Loveday}, {Lucinio}, {Lupton}, {MacKinnon}, {Mannery}, {Mantsch}, {Margon}, {McGehee}, {McKay}, {Meiksin}, {Merelli}, {Monet}, {Munn}, {Narayanan}, {Nash}, {Neilsen}, {Neswold}, {Newberg}, {Nichol}, {Nicinski},
  {Nonino}, {Okada}, {Okamura}, {Ostriker}, {Owen}, {Pauls}, {Peoples}, {Peterson}, {Petravick}, {Pier}, {Pope}, {Pordes}, {Prosapio}, {Rechenmacher}, {Quinn}, {Richards}, {Richmond}, {Rivetta}, {Rockosi}, {Ruthmansdorfer}, {Sandford}, {Schlegel}, {Schneider}, {Sekiguchi}, {Sergey}, {Shimasaku}, {Siegmund}, {Smee}, {Smith}, {Snedden}, {Stone}, {Stoughton}, {Strauss}, {Stubbs}, {SubbaRao}, {Szalay}, {Szapudi}, {Szokoly}, {Thakar}, {Tremonti}, {Tucker}, {Uomoto}, {Vanden Berk}, {Vogeley}, {Waddell}, {Wang}, {Watanabe}, {Weinberg}, {Yanny}, {Yasuda}, \& {SDSS Collaboration}}]{york20}
{York}, D.~G., {Adelman}, J., {Anderson}, John~E., J., {et~al.} 2000, \aj, 120, 1579, \dodoi{10.1086/301513}

\end{thebibliography}





\end{CJK*}
\end{document}